\documentclass[12pt]{article}
\usepackage[doublespacing]{setspace}
\usepackage[pdftex]{graphicx}
\usepackage[pdftex]{color}
\usepackage{url}
\usepackage{epsfig,longtable}
\usepackage{makeidx}
\usepackage{varioref}
\usepackage{xr}
\usepackage{amsmath}
\usepackage{amsthm}
\usepackage{amssymb}
\usepackage{natbib}
\usepackage{setspace}
\usepackage{subfig}
\usepackage{multirow}
\usepackage{dsfont}
\usepackage{enumitem}
\usepackage{booktabs}
\usepackage{array}
\usepackage{caption}

\newtheorem{theorem}{Theorem}[section]

\newtheorem{remark}{Remark}[section]

\numberwithin{equation}{section}

\newcommand{\Ical}{\mathcal{I}}

\newcommand{\Ccal}{\mathcal{C}}
\newcommand{\Acal}{\mathcal{A}}

\def\ibinom#1#2{{\binom{#1}{#2}}}

\input{tcilatex}

\begin{document}

{\sffamily\bfseries\Large  Consistent distribution-free $K$-sample and independence tests for univariate random variables}

\noindent%

\textsf{ Ruth Heller, Yair Heller, Shachar Kaufman, Barak Brill and Malka Gorfine}%
\footnote{\textit{Address for correspondence:} Department of
Statistics and Operations Research, Tel-Aviv university, Tel-Aviv,
Israel.  \ \textsf{E-mail:} ruheller@post.tau.ac.il. \ The work of Ruth Heller and Barak Brill was supported by grant no. 2012896 from the
Israel Science
Foundation (ISF), and the work of Shachar Kaufman was supported by a fellowship from the Edmond J. Safra Center for Bioinformatics at Tel-Aviv university. }\\

\textsf{Abstract. \
A popular approach for testing if two univariate random variables are statistically independent consists of partitioning the sample space into bins, and evaluating a test statistic on the binned data. The partition size matters, and the optimal partition size is data dependent. While for detecting simple relationships coarse partitions may be best, for detecting complex relationships a great gain in power can be achieved by considering finer partitions.
We suggest novel consistent distribution-free tests that are based on summation or maximization aggregation of scores over all partitions of a fixed size. We show that our test statistics based on summation can serve as good estimators of the mutual information. Moreover, we suggest regularized tests that aggregate over all partition sizes, and prove those are consistent too. We provide polynomial-time algorithms, which are critical for computing the suggested test statistics efficiently.
We show that the power of the regularized tests is excellent compared to existing
tests, and almost as powerful as the tests based on the optimal (yet unknown
in practice) partition size, in simulations as well as on a real data example.
\medskip }

\noindent

\textsf{Keywords: \ Bivariate distribution;  Nonparametric test; Statistical independence; Mutual information; Two-sample test.}\newpage

\setcounter{page}{1}

\newpage
\section{Introduction}
\label{sec:intro}

Testing if two univariate random variables $X$ and $Y$ are independent of one another, given a random paired sample ${(x_i, y_i)}_{i=1}^N$, is a fundamental and extensively studied problem in statistics. Classical methods have focused on testing linear (Pearson's correlation coefficient) or monotone (Spearman's $\rho$, Kendall's $\tau$) univariate dependence, and  have little power to detect non-monotone relationships. 
 Recently, there has been great interest in developing methods to capture complex dependencies between pairs of random variables. This interest follows from the recognition that in many modern applications, dependencies of interest may not be of simple forms,  and therefore the classical methods cannot capture them. Moreover, in modern applications, thousands of variables are measured simultaneously, thus making it impossible to view the scatter-plots of all the potential pairs of variables of interest. For example, \cite{steuer2002mutual} searched for pairs of genes that are co-dependent, among thousands of genes measured, using the estimated mutual information as a dependence measure. 
   \cite{Reshef11} searched for any type of relationship, not just linear or monotone, in large datasets from global health, gene expression, major-league baseball, and the human gut microbiota.    They proposed a novel criterion  which generated much interest but has also been criticised for lacking power by \citet{Simon11} and \citet{Gorfine11}, and for other theoretical grounds by \citet{kinney2013equitability}.

A special important case is when $X$ is categorical. In this case, the 
 problem reduces to that of testing the equality of distributions, usually referred to as the $K$-sample problem (where $K$ is the number of categories $X$ can have).  \cite{Jiang14} searched for genes that are differentially expressed across two conditions (i.e., the 2-sample problem),  using a novel test that has higher power over traditional methods such as Kolmogorov--Smirnov tests \citep{Darling57}. 

For modern applications, where all types of dependency are of interest, a desirable property for a test of independence is consistency against any alternative. A consistent test will have power increasing to one as the sample size increases, for any type of dependency between $X$ and $Y$. Recently, several  consistent tests of independence between univariate or multivariate random variables were proposed.  \cite{Szekely07} suggested the distance covariance test statistic, that is  the distance (in weighted $L_2$ norm) of the joint empirical characteristic function from the product of marginal characteristic functions.  \cite{Gretton08} and \cite{Gretton10} considered a family of kernel based methods, and \cite{Sejdinovic12} elegantly showed that the test of \cite{Szekely07} is a kernel based test with a particular choice of kernel. \cite{Heller12} suggested a permutation test for independence between two random vectors $X$ and $Y$, which uses as test statistics the sum over all pairs of points $(i,j), i\neq j$, of a score that tests for association between the two binary variables $I\{d(x_i, X)\leq d(x_i,x_j) \}$ and $I\{d(y_i, Y)\leq d(y_i,y_j) \}$, where $I(\cdot)$ is the indicator function and $d(\cdot,\cdot)$ is a distance metric, on the remaining $N-2$ sample points.  \citet{Gretton10} also considered dividing the underlying space into partitions that are refined with increasing sample size.  For the $K$-sample problem, \cite{Szekely05} suggested the energy test. This test was also proposed by \cite{Baringhaus04} and mentioned in \cite{Sejdinovic12} to be related to the MMD test proposed in \cite{Gretton06} and \cite{Gretton12b}.  \citet{Harouchi08} adopted the kernel approach of \cite{Gretton06} and incorporated the covariance into the test statistic by using the kernel Fisher discriminant.

The tests in the previous paragraph are not distribution-free, i.e., the null distribution of the test statistics depends on the marginal distributions of  $X$ and $Y$. Therefore, the computational burden of applying these tests to a large family of hypotheses may be great. For example, the yeast gene expression dataset from \citet{hughes2000functional} contained $N=300$ expression levels for each of $6,325$ {\em Saccharomyces cerevisiae} genes. In order to test each pair of genes for co-expression, it is necessary to account for multiplicity of $M = 2\times 10^7$ hypotheses. For the permutation tests of \cite{Heller12} and \cite{Szekely07}, the number of permutations required for deriving a $p$-value that is below $0.05/M$ is therefore of the order of $10^{10}$. Since these test statistics are relatively costly to compute for each hypothesis, e.g., $O(N^2)$ in \citet{Szekely07}, and $O(N^2 \log N)$ in \citet{Heller12}, applying them to the family of $M = 2\times 10^7$ hypotheses is computationally very challenging, even with sophisticated resampling approaches such as that of 
~\citet{yu2011efficient}.
Distribution-free tests have the advantage over non-distribution-free tests, that quantiles of the null distribution of the test statistic can be tabulated once per sample size, and repeating the test on new data for the same sample size  will not require recomputing the null distribution.  Therefore, the computational cost is only that of computing the test statistic for each of the hypotheses. 

We note that for univariate random variables \citet{Szekely09} considered using the ranks of each random variable instead of the actual values in the test of distance covariance~\citep{Szekely07}, resulting in a distribution-free test. Similarly, for the test of \citet{Heller12} replacing data with ranks results in a distribution-free test.
An earlier work by \cite{Feuerverger93} defined another test based on  the empirical characteristic functions for univariate random variables. The test statistic of \cite{Feuerverger93}
was based on a different distance metric of the joint empirical characteristic function from the product of marginal characteristic functions than that of \cite{Szekely07}. Moreover, in Feuerverger's test the $X$'s and $Y$'s are first replaced by their normal scores, where the normal scores of the $X$'s depend on the data only through their ranks among the $X$'s, and similarly the normal scores of the $Y$'s depend on the data only through their ranks among the $Y$'s, thus making this test distribution-free.  

A popular approach for developing distribution-free tests of independence considers partitioning the sample space, and evaluating a test statistic on the binned data. A detailed review of distribution-free partition-based tests is provided in Section \ref{subsec-revdfind} for the independence problem, and in Section \ref{subsec-revksample} for the $K$-sample problem. In \ref{subsec-goal} we describe our goals and the outline of the present paper.
\subsection{Review of distribution-free tests of independence based on sample space partitions}\label{subsec-revdfind}
For detecting any type of dependence between $X$ and $Y$, the null hypothesis states that $X$ and $Y$ are independent, $H_0: F_{XY} = F_X F_Y,$ where the joint distribution of $({X},{Y})$ is denoted by $F_{XY}$, and the marginal distributions of ${ X}$ and ${ Y}$, respectively, are denoted by $F_X$ and $F_Y$. The alternative is that $X$ and $Y$ are dependent, $H_1: F_{XY} \neq F_X F_Y.$

\begin{figure}[htbp]
  \centering
  \begin{tabular}{cc}
  \includegraphics[page=1, width=0.35\textwidth, trim=0.6in 0.6in 0.6in 0.6in, clip]{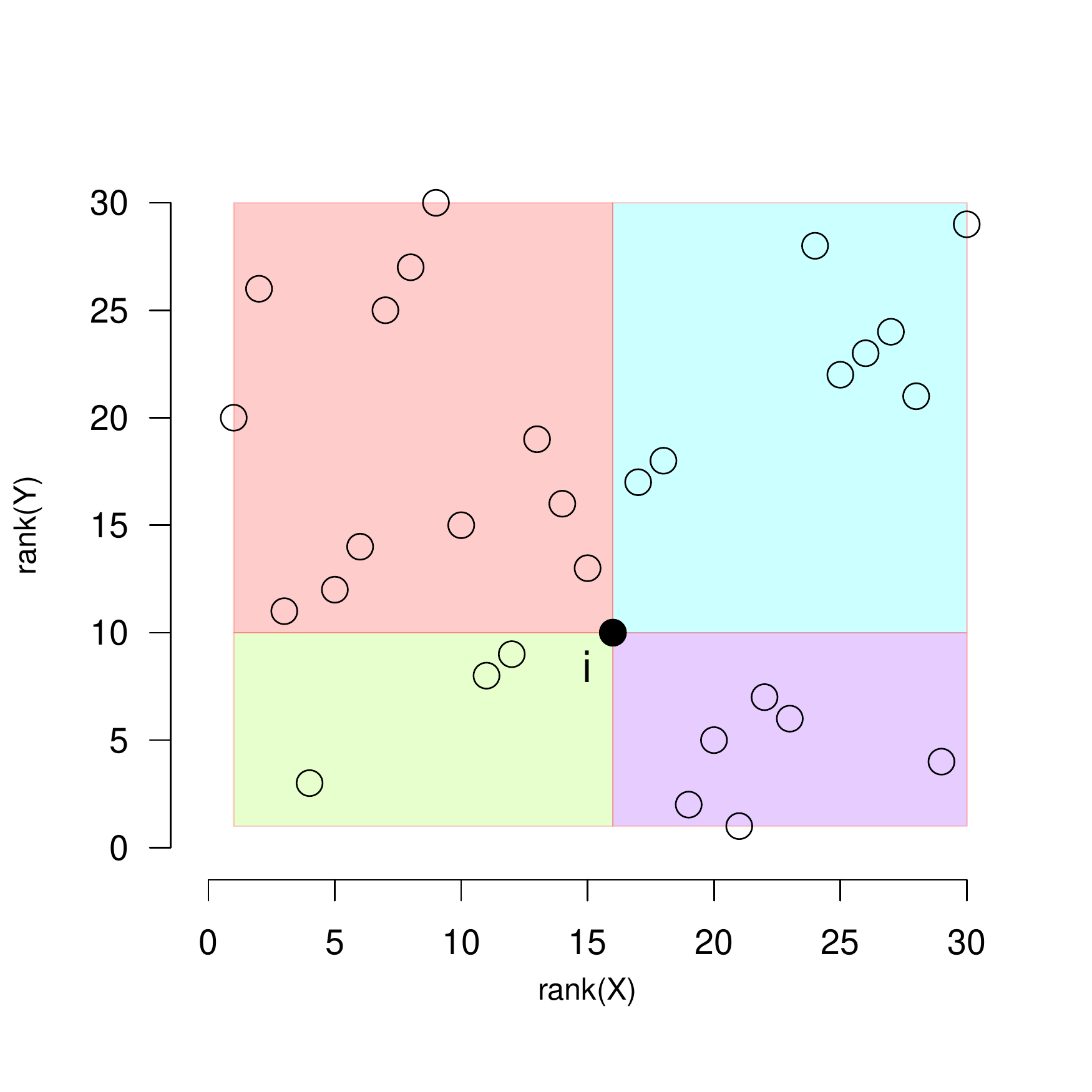} &
  \includegraphics[page=1, width=0.35\textwidth, trim=0.6in 0.6in 0.6in 0.6in, clip]{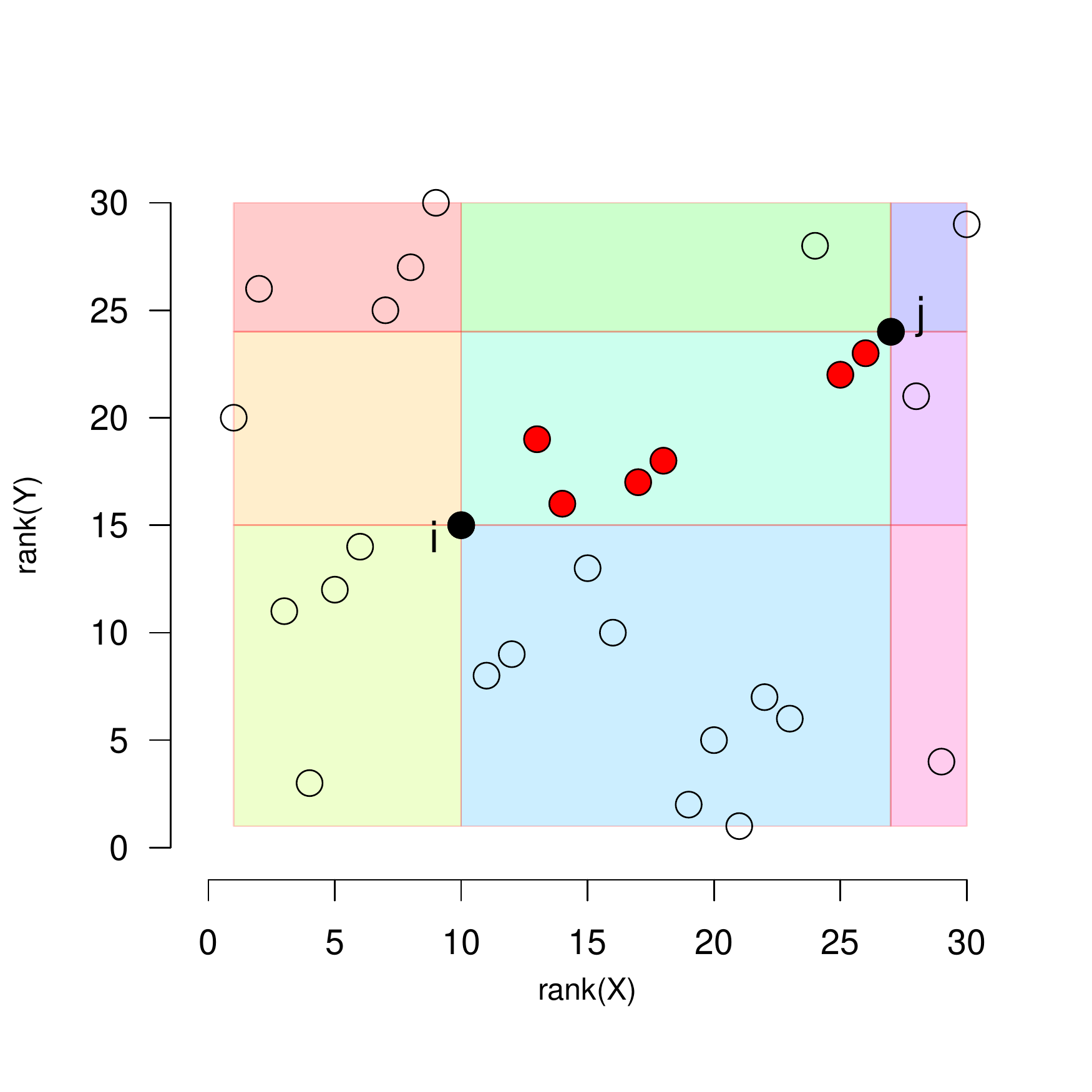} \\
  \includegraphics[page=1, width=0.35\textwidth, trim=0.6in 0.6in 0.6in 0.6in, clip]{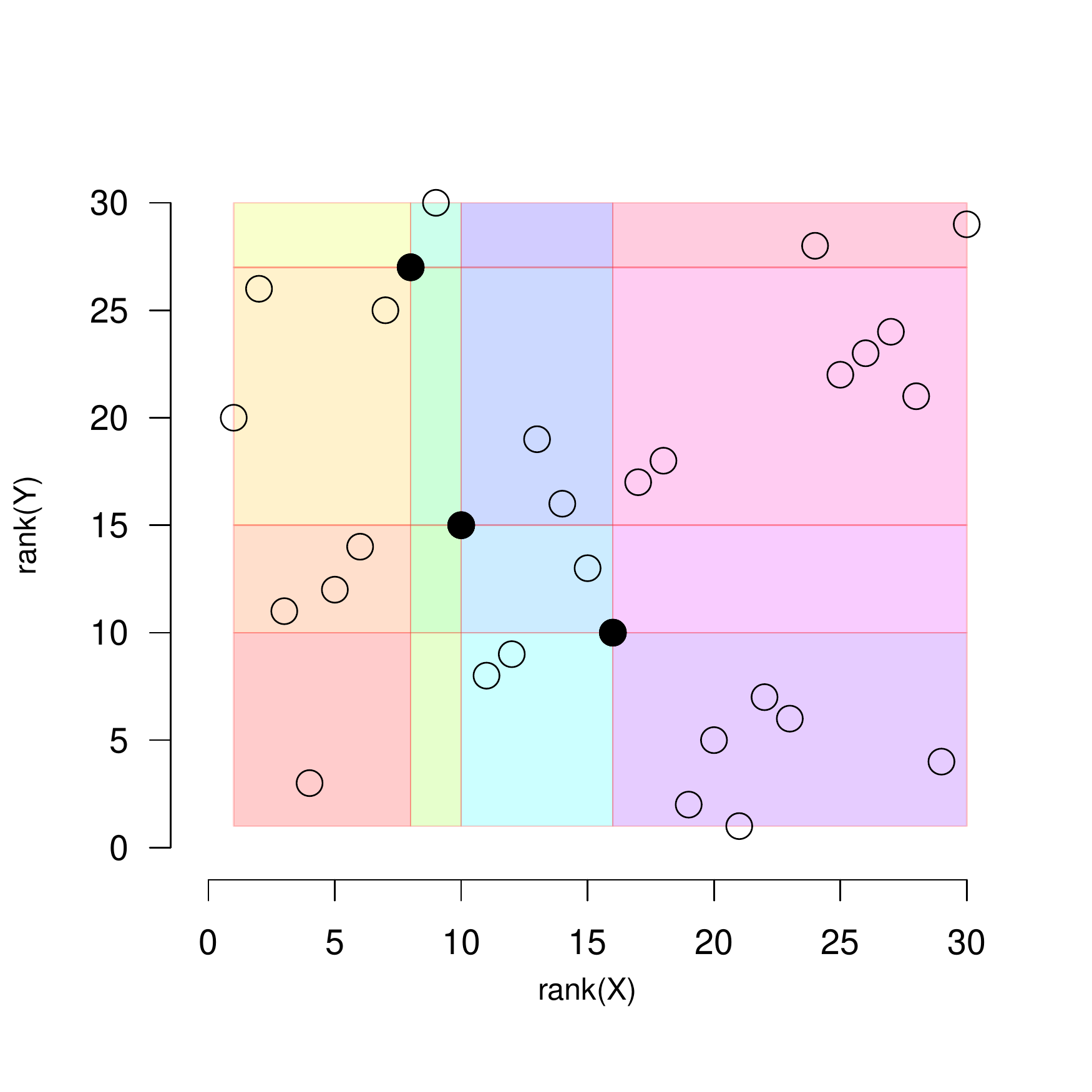} &
  \includegraphics[page=1, width=0.35\textwidth, trim=0.6in 0.6in 0.6in 0.6in, clip]{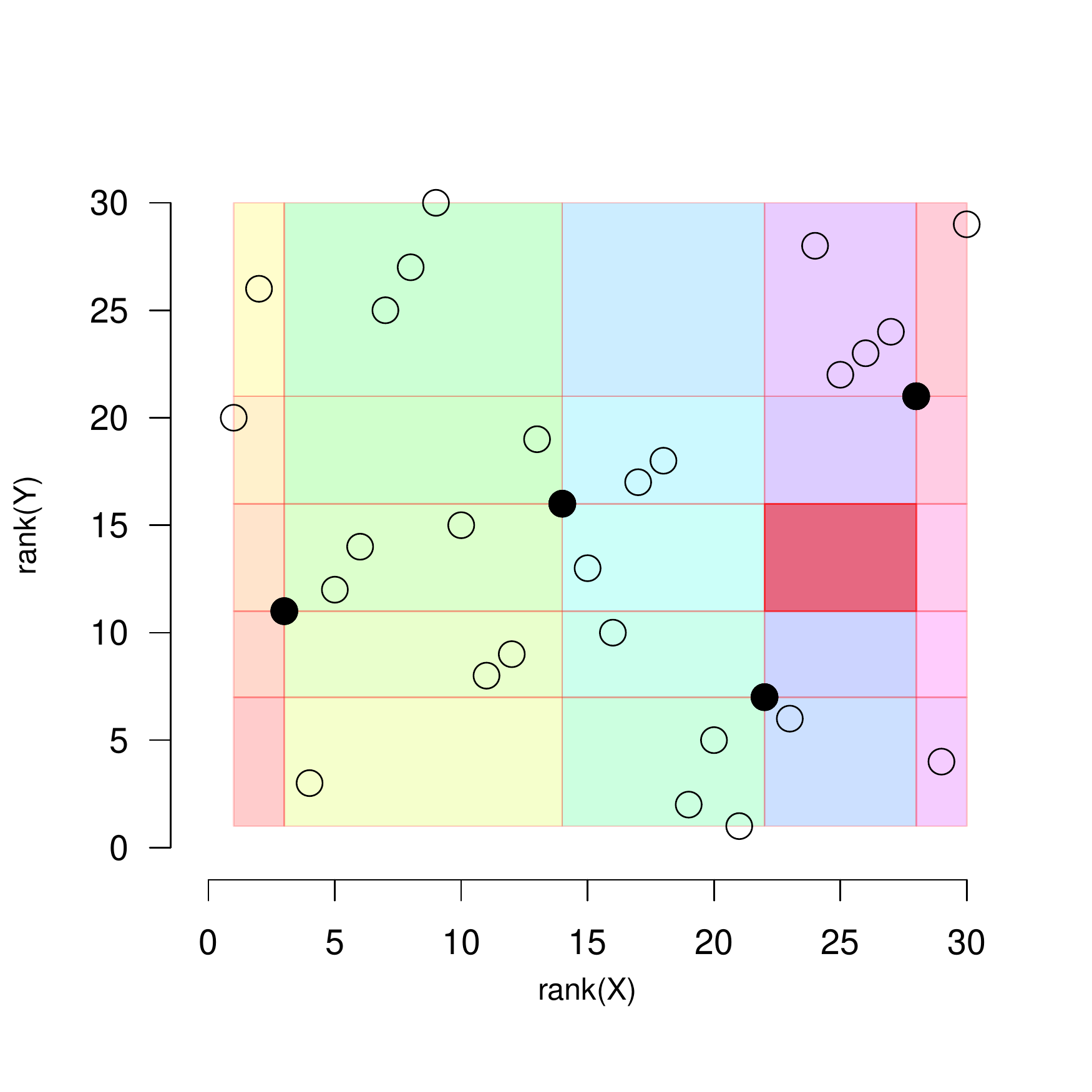}
  \end{tabular}
  \caption[Visualization of the partitiong of the rank-rank plane]{A visualization of the partitioning of the rank--rank plane which is at the basis of the data derived partitions (DDP) tests. Here, $N = 30$, and circles represent observed points. Full black circles represent those observations that were chosen to induce the partition, and different shades represent partition cells. With $m = 2$, all cells are corner cells (top-left);  with $m = 3$,  the center cell has two vertices which are observed sample points (top-right); with $m = 4$, all internal cells, i.e., cells that are not on the boundary, have at least one observed point vertex (bottom-left); only with $m \geq 5$, there exists at least one internal cell free of observed vertices (bottom-right, internal cell with no vertex which is a sample point is marked in red).}
  \label{fig:visualization}
\end{figure}

Figure~\ref{fig:visualization} shows example partitions of the sample space based on the ranked observations, $rank(Y)$ versus $rank(X)$, where a $m\times m$ partition is based on $m-1$ observations. We refer to such partitions as data derived partitions (DDP). The dependence in the data can be captured by many partitions, and some partitions are better than others.

\citet{hoeffding1948non} suggested a test based on summation of a score over all $N$ $2 \times 2$  DDP of the sample space, which is consistent against any form of dependence if the bivariate density is continuous. Hoeffding's test statistic is 
$$ \iint N \left\{\hat F_{XY}(x,y) -\hat F_{X}(x)\hat F_{Y}(y)\right\}^2d\hat F_{XY}(x,y), $$
where $\hat F$ denotes the empirical cumulative distribution function. \citet{blum1961distribution} showed that Hoeffding's test statistic is asymptotically equivalent to
$\sum_{i=1}^N(o^i_{1,1}o^i_{2,2} -o^i_{1,2}o^i_{2,1} )^2/N^4,$
where $o^i_{u,v}$, $u,v\in \{1,2\}$, is the observed count of cell $(u,v)$ in the $2\times 2$ contingency table defined by the $i$th observation. 
\citet{Thas04} noted that by appropriately normalizing each term in the sum, the test statistic becomes the average of all Pearson statistics for independence applied to the contingency tables that are induced by $2\times 2$ sample space partitions centered about observation $i\in \{1,\ldots,N \}$. They proved that the weighted version of Hoeffding's test statistic is still consistent.

Partitioning the sample space into finer partitions than the $2 \times 2$ quadrants of the classical tests, based on the observations, was also  considered in \citet{Thas04}. They suggested that the average of all Pearson statistics on finer partitions of fixed size $m\times m$ may improve the power, but did not provide a proof that the resulting tests are consistent. They examined in simulations only $3 \times 3$ and $4 \times 4$ partitions.
\citet{Reshef11} suggested the maximal information coefficient, which is a test statistic based on the maximum over dependence scores taken for partitions of various sizes, after normalization by the partition size, where the purpose of the normalization is equitability rather than power. Since computing the statistic exactly is often infeasible, they resort to a heuristic for selecting which partitions to include. Thus, in practice, their algorithm goes over only a small fraction of the partitions they set out to examine. In Section~\ref{sec:simulation} we show that the power of this test is typically low.

\subsection{Review of the $K$-sample problem}\label{subsec-revksample}
As in Section \ref{subsec-revdfind}, we focus on consistent partition-based distribution-free tests. For testing equality of distributions, i.e., for a categorical $X$,  one of the earliest and still very popular distribution-free consistent tests is the Kolmogorov--Smirnov test \citep{Darling57}, which is based on the maximum score of all $N$ partitions of the sample space based on an observed data point. Aggregation by summation over  all $N$ partitions has been considered by Cramer and von Mises \citep{Darling57}, Pettitt \citep{Pettitt76} who constructed a test-statistic of the Anderson and Darling family \citep{Anderson52}, and \cite{scholz87}.

\cite{Thas07} suggested the following extension of the Anderson--Darling type test. For random samples of size $N_1$ and $N-N_1$, respectively, from two continuous densities, for a fixed $m$, they consider all possible partitions into $m$ intervals of the sample space of the univariate continuous random variable.  They compute Pearson's chi-square score  for the observed versus expected (under the null hypothesis that the two samples come from the same distribution) counts, then aggregate by summation  to get their test statistics. A permutation test is applied on the resulting test statistic, since under the null all $\binom{N}{N_1}$ assignments of the group labels are equally likely.  They show that the suggested statistic for $m=2$ is the Anderson--Darling test. They examined in simulations only partitions into $m\leq 4$ intervals.

\cite{Jiang14} suggested a penalized score that aggregates by maximization the penalized log likelihood ratio statistic for testing equality of distributions, with a penalty for fine partitions. They developed an efficient dynamic programming algorithm to determine the optimal partition, and suggested a distribution-free permutation test to compute the $p$-value.

 Although there are many additional tests for the two-sample problem, the list above contains the most common as well as the most recent developments in this field. Interestingly, when working with ranks,  the energy test of \cite{Szekely05}  and the Cramer--von Mises test turn out to be equivalent. 

\subsection{Overview of this paper}\label{subsec-goal}
In this work, we suggest several novel distribution-free tests that are based on sample space partitions. The novelty of our approach lies in the fact that we consider aggregation of scores over all partitions of size $m\times m$ (or $m$ for the $K$-sample case), where $m$ can increase with sample size $N$, as well as consideration of all $m$s simultaneously without any assumptions on the underlying distributions. In Section \ref{sec:new_tests} we present the new tests both for the independence problem and for the $K$-sample problem, with a focus on our regularized scores (that consider all $m$s) in Section \ref{subsec-regstat}. We prove that all suggested tests are consistent, including those presented in \cite{Thas04}, and show the connection between our tests and mutual information (MI). In Section \ref{sec-algorithms} we present innovative algorithms for the computation of the tests, which are essential for large $m$ since the computational complexity of the naive algorithm is exponential in $m$. Simulations are presented in Section \ref{sec:simulation}. Specifically, in Section \ref{subsec-sim2sample} we show that for the two-sample problem for complex distributions there is a clear advantage for fine partitions, while for simple distributions rougher partitions have an advantage. In Section \ref{subsec-simind} we show that for the independence problem there typically is a clear advantage for finer partitions for complex non-monotone relationships while for simpler relationships there is an advantage for rougher partitions. We further demonstrate  the ability of our regularized method (which aggregates over all partitions) to adapt and find the best partition size.   Moreover, in simulations we show that for complex relationships all these tests are more powerful than other existing distribution-free tests. In Section~\ref{sec:real_data} we analyze the yeast gene expression dataset from \citet{hughes2000functional}. With our distribution-free tests, we discover interesting non-linear relationships in this dataset that could not have been detected by the classical tests, contrary to the conclusion in \citet{steuer2002mutual} that there are no non-monotone associations.
Efficient implementations of all statistics and tests described herein are available in the $R$ package \emph{HHG}, which can be freely downloaded from the Comprehensive $R$ Archive Network, \url{http://cran.r-project.org/}. Null tables can be downloaded from the first author's web site.

\section{The proposed statistics}\label{sec:new_tests}
We assume that $Y$ is a continuous random variable, and that $X$ is either continuous or discrete. We have $N$ independent realizations $(x_1, y_1), \ldots, (x_N, y_N)$ from the joint distribution of $X$ and $Y$. Our test statistics will only depend on the marginal ranks, and therefore are distribution free, i.e., their null distributions are free of the marginal distributions $F_X$ and $F_Y$. 

\begin{paragraph}{Test statistics for the $K$-sample problem}
We first consider the case that $X$ is categorical with $K\geq 2$ categories. In this case, a test of association is also a $K$-sample test of equality of distributions.  For $N$ observations, there are $\ibinom{N+1}{2}$ possible cells, and $\ibinom{N-1}{m-1}$ possible partitions of the observations  into $m$ cells, where a cell is an interval on the real line. Since the cell membership of observations is the same regardless of whether the partition is defined on the original observations or on the ranked observations, and the statistics we suggest only depend on these cell memberships, we describe the proposed test statistics on the ranked observations, $rank(Y) \in \{1,...,N\}$. Let $\Pi_m$ denote the set of partitions into $m$ cells. For any fixed partition $\Ical = \{i_1,\ldots,i_{m-1}\}\subset \{1.5,\ldots, N-0.5 \}$, $i_1<i_2<\ldots<i_{m-1}$,  $\Ccal(\Ical)$ is the set of $m$ cells defined by the partition. 
For a cell $C\in \Ccal(\Ical)$, let $o_C(g)$ and $e_C(g)$ be the observed and expected counts inside the cell for distribution $g\in \{1,\ldots,K \}$, respectively. 
The expected count $e_C(g)$ is the width of cell $C$ based on ranks multiplied by $N_g/N$, where $N_g$ is the total number observations from distribution $g$: $e_{[i_l,i_{l+1}]}(g) = (i_{l+1}-i_{l})\times N_g/N$, where  $l\in \{0,\ldots,  m-1\}$, $i_0=0.5$ and $i_m = N+0.5$.
We consider either Pearson's score or the likelihood ratio score for a given cell $C$,
\begin{equation}
  \label{eq:ksample_cell_score}
  t_{C} \in \left\{{\sum_{g=1}^K\frac{[o_{C}(g) - e_{C}(g)] ^ 2}{e_{C}(g)}},\; \sum_{g=1}^K o_{C}(g)\log{\frac{o_{C}(g)}{e_{C}(g)}} \right\}.
\end{equation}
For a given partition $\Ical$, the score is
$T^{\Ical} = \sum_{C\in \Ccal(\Ical) }t_C$ (where if $t_C = \sum_{g=1}^Ko_{C}(g)\log{\frac{o_{C}(g)}{e_{C}(g)}}$ then $T^{\Ical}$ is the likelihood ratio given the partition). Our test statistics aggregate over all partitions by summation (Cramer--von Mises-type statistics) and by maximization (Kolmogorov--Smirnov-type statistics):
\begin{equation}
  \label{eq:ksample_score}
  S_{m } = \sum_{\Ical\in \Pi_m }{T^{\Ical}}, \quad M_{m }= \max_{\Ical\in \Pi_m }{T^{\Ical}}.
\end{equation}
\end{paragraph}

Tables of critical values for given sample sizes $N_1, \ldots, N_K$ can be obtained for (very) small sample sizes by generating all possible $N!/(\Pi_{g=1}^K N_g!)$ reassignments of ranks $\{1,\ldots,N\}$ to $K$ groups of sizes  $N_1, \ldots, N_K$ and computing the test statistic for each reassignment. The $p$-value is the fraction of reassignments for which the computed test statistics are at least as large as observed. 
When the number of possible reassignments is large, the null tables are obtained by large scale Monte Carlo simulations (we used $B=10^6$  replicates for each given sample size $N_1, \ldots, N_K$). For each of the $B$ reassignment selected at random from all possible reassignments, the test statistic is computed.  Clearly, the $B$ computations do not depend on the data, hence the tests based on these statistics are distribution free. 
Again, the $p$-value is the fraction of reassignments for which the computed test statistics are at least as large as the one observed, but here the fraction is computed out of the $B+1$ assignments that include the $B$ reassignments selected at random and the one observed assignment, see Chapter 15 in \cite{Lehmann05}. 
The  test based on each of these statistics is consistent: 

\begin{theorem}
  \label{thm-consistency2sample}
Let $Y$ be continuous, and $X$ categorical with $K$ categories. Let $N_g$ be the total number of observations from distribution $g\in \{1,\ldots,K\}$, and $N = \sum_{g=1}^K N_g$.  If the distribution of $Y$ differs at a continuous density point $y_0$ across values of $X$ in at least two categories, label these 1 and 2,   $\lim_{N\rightarrow \infty}\frac{\min(N_1,N_2)}{N}>0$, and $m$ finite or $\lim_{N\rightarrow\infty}m/N = 0$, then the distribution-free permutation tests based on $S_m$ and $M_m$ are consistent.
\end{theorem}
We omit the proof, since it is similar to (yet simpler than) the proof of Theorem \ref{thm-consistency} below.

\begin{paragraph}{Test statistics for the independence problem}
We now consider the case that $X$ is continuous. For $N$ pairs of observations,  there are $\ibinom{N-1}{m-1}\times \ibinom{N-1}{m-1}$ partitions of the sample space into $m\times m$ cells, where a cell is a rectangular area in the plane. We refer to these partitions as the all derived partitions (ADP) and denote this set by $\Pi^{ADP}_m$. Since the cell membership of observations is the same regardless of whether the partition is defined on the original observations or on the ranked observations, and the statistics we suggest only depend on these cell memberships, we describe the proposed test statistics on the ranked observations, so the $N$ pairs of observations  are on the grid $\{1, \ldots, N\}^2$.
For any fixed partition $\mathcal{I}= \{(i_1,j_1),\ldots, (i_{m-1},j_{m-1}) \}\subset \{1.5,\ldots, N-0.5 \}^2$, , $i_1<i_2<\ldots<i_{m-1}$, $j_1<j_2<\ldots<j_{m-1}$, $\mathcal{C}(\mathcal{I})$ is the set of $m \times m$ cells defined by the partition.
For a cell $C \in \Ccal(\mathcal{I})$, let $o_C$ and $e_C$ be the observed and expected counts inside the cell, respectively. The expected count in cell $C$ with boundaries $[i_k,i_{k+1}]\times [j_l,j_{l+1}]$ is $e_C = (i_{k+1}-i_k)\times (j_{l+1}-j_l)/N$, where $k,l \in \{0,\ldots, m-1\}$, $i_0=j_0=0.5, i_m=j_m = N+0.5$. 
 As with the $K$-sample problem, we consider either Pearson's score or the likelihood ratio score for a given cell $C$,
\begin{equation}
  \label{eq:cell_score}
  t_{C} \in \left\{{\frac{(o_{C} - e_{C}) ^ 2}{e_{C}}},\; o_{C}\log{\frac{o_{C}}{e_{C}}} \right\}.
\end{equation}
For a given partition $\Ical$, the score is
$T^{\Ical} = \sum_{C\in \Ccal(\Ical) }t_C$ (where if $t_C =  o_{C}\log{\frac{o_{C}}{e_{C}}}$ then $T^{\Ical}$ is the likelihood ratio given the partition).
 As above, we consider as test statistics aggregation  by summation and by maximization:
\begin{equation}
  \label{eq:adp_score}
  S_{m \times m}^{ADP} = \sum_{\Ical\in \Pi^{ADP}_m}{T^{\Ical}}, \quad M_{m \times m}^{ADP}= \max_{\Ical \in \Pi^{ADP}_m}{T^{\Ical}}.
\end{equation}
We consider another test statistic based on DDP, where each set of $m-1$ observed points in their turn define a partition (see Figure \ref{fig:visualization}). 
This variant has a computational advantage over the ADP statistic for $m<5$, see Remark \ref{rem-smallmalgs}.
Since all observations have unique  values, the remaining $N-(m-1)$ points are inside cells defined by the partition. There are $\ibinom{N}{m-1}$ partitions, denote this set of partitions by $\Pi^{DDP}_m$. As before, since the cell membership of observations is the same regardless of whether the partition is defined on the original observations or on the ranked observations, and the statistics we suggest only depend on these cell memberships, we describe the proposed test statistics on the ranked observations. For a cell $C\in \Ccal(\mathcal{I})$, where $\mathcal{I}\in \Pi^{DDP}_m$,  the boundaries of $C$ are not necessarily defined by two sample points, as depicted at the bottom right panel of Figure \ref{fig:visualization}. We refer to $r_l$ and $r_h$ as the lower and upper values of the ranks of $X$ in $C$, and to $s_l$ and $s_h$ as the lower and upper values of the ranks of $Y$ in $C$,  where $r_l, r_h, s_l,s_h \in \{1, \ldots, N\}$.  Let $o_C$ and $e_C$ be the observed and expected counts strictly inside the cell, respectively. The expected count in cell $C$ with rank range $[r_l, r_h]\times [s_l, s_h]$ is  $e_C = (r_h-r_l-1)(s_h-s_l -1)/[N-(m-1)]$. 
  We consider Pearson's score or the likelihood ratio score for a given cell $C$, and define $t_C$ as in~\eqref{eq:cell_score}. For a given partition $\Ical$, the score is $T^{\Ical} = \sum_{C\in \Ccal(\Ical) }t_C$, and similarly to~\eqref{eq:adp_score} we define

\begin{equation}
  \label{eq:ddp_score}
 S_{m \times m}^{DDP} = \sum_{\Ical\in \Pi^{DDP}_m}{T^{\Ical}}, \quad M_{m \times m}^{DDP}= \max_{\Ical \in \Pi^{DDP}_m}{T^{\Ical}}.
\end{equation}

For each of the test statistics in (\ref{eq:adp_score}) and (\ref{eq:ddp_score}), tables of exact critical values for a given sample size $N$ can be obtained for small $N$ by generating all possible $N!$ permutations of $\{1,\ldots,N\}$. For each permutation $(\pi(1), \ldots, \pi(N))$, the test statistic is computed for the reassigned pairs $(1,\pi(1)), \ldots, (N, \pi(N))$. Clearly, the computation of these null distributions  does not depend on the data, hence the tests based on these statistics are distribution free. As in the case of the $K$-sample problem, the $p$-value is the fraction of permutations for which the computed test statistics are at least as large as the one observed, 
and when the number of possible permutations is large, the critical values are obtained by large scale Monte Carlo simulations.
The test based on each of these statistics is consistent:
\begin{theorem}
  \label{thm-consistency}
  Let the joint density of $X$ and $Y$ be $h(x,y)$, with marginal densities  $f(x)$ and $g(y)$. If there exists a point $(x_0, y_0)$ such that $h(x_0, y_0)$ is continuous and $h(x_0, y_0) \neq f(x_0) g(y_0)$, i.e., there is local dependence at a continuous density point, and if $m$ is finite or $\lim_{N\rightarrow\infty}m/\sqrt{N} = 0$, then the distribution-free permutation tests based on the following test statistics are consistent:
  \begin{enumerate}
    \item The test statistics aggregated by summation:  $S^{DDP}_{m\times m}$ and $S^{ADP}_{m\times m}$.
    \item The test statistics aggregated by maximization:  $M^{DDP}_{m\times m}$ and $M^{ADP}_{m\times m}$.
  \end{enumerate}
\end{theorem}
A proof is given in Appendix~\ref{sec:test_consistency_proof}.
\end{paragraph}

We note that Thas and Ottoy suggested $S_m$ with $t_{C}=\sum_{g=1}^2\frac{(o_{C}(g) - e_{C}(g)) ^ 2}{e_{C}(g)}$ in \cite{Thas07}, and  $S_{m \times m}^\mathrm{DDP}$ using Pearson's score for finite $m$ in  \citet{Thas04}. However, they examined in simulations only $m\leq 4$.
Thanks to the efficient algorithms we developed, detailed in Section \ref{sec-algorithms},  we are able to test for any $m\leq N$ in the $K$-sample problem, and for aggregation by summation in the test of independence. If the aggregation is by maximization in the test of independence, the algorithm, detailed in Section \ref{sec-algorithms}, is exponential in $m$ and thus the computations are feasible only for $m\leq 4$.

We shall show in Section~\ref{sec:simulation} that the power of the test based on a summation statistic can be different from the power of the test based on a maximization statistic, and which is more powerful depends on the joint distribution. However, for both aggregation methods, using $m>3$ partitions improves power considerably for complex settings. Therefore, in complex settings our tests with $m>3$ have a power advantage over the classical distribution-free tests, which focused on rough partitions, typically $m=2$.

\begin{paragraph}{Connection to the MI}
An attractive feature of the statistics $S_m$ and $S_{m\times m}^{ADP}$, for $m$ large enough, is that they are directly associated with the MI. MI
(defined as $I_{XY} = \int h(x,y)\log[h(x,y)/\{f(x)g(y)\}]dxdy$ for continuous $X$ and $Y$)
is a useful measure of statistical dependence. The variables $X$ and $Y$ are independent if and only if the MI is zero. 
Estimated MI is used in many applications to quantify the relationships between variables, see  \citet{steuer2002mutual}, \citet{paninski2003estimation}, \citet{kinney2013equitability} and references within. Although many works on MI estimation exist, no single one has been accepted as a state-of-the-art solution in all situations~\citep{kinney2013equitability}.  A popular estimator among practitioners due to its simplicity and consistency is the {\em histogram} estimator, where the data are binned according to some scheme and the empirical mutual information of the resulting partition, i.e, the likelihood ratio score, is computed. Intuitively, one can expect that the statistic $S_{m \times m}^{ADP}$, properly normalized, can also serve as a consistent estimator of the mutual information, when the contingency tables are summarized by the likelihood ratio statistic, since it is the average of histogram estimators, over all partitions. This intuition is true despite the fact that the number of partitions goes to infinity, since we show that the convergence is uniform and that the fraction of ``bad" partitions (i.e., partitions with cells that are too big or too small) is small, as long as $m$ goes to infinity at a slow enough rate.
\begin{theorem}
  \label{thm:kl}
  Suppose $X$ is categorical with $K$ categories and $Y$ is continuous. Let $N_g$ be the total number of observations from distribution $g\in \{1,\ldots,K\}$, and $N = \sum_{g=1}^K N_g$. If $\lim_{N\rightarrow \infty} \frac{N_g}N>0$ for $g=1,\ldots, K$,  $\lim_{N \rightarrow\infty}\frac mN = 0$, and $\lim_{N \rightarrow \infty} m= \infty$, then $\frac{S_m}{N\ibinom{N-1}{m-1}}$ is a consistent estimator of the MI.
\end{theorem}
\begin{theorem}
  \label{thm:mi}
  Suppose the bivariate density of $(X,Y)$ is continuous with bounded mutual information. If $\lim_{N\rightarrow \infty} m/\sqrt{N} = 0$, and $\lim_{N\rightarrow \infty} m= \infty$, then $\frac{S^{ADP}_{m\times m}}{N \ibinom{N-1}{m-1}\times \ibinom{N-1}{m-1}}$ is a consistent estimator of the MI.
\end{theorem} 
See Appendix \ref{sec:mi_proof} for a proof of Theorem \ref{thm:mi}. The proof of Theorem \ref{thm:kl} is omitted since it is similar to that of Theorem \ref{thm:mi}. See Appendix \ref{sec:mi example} for a simulated example of MI estimation using $S^{DDP}_{m\times m}$, $S^{ADP}_{m\times m}$, and the histogram estimator. The ADP estimator is the least variable, as is intuitively expected since it is the average over many partitions.
\end{paragraph}

\begin{remark}
  \label{rem:ties}
  In this work we assume there are no ties among the continuous variables. In our software, tied data are broken randomly, so that our test remains distribution free. An alternative approach, which is no longer distribution free, is a permutation test on the ranks, with average ranks for ties. Then a tied observation, that falls on the border of a contingency table cell, receives equal weight in each of the cells it borders with.
\end{remark}

\subsection{The proposed regularized statistics}\label{subsec-regstat}
An important parameter of the statistics proposed above is $m$, the partition size. 
A poor choice of $m$ may lead to substantial power loss:  if $m$ is too small or too large, it may lack power to discover complex non-monotone relationships. For example, consider the three simulation settings for the two-sample problem in the first row of Figure \ref{fig:setups2sample}. The best partition for setting 1, ``normal vs. normal with delta", for small sample sizes, is intuitively to divide the real line into three cells: until the start of the narrow peak, the support of the narrow peak, and after the peak ends. Moreover, the best aggregation method is by maximization, not summation, since there are very few good partitions that capture the peak and aggregation by summation using $m=3$ will aggregate many bad partitions that miss the peak.  Therefore, we expect that $M_3$ will be the most powerful test statistic for setting 1. For setting 2, ``Mix. vs. Mix. 3 vs. 4 components", intuitively it seems best to partition into more than seven cells, and that many partitions will work well. For setting 3, ``normal vs. normal with many deltas", it seems best to partition into many cells.
Indeed, the power curves in Figure \ref{fig:pwr2sample} show that for the first setting, $M_m$ is optimal at value $m=3$, yet if we use this value for the second setting, the test has  20\% lower power than optimal power (which is 86\% at $m=10$), and if we use this value for the third setting, the test has  58\% less power than the optimal power (which is 88\% at $m=34$).

Since the optimal choice of $m$ is unknown in practice, we suggest two types of regularizations which take into consideration the scores from all partition sizes.
The first type of regularization we suggest is to combine the $p$-values from each $m$,  so that the test statistic becomes the combined $p$-value. Specifically, let $p_m$ be the $p$-value from a test statistic based on partition size $m$, be it $S_m$ or $M_m$ for the $K$-sample problem, or  $S_{m\times m}^{ADP}$ or $S_{m\times m}^{DDP}$ for the independence problem. Due to the computational complexity, we do not consider a regularized score for $M_{m\times m}$.
We consider as test statistics the minimum $p$-value,
$\min_{m\in \{2, \ldots, m_{\max} \}}p_m$, as well as the Fisher combined $p$-value, $-\sum_{m=2}^{m_{\max}} \log p_m$. These combined $p$-values are not $p$-values in themselves, but their null distribution can be easily obtained from the null distributions of the test statistics for fixed $m$s, as follows: (1) for each of $B$ permutations, compute the test statistics for each $m\in \{2,\ldots,m_{\max}\}$; (2) compute the $p$-value of each of the resulting statistics, so for each permutation, we have a set of $p$-values $p_2,\ldots, p_{m_{\max}}$ to combine; (3) combine the $p$-values for each of the $B$ permutations. Choose $B$ to be large enough for the desired accuracy of approximation of the quantiles of the null distribution of the combined $p$-values used for testing.
Obviously, since the combined $p$-values are based on the ranks of the data, they are distribution-free. Since they do not require fixing $m$ in advance, they are a practical alternative to the tests that require $m$ as input.

In order to examine how close this regularized score is to the optimal $m$ (i.e., the $m$ with highest power), we looked at the distribution of the $m$s with minimum $p$-values in 20000 data samples from the above-mentioned three simulation settings. For these settings, using the aggregation by maximization statistic, the median $m$ of the minimal $p$-value was: 3 for the first setting, 9 for the second setting, and 33 for the third setting. Moreover, the first and third quartiles were 3 to 5 for the first setting, 7 to 14 for the second setting, and 19 to 60 for the third setting.  We conclude that for these examples,
 the $m$ that achieves the minimum $p$-values in most runs was remarkably close to the optimal $m$ (which was 3, 10, and 34 for settings 1,2, and 3, respectively), suggesting that the power of the minimum $p$-value statistic is close to that of the statistic with optimal $m$. Indeed, the power of the minimum $p$-values in settings 1-3 using aggregation by maximum was 0.825, 0.799, and 0.785, whereas the power using the (unknown in practice) optimal $m$ in settings 1-3 was 0.894, 0.86, and 0.88, respectively. Further empirical investigations detailed in Section \ref{sec:simulation} give additional support to this regularization method.

The second type of regularization adds a penalty to the statistic, so that the test statistic becomes the maximum (over all $m$s) of the statistic plus penalty.
For the $K$-sample problem, \cite{Jiang14} suggested assigning a prior on the partition scheme and they regularized the likelihood ratio score using this prior. Specifically, they assumed the partition size is Poisson and the conditional distribution on the $m$ partition widths (normalized to sum to one) is $Dirichlet(1,\ldots, 1)$. This led to their penalty term $-\lambda_0 (\log N) (m-1)$, where $\lambda_0>0$ has to be fixed.  We assume that the marginal distribution on the partition size is $\pi(m)$ (e.g., Poisson or Binomial),  and that the prior probability of selecting $\mathcal I$ given $m$, $\pi(\mathcal I|m)$,  is uniform. There is an important difference between our uniform discrete prior distribution on partitions of size $m$ and the continuous Dirichlet uniform prior of \cite{Jiang14}. Our prior is uniform on all partitions that truly divide the sample space into $m$ cells,  i.e., we cannot have two partition lines between two consecutive samples, since this is actually an $m-1$ partition. Using the continuous Dirichlet prior results in practice in at most $m$ partitions, but the partition size may also be strictly smaller than $m$ if two partition points lie between two sample points. Therefore, their conditional distribution given the partition size parameter  is not necessarily the true size of the partition. Their penalty translates to a conditional probability given a true partition size $m$ of $\frac{(m-1)!}{(N-1)^{(m-1)}}$, compared to our $\pi(\mathcal I|m) = 1/\ibinom{N-1}{m-1}$. Therefore, their score penalizes more severely large $m$s, and their regularized test statistic has less power when the optimal $m$ is large in our simulations.

For aggregation by maximum in the $K$-sample problem, we consider the regularized statistic,
\begin{equation}\label{eq-LRregularizedscore}
\max_{m\in\{2,\ldots,m_{\max}\} } \{ M_{m }+\log [\pi(\mathcal I|m)\pi(m)]\},
\end{equation}
where we use the likelihood ratio score per partition. Due to the computational complexity, we do not consider a regularized score for $M_{m\times m}$.
For aggregation by summation, our efficient algorithms described in Section \ref{sec-algorithms} enable us to consider the penalized average score per $m$,
\begin{eqnarray}\label{eq-regularizedsum}
\max_{m\in\{2,\ldots,m_{\max}\} } \{ SLR_{m} \pi(\mathcal I|m)+\log \pi(m)\}
\end{eqnarray}
where $SLR_{m} \pi(\mathcal I|m)$ is $S_m$ divided by the number of partitions of size $m$ for the $K$-sample test, and $S^{ADP}_{m\times m}$ (or $S^{DDP}_{m\times m}$) divided by the number of partitions of size $m\times m$ for the test of independence, using the likelihood ratio score per partition.  The null distribution of these regularized statistics is computed by a permutation test, and they are distribution free.

 An extensive numerical investigation, partially summarized in Appendix \ref{app:additional simulations}, led us to choose the minimum $p$-value as the preferred regularization method. Between the two combining functions, we preferred the minimum  over Fisher, since Fisher was far more sensitive to the choice of the range of $m$ for combining (see Table \ref{tab:FisherVsMinP2sample}). Regularization using priors was less effective, except when the Poisson prior was used with parameter $\lambda = \sqrt N$ (see Table \ref{tab:priors2sample}). We preferred the first type of regularization since it was at least as effective as  regularizing by a Poisson prior, without requiring setting any additional parameters.
This regularized statistic is consistent, as the next theorems show.

\begin{theorem}
  \label{thm-consistency2sampleminp}
Let $Y$ be continuous, and $X$ categorical. Let $N_g$ be the total number of observations from distribution $g\in \{1,\ldots,K\}$, and $N = \sum_{g=1}^K N_g$. If the distribution of $Y$ differs at a continuous density point $y_0$ across values of $X$ in at least two categories, label these 1 and 2,   $\lim_{N\rightarrow\infty}\frac{\min(N_1,N_2)}{N}>0$ , then 
the permutation test based on $\min_{m\in \{2, \ldots, m_{\max} \}}p_m$ is consistent, if:
  \begin{enumerate}
    \item it is based on  $S_m, m\in \{2,\ldots,m_{\max}\}$, and $\lim_{N\rightarrow\infty}m_{\max}/\sqrt{N} = 0$ or $m_{\max}$ is finite. \vspace{3mm}
    \item it is based on  $M_m, m\in \{2,\ldots,m_{\max}\}$ and $\lim_{N\rightarrow\infty} m_{\max}/N = 0$ or $m_{\max}$ is finite.
  \end{enumerate}
\end{theorem}

\begin{theorem}
  \label{thm-consistencyminp}
 Let the joint density of $X$ and $Y$ be $h(x,y)$, with marginal densities  $f(x)$ and $g(y)$. If there exists a point $(x_0, y_0)$ such that $h(x_0, y_0)$ is continuous and $h(x_0, y_0) \neq f(x_0) g(y_0)$, i.e., there is local dependence at a continuous density point,  then the permutation test based on $\min_{m\in \{2, \ldots, m_{\max} \}}p_m$ is consistent, if 
    \begin{enumerate}
    \item it is based on $S^{DDP}_{m\times m}$ or $S^{ADP}_{m\times m}$, and $\lim_{N\rightarrow \infty} m_{\max}/N^{1/3}=0$ or $m_{\max}$ is finite. \vspace{3mm}
    \item it is based on $M^{DDP}_{m\times m}$ and $M^{ADP}_{m\times m}$, and  $\lim_{N\rightarrow\infty}m_{\max}/\sqrt{N} = 0$ or $m_{\max}$ is finite.
\end{enumerate}     
\end{theorem}
The proof of Theorem \ref{thm-consistencyminp} follows in a straightforward way from the proofs of Theorem \ref{thm-consistency}, see Appendix \ref{app:proofconsistencyminp} for details. The proof of Theorem \ref{thm-consistency2sampleminp} follows similarly from the proof of Theorem \ref{thm-consistency2sample}, and it is omitted.

\section{Efficient algorithms}\label{sec-algorithms}
For computing the above test statistics for a given $N$ and partition size $m$, the computational complexity of a naive implementation is exponential in $m$.
We show in Section \ref{subsec-sumalgs} more sophisticated algorithms for computing the aggregation by sum statistics for all $m\in \{2, \ldots, N\}$ at once that have complexity $O(N^2)$  for the $K$-sample problem, and $O(N^4)$ for the independence problem. This is possible since instead of iterating over partitions, the algorithms iterate over cells. Moreover, the algorithms also enable calculating the regularized sum statistics of Section \ref{subsec-regstat} in $O(N^2)$ and $O(N^4)$ for the $K$-sample and independence problems, respectively, since we just need to go over the list of $m$ scores and for each score $S_m$ check its p-value in our pre-calculated null tables, which requires just an additional $O(N\log(B))$, where $B$ is the null-table size.

We show in Section \ref{subsec-maxalgs} an algorithm with complexity $O(N^3)$ for the $K$-sample problem for computing the aggregation by maximum for all $m$ at once. This algorithm also enables calculating the regularized maximum statistic   of Section \ref{subsec-regstat} in $O(N^3)$.
The algorithms for aggregating by maximum in the independence problem are exponential in $m$, and therefore infeasible for modest $N$ and $m>4$. However, for $m=3$ and $m=4$ we provide efficient algorithms with $O(N^2)$ and $O(N^3)$ complexity, respectively, for the DDP test statistics.

\subsection{Aggregation by summation}\label{subsec-sumalgs}
The algorithms for aggregation by summation are efficient due to two key observations. First, because the score per partition is a sum of contributions of individual cells, and the total number of cells is much smaller than the number of partitions (unless $m=2$ in the $K$-sample problem, and $m\leq 4$ when using DDP in the independence problem, see Remark \ref{rem-smallmalgs} below). Therefore, we can interchange the order of summation between cells and partitions and thus achieve a big gain in computational efficiency, since it is easy to calculate in how many partitions each cell appears, see equations (\ref{eq-agg-sum-2sample}) and (\ref{eq-agg-sum-ind}).

Second, because for a fixed $m$ the number of partitions in which a specific cell appears depends only on the width (and for independence testing, also length) of the cell, the data-dependent computations do not depend on $m$: the test statistics are the sum of cell scores for every width for the $K$-sample test, and for every combination of width and length for the independence test, see equations (\ref{eq-agg-sum-2sample-2}) and (\ref{eq-agg-sum-ind-2}). The complexity of the algorithm remains the same even when the scores are computed for all $m$s, since the complexity is determined by a preprocessing phase which is shared by all $m$s. Therefore, the complexity for the regularized scores is the same as the complexity for a single $m$.

\subsubsection{Algorithm for the $K$-sample problem}
For $g=1,\ldots,K$ (the categories of $X$) and $r = 1, \ldots, N$ (the ranks of $Y$), we first compute in $O(N)$ $A$ as follows:
\begin{equation}
A(g, r) = \sum_{i=1}^r{I(g_i = g)}, \nonumber
\end{equation}
and let $A(g, 0) = 0$.
For a cell with rank range $[r_l, r_h]$, where $r_l, r_h \in \{1, \ldots, N\}$, using $A$, the count of observations in  category $g$ that fall  inside the cell can be computed in $O(1)$ operations as $o_C(g) = A(g, r_h) - A(g, r_l-1)$. Therefore, for each cell $C$ the contribution of the cell, $t_C$, can be computed in $O(1)$ time. 

Because the score per partition is a sum of contributions of individual cells, $S_m$ is the sum over the score per cell, multiplied by the number of times the cell appears in a partition of size $m$. Considering further summing cells of width 1 to $N$, we may write $S_m$ as follows:
\begin{equation}\label{eq-agg-sum-2sample}
S_m=\sum_{\Ical\in \Pi_m}{T^{\Ical}} = \sum_{C\in \Ccal}t_C\sum_{\Ical \in \Pi_m}I[C\in \Ccal(\Ical)] =    \sum_{w=1}^N\sum_{C\in \Ccal(w)}t_C n(w,m,C),
\end{equation}
where $\Ccal(w)$ is the collection of cells of width $w$ and $n(w,m,C)$ is the number of partitions that include $C$.
For computing $n(w,m,C)$, we differentiate between two possible types of cells: edge cells and internal cells. Edge cells differ from internal cells by having either $r_l = 1$ or $r_h = N$. The number of partitions of order $m$ that include an edge cell of width $w = r_h - r_l + 1$ is given by $\ibinom{N - 1 -  w }{m - 2}$. The number of partitions including a similarly wide internal cell is $\ibinom{N -2-w}{m - 3}$.  Therefore, we may write $S_m$ as follows:
\begin{equation}\label{eq-agg-sum-2sample-2}
S_m=\sum_{w=1}^N \ibinom{N -2- w}{m - 3}T_i(w)  + \sum_{w=1}^N \ibinom{N -1- w }{m - 2} T_e(w) ,
\end{equation}
where $T_i(w) =\sum_{\substack{C\in \Ccal(w),\\ r_l\neq 1 \cap r_h\neq N}}t_C$ and $T_e(w) =  \sum_{\substack{C\in \Ccal(w),\\ r_l = 1 \cup r_h =  N}}t_C$.
The algorithm proceeds as follows. 
First, in a preprocessing phase, we calculate $T_i(w)$ and  $T_e(w)$ for all $w\in \{1,\ldots,N\}$. Since $t_C$ can be calculated in $O(1)$, as described above, the calculation of $T_i(w)$ and  $T_e(w)$ for a fixed $w$ takes $O(N)$. Since there are $N$ values for $w$, we can compute and store all values of $T_i(w)$ and  $T_e(w)$ in $O(N^2)$. Also in the preprocessing phase, for all $u, v\in \{0, \ldots,N\}$ we calculate and store all $\binom{u}{v}$. This can be done in $O(N^2)$  using Pascal's triangle method.

Given $T_i(w), T_e(w), w=1,\ldots, N-1$ (which are independent of m!), and all $\binom uv$,  we can clearly calculate $S_m$ according to equation (\ref{eq-agg-sum-2sample-2}) for any $m$ in $O(N)$ and therefore for all $m$s in $O(N^2)$, since $m<N$. Therefore the overall complexity of computing the scores for all $m$s is $O(N^2)$.

\subsubsection{Algorithm for the independence problem}
Let $r_i$ be the rank of $x_i$ among the observed $x$ values, and $s_i$ is the rank of $y_i$ among the $y$ values
The algorithm first computes the empirical cumulative distribution in $O(N^2)$ time and space,
\begin{equation}\label{eq-agg-sum-ind-A}
A(r, s) = \sum_{i=1}^NI(r_i\leq r \ \textrm{and} \  s_i\leq s), \quad (r,s) \in \{0,1,\ldots,N \}^2
\end{equation}
where $A(0,s) =0, A(r,0)=0$ and $\hat F(r,s) = A(r,s)/N$. First, let $B$ be the $(N+1)\times (N+1)$  zero matrix, and initialize to one $B(r_i, s_i)$ for each observation $i=1,\ldots,N$. Next, go over the grid in $s$-major order, i.e., for every $s$ go over all values of $r$, and compute:
\begin{enumerate}
\item $A(r, s) = B(r , s - 1) + B(r-1, s ) - B(r - 1, s - 1) + B(r, s)$, and
\item $B(r,s) = A(r,s)$.
\end{enumerate}

We describe the algorithm for the ADP statistic,  which selects partitions on the grid $\{1.5, \ldots, N-0.5\}^2$ on ranked data.  The main modifications for the DDP statistic
are provided in Appendix \ref{app-DDPalgorithm}.
The count of samples  inside a cell with rank ranges $r \in [r_l, r_h]$ and $s \in [s_l, s_h]$ can be computed in $O(1)$ operations via the inclusion-exclusion principle: $$o_C = A(r_h, s_h ) - A(r_{l}-1, s_h ) - A(r_h, s_{l}-1) + A(r_l-1, s_{l}-1).$$ Therefore, for each cell $C$ the contribution of the cell $t_C$ can be computed in $O(1)$ time.
Because the score per partition is a sum of contributions of individual cells, we may write $S_{m\times m}^{ADP}$ as follows:
\begin{equation}\label{eq-agg-sum-ind}
\sum_{\Ical \in \Pi_m^{ADP}}{T^{\Ical}} = \sum_{C\in \Ccal}t_C\sum_{\Ical \in \Pi_m^{ADP}}I[C\in \Ccal(\Ical)] =    \sum_{w=1}^{N-2}\sum_{l=1}^{N-2}\sum_{C\in \Ccal(w,l)}t_C n(w,l,m,C),
\end{equation}
where $\Ccal(w,l)$ is the collection of cells of width $w$ and length $l$ and $n(w,l,m,C)$ is the number of partitions that include $C$. As in the algorithm for the $K$-sample problem, $n(w,l,m,C)$ depends only on $w$, $l$, $m$, and whether the cell is an internal cell or an edge cell. For simplification, we discuss only the computation of the contribution of internal cells to the sum statistic, and non-internal cells can be handled similarly  (as discussed in the algorithm for the $K$-sample problem). Therefore, our aim is to compute:
\begin{equation}\label{eq-agg-sum-ind-2}
\sum_{w=1}^{N-2}\sum_{l=1}^{N-2}n(w,l,m)T(w,l),
\end{equation}
where $T(w,l) = \sum_{C\in \Ccal(w,l)}t_C$ and $n(w,l,m)$ is the number of partitions that include an internal cell of width $w$ and length $l$ when the partition size is $m$ and $\Ccal(w,l)$ is relabelled to be the collection of internal cells of width $w$ and length  $l$.

The algorithm proceeds as follows. First in a preprocessing phase we perform two computations:
1) calculate and store $T(w,l)$ for all pairs $(w,l)\in \{1...N-2\}^2$. Since $t_C$ can be calculated in $O(1)$, as described above,  the calculation of $T(w,l)$ for a fixed $(w,l)$ takes $O(N^2)$ and since there are $(N-2)^2$ pairs $(w,l)$ the total preprocessing phase takes $O(N^4)$; 
2) for all $u, v\in \{0, \ldots,N\}$ we calculate and store all $\binom{u}{v}$  in $O(N^2)$ steps  using Pascal's triangle method. 

Given $T(w,l)$, and all $\binom uv$, since $n(w,l,m)=\ibinom{N-2-w}{m-3}\ibinom{N-2-l}{m-3}$,  we can clearly calculate equation (\ref{eq-agg-sum-ind-2}) for a fixed $m$ in $O(N^2)$ and therefore for all $m$s in $O(N^3)$. Due to the preprocessing phase the total complexity is $O(N^4)$.

\begin{remark}\label{rem-smallmalgs}
When it is desired to only compute the statistic for very small $m$, faster alternatives exist.
For the two-sample problem, for $m=2$, the number of partitions is $O(N)$ and therefore an $O(N\log N)$ algorithm can be applied that aggregates over all partitions, and the complexity is dominated by the sorting of the $N$ observations (for $m=3$, the number of partitions is already $O(N^2)$).
Similarly, for the test of independence, the ADP statistic can be calculated in $O(N^2)$ steps for $m=2$, and the DDP statistic  in $O(N^2)$ for $m=3$, and in $O(N^3)$ for $m=4$, since this is the order of the number of partitions.  Per partition, the computation of the score for $m\leq 4$ is computed in $O(1)$ time since the contribution of the cell can be computed in $O(1)$ time (as shown above).
The DDP statistic for $m=2$ can be computed in $O(N\log N)$, using a similar sorting scheme as that detailed in \cite{Heller12}.
\end{remark}
\subsection{Aggregation by maximization}\label{subsec-maxalgs}
\paragraph{Algorithm for the $K$-sample problem}
\cite{Jiang14} suggested an elegant and simple dynamic programming algorithm for calculating $\max_{m} \{M_m-m\lambda(N)\}$ for any function $\lambda(\cdot)$ in $O(N^2)$. We present a modification of their algorithm that enables us to calculate $M_m$ for all $m$s in $O(N^3)$.
As a first step, for all $i\leq N$ and for all $j<i$ we calculate iteratively $M(i,j)$, the maximum score which partitions the first $i$ samples into $j$ partitions.
We compute $M(i+1,j)$ from $M(a,j-1), a\leq i$ using:
$$M(i+1,j)=\max_{a\in\{2,\ldots,i\} } \{M(a,j-1)+ t_{[a+0.5,i+1+0.5]}\}, $$
where $t_{[a+0.5,i+1+0.5]}$ is the score of the cell from $a+0.5$ to $i+1+0.5$.
This calculation takes $O(N)$, and since we have $O(N^2)$ such items to calculate, this step takes $O(N^3)$. Since $M_m =M(N,m)$, the overall complexity for computing the scores for all $m$s is $O(N^3)$.
Note that this algorithm enables us to calculate $\max_{m\in\{2,\ldots,m_{\max}\} } \{ M_{m }+\log [\pi(\mathcal I|m)\pi(m)]\}$ in $O(N^3)$ for any function $\pi(m)$,  thus the regularized test statistic in Section \ref{subsec-regstat} can also be computed in $O(N^3)$.

\paragraph{Algorithm for the  independence problem}
The algorithm is the same as described in Remark \ref{rem-smallmalgs} for the ADP statistic for $m=2$ and for the DDP statistic for $m=3$ and $m=4$, with the difference that the aggregation is by maximization (not summation) over the scores per partition. We are not aware of a polynomial-time algorithm for arbitrary $m$. We discuss ways to reduce the computational complexity in Section \ref{sec:Discussion}.

\begin{remark}
We 
show in Appendix \ref{supp:sec-fastHHG} that for univariate data the test of \citet{Heller12} with an arbitrary distance metric, with or without ties, can be computed in $O(N^2)$ in a similar fashion, thus improving their algorithm by a factor of $\log N$ when $X$ and $Y$ are univariate.
\end{remark}

\section{Simulations}\label{sec:simulation}
In simulations, we compared the power of our different test statistics in a wide range of scenarios.   All tests were performed at the 0.05 significance level. Look-up tables of the quantiles of the null distributions of the test statistics for a given $N$ were stored. Power was estimated by the fraction of test statistics that were at least as large as the $95$th percentile of the null distribution. The null tables were based on $10^6$ permutations.

 The noise level was chosen separately for each configuration and sample size, so that the power is reasonable for at least some of the variants. This enables a clear comparison using a range of scenarios of interest. Since the power was very similar for the Pearson and likelihood ratio test statistics, only the results of the likelihood ratio test statistic are presented.

 The simulations for the two-sample problem are detailed in \ref{subsec-sim2sample}, and for the independence problem in \ref{subsec-simind}. The analysis was done with the R package \emph{HHG}, now available on CRAN.
\subsection{The two-sample problem}\label{subsec-sim2sample}
We examined the power properties of the statistic aggregated by summation as well as by maximization for $m\in \{2,\ldots, N/2\}$, as well as the minimum $p$-value statistic, $\min_{m\in \{2, \ldots, m_{\max} \}}p_m$. We display here the results for $m_{\max}=149$ and $N=500$.  However, the choice of $m_{\max}$ has little effect on power, see Appendix \ref{app:additional simulations} for results with other values of $m_{\max}$. Also, see Appendix \ref{app:additional simulations} for the results for $m_{\max}=29$ and $N=100$.

We compared these tests to six two-sample distribution-free tests suggested in the literature. We compared to Wilcoxon's rank sum test, since it is one of the most widely used  tests to detect location shifts. We compared to two consistent tests suggested recently in the literature, the test of \cite{Jiang14}, referred to as DS, and the test of  \citet{Heller12} on ranks, referred to as HHG on ranks. Finally, we compared to the classical consistent tests of Kolmogorov and Smirnov, referred to as KS, of Cramer and von Mises (which is equivalent to the energy test of \cite{Szekely05} on ranks), referred to as CVM, and of Anderson and Darling, referred to as AD.

\begin{figure}[htbp]
  \centering
  \includegraphics{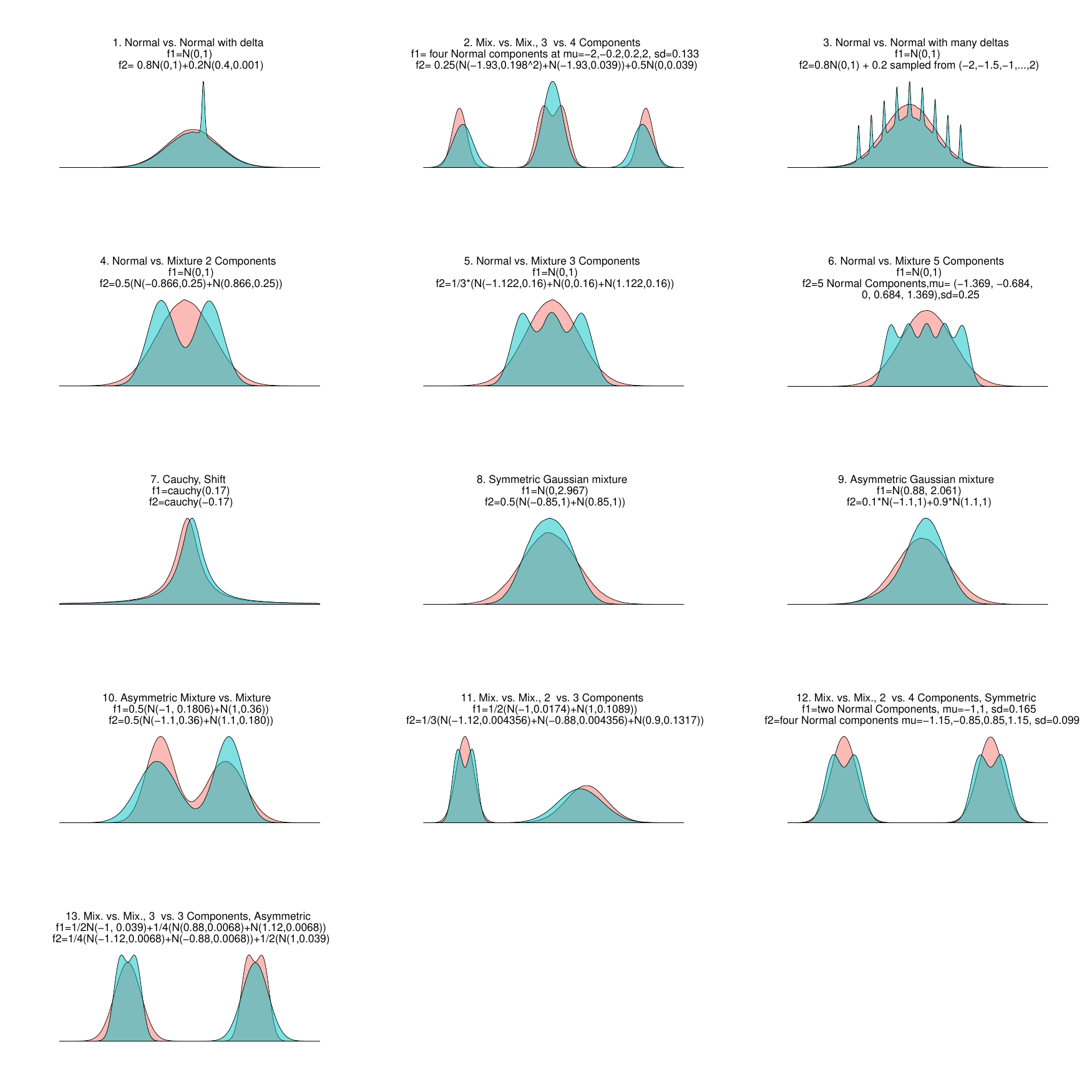}
  \caption{The two-sample problem in 13 different setups considered for $N=500$, which differ in the number of intersections of the densities, the range of support where the differences lie, and whether they are symmetric or not.}
  \label{fig:setups2sample}
\end{figure}

We examined the distributions depicted in Figure~\ref{fig:setups2sample}. The three scenarios in the third row were examined in \citet{Jiang14}. The remaining scenarios were chosen to have different numbers of intersections in the densities, ranging from 2 to 18, in order to examine the effect of partition size $m$ on power when the optimal partition size increases, as well as verify that the regularized statistic has good power. The scenarios also differ by the range of support of where the differences in the distributions lie (specifically, in the first and third scenario in the first row the difference between the distributions is very local), since this makes the comparison between the two aggregation methods interesting. We considered symmetric as well as asymmetric distributions.
 Gaussian shift  and scale setups were considered in Appendix \ref{app:additional simulations}. Such setups are less interesting in the context of this work, because if the two distributions differ only in shift or scale then specialized tests such as Wilcoxon rank-sum for shift will be preferable, but it is important to know that the suggested tests do not break down in this case. We used 20000 simulated data sets, in each of the configurations of Figure~\ref{fig:setups2sample}.

\begin{figure}[htbp]
  \centering
  \includegraphics[page=1, width=1\textwidth, trim=0in 0in 0in 0in, clip]{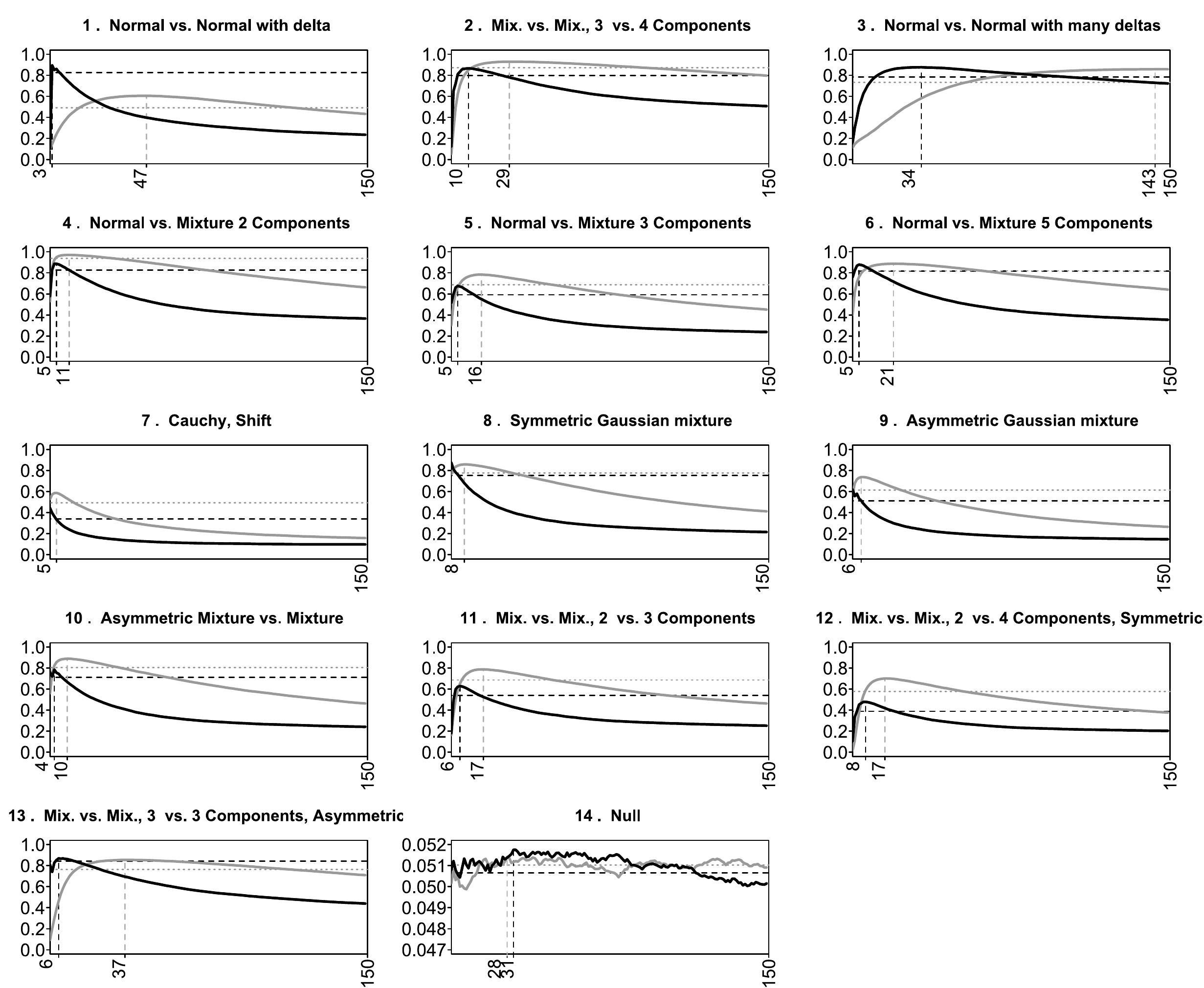}
  \caption{Estimated power with $N=500$ sample points for the $M_m$ (black) and $S_m$ (grey) statistics for $m\in \{2,\ldots, 149\}$  for the setups of Figure~\ref{fig:setups2sample}. The score per partition was the likelihood ratio test statitsic. The power of the minimum $p$-value is the horizontal dashed black line when it combines the $p$-values based on $M_m$, and the horizontal dotted grey line when it combines the $p$-values based on $S_m$. The vertical lines show the optimal $m$ for $M_m$ (black) and $S_m$ (grey).}
  \label{fig:pwr2sample}
\end{figure}

Table \ref{tab:competitors2sample} and Figure \ref{fig:pwr2sample} show the power for the setups in Figure \ref{fig:setups2sample}. These results show that if the number of intersections of the two densities is at least four, tests statistics with $m\geq 4$ have an advantage. Since the classical competitors, KS, CVM and AD, are based on $m=2$, they perform far worse in these setups. Moreover, although HHG and DS have better power than the classical tests, HHG is essentially an $m\leq 3$ test, and DS penalizes large $m$s severely, therefore their power is still too low when fine partitioning is advantageous. The minimum $p$-value statistic, which does not require to preset $m$, is remarkably efficient: in Figure \ref{fig:setups2sample} we see that in all settings considered, it is close to the power of the optimal $m$.

The choice of aggregation by maximization versus summation depends on how local the differences are between the distributions. In Figure  \ref{fig:pwr2sample} we see clearly that  when the differences are in very local areas, maximization achieves the greatest power and the test based on minimum $p$-value has more power if the aggregation is by maximization rather than by summation (setups 1, 2, and 13), and aggregation by summation is better otherwise. Note that the optimal $m$ for aggregation by summation is always larger than for aggregation by maximization. The reason is that in order to have a powerful statistic aggregated by maximization, it is enough to have one good partition (i.e., contain cells where the distributions clearly differ) for a fixed $m$, whereas by summation it is necessary to have a large fraction of good partitions among all partitions of size $m$.

\begin{table}[ht]
\caption{Power of competitors (columns 4-9), along with  the minimum $p$-value statistic using the $M_m$ $p$-values (column 2) and the $S_m$ $p$-values (column 3), for $N=500$. The score per partition was the likelihood ratio test statitsic.  The standard error was at most 0.0035. The advantage of the test based on the minimum $p$-value is large when the number of intersections of the two densities is at least four (setups 2,3,4,5,6,10,11,12, and 13). The best competitors are HHG and DS, but HHG is essentially an $m\leq 3$ test, and DS penalizes large $m$s severely, therefore in setups where $m\geq 4$ partitions are better they can perform poorly. Among the two variants in columns 2 and 3, the better choice clearly depends on the range of support in which the differences in distributions occur:  aggregation by maximum has better power when the difference between the distributions is very local (setups 1, 3, and 13), and aggregation by summation has better power otherwise. The highest power per row is underlined. }\label{tab:competitors2sample}
\scriptsize
\centering
\vspace{0.5cm}
\begin{tabular}{llcccccccc}
 & &\multicolumn{2}{c}{Min $p$-value aggreg.} &  &  &  &  &  &  \\
& Setup & by Max & by Sum& Wilcoxon & KS & CVM & AD & HHG & DS \\
& & & & & & & &   \\

1 & Normal vs. Normal with delta & 0.825 & 0.491 & 0.072 & 0.149 & 0.108 & 0.099 & 0.175 & \underline{0.849} \\ 
2  & Mix. Vs. Mix., 3  Vs. 4 Components & 0.799 & \underline{0.873} & 0.000 & 0.020 & 0.001 & 0.021 & 0.344 & 0.560 \\ 
3  & Normal vs. Normal with many deltas  & \underline{0.785} & 0.733 & 0.051 & 0.078 & 0.073 & 0.099 & 0.142 & 0.245 \\ 
4  & Normal vs. Mixture 2 Components& 0.827 & \underline{0.937} & 0.053 & 0.531 & 0.458 & 0.495 & 0.855 & 0.796 \\ 
 5  & Normal vs. Mixture 5 Components & 0.592 & \underline{0.686} & 0.048 & 0.238 & 0.179 & 0.246 & 0.484 & 0.556 \\ 
6  & Normal vs. Mixture 10 Components & 0.818 & \underline{0.820} & 0.048 & 0.240 & 0.211 & 0.310 & 0.561 & 0.789 \\ 
7  & Cauchy, Shift & 0.339 & 0.492 & 0.542 & 0.620 & 0.627 & 0.577 & \underline{0.641} & 0.436 \\ 
8  & Symmetric Gaussian mixture & 0.752 & 0.775 & 0.033 & 0.194 & 0.242 & 0.617 & 0.749 & \underline{0.835} \\ 
9  & Asymmetric Gaussian mixture & 0.512 & 0.613 & 0.050 & 0.253 & 0.277 & 0.469 & \underline{0.678} & 0.599 \\ 
10  & Asymmetric Mixture vs. Mixture   & 0.711 & \underline{0.806} & 0.000 & 0.159 & 0.119 & 0.395 & 0.690 & 0.747 \\ 
11  & Mix. Vs. Mix., 2  Vs. 3 Components & 0.540 & \underline{0.686} & 0.004 & 0.093 & 0.057 & 0.116 & 0.302 & 0.440 \\ 
 12  & Mix. Vs. Mix., 2  Vs. 4 Components, Symmetric & 0.390 & \underline{0.577} & 0.000 & 0.005 & 0.000 & 0.005 & 0.079 & 0.270 \\ 
13  & Mix. Vs. Mix., 3  Vs. 3 Components, Asymmetric & \underline{0.844} & 0.764 & 0.000 & 0.001 & 0.000 & 0.013 & 0.042 & 0.780 \\ 
14  & Null & 0.051 & 0.051 & 0.050 & 0.043 & 0.050 & 0.050 & 0.050 & 0.050 \\

\end{tabular}
\end{table}

\subsection{The independence problem}\label{subsec-simind}
We examined the power properties of the ADP and DDP statistics, aggregated by summation for $m\in \{1, \ldots, \sqrt{N}\}$, aggregated by maximization for $m\leq 4$, as well as the minimum $p$-value statistic $\min_{\{m\in 2,\ldots, m_{max}\}}p_m$ based on aggregation by summation. We display here the results for $m_{\max}=10$ and $N=100$.

We compared these tests to seven tests of independence suggested in the literature. We compared to Spearman's $\rho$, since it is perhaps the most widely used test to detect monotone associations. 
We also compared to previous tests suggested in the literature with the same two important properties as our suggested tests, namely proven consistency and distribution-freeness, as well as an available implementation: the test of \citet{hoeffding1948non}, referred to as Hoeffding; the test of \citet{Reshef11}, referred to as MIC; the tests of \citet{Szekely07} and \citet{Heller12} that first transform the observations of each variable into ranks, referred to as dCov and HHG, respectively. We note that the power of the original dCov and HHG was fairly similar to the power of their distribution-free variants, see Appendix \ref{app-dcovHHGondata}.

\begin{figure}[htbp]
  \centering
  \includegraphics[page=1, width=0.8\textwidth, trim=0in 0in 0in 0in, clip]{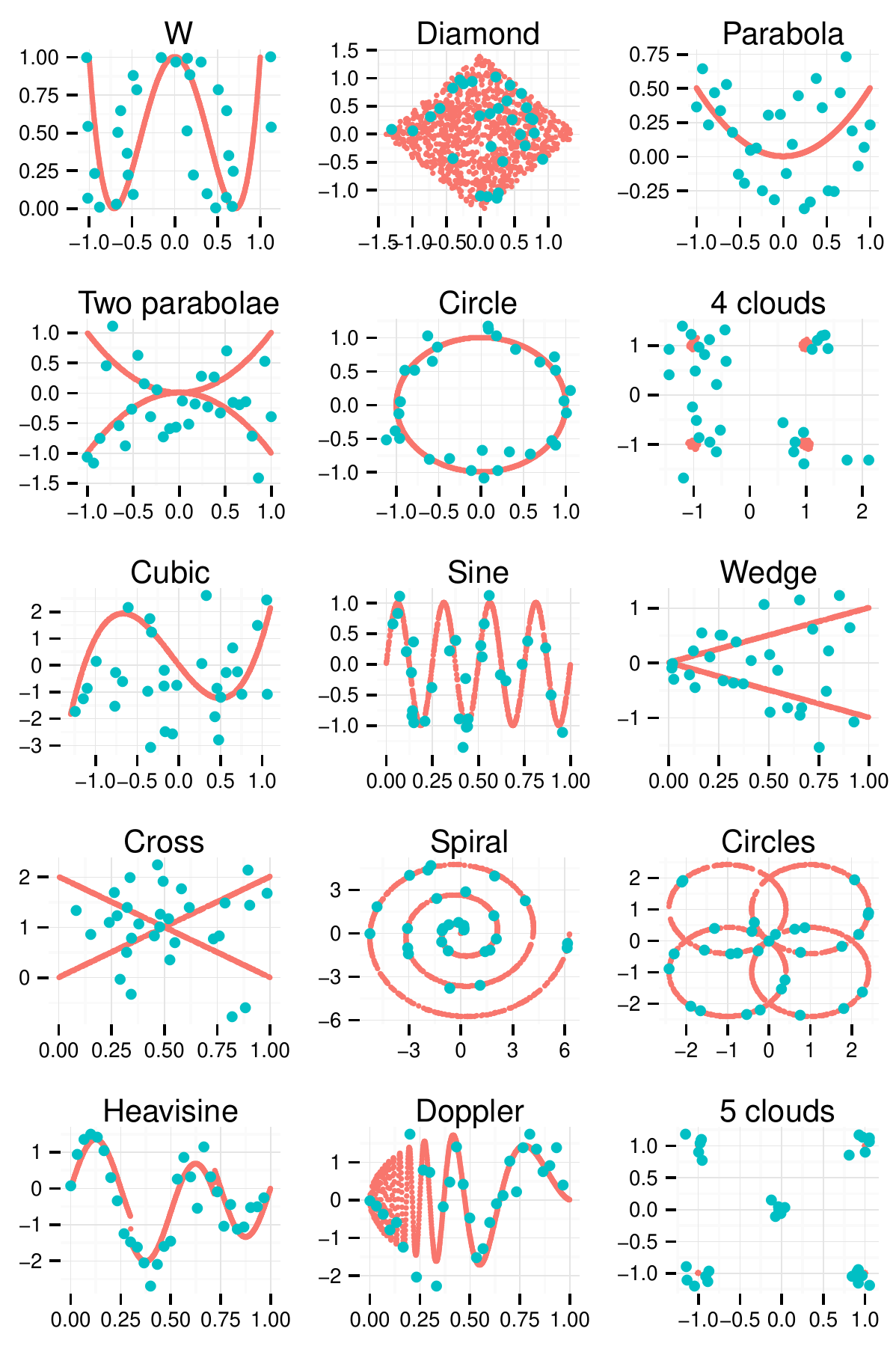}
  \caption[Nonmonotonic simulation setups]{Bivariate relationships (in red), along with a sample $N=100$ noisy observations (in blue). The ``four clouds'' relationship is a null relationship, where the two random variables are independent.}
  \label{fig:setups}
\end{figure}

We examine complex bivariate relationships depicted in Figure~\ref{fig:setups}. Most of these scenarios were collected from the literature illustrating the performance of other methods. Specifically, the first two rows were examined in \citet{Newton09},  the next two rows are similar to the relationships examined in \citet{Reshef11}, and the Heavisine and Doppler examples in the last row were used extensively in the literature on denoising, see e.g., \citet{Donoho95}. In all but the 4 Independent Clouds setup, there is dependence. The 4 Independent Clouds setup allows us to verify that the tests maintain the nominal level. We used 2000 simulated data sets for $N=100$ and $N=300$, in each of the configurations of Figure~\ref{fig:setups}.
Monotone setups are presented in Appendix \ref{sec:mono_simulations} in Figure~\ref{fig:mic_setups_mono}. Monotone setups are less interesting in the context of this work, because if there is reason to believe that the dependence is monotone, specialized tests such as Spearman's $\rho$ or Kendall's $\tau$ will be preferable, but it is important to know that they still have reasonable power, as demonstrated in the results in Appendix \ref{sec:mono_simulations}, Figure~\ref{fig:mic_results_mono} and Table~\ref{tbl:mono}.

\begin{figure}[htbp]
  \centering
  \includegraphics[page=1, width=1\textwidth, trim=0in 0in 0in 0in, clip]{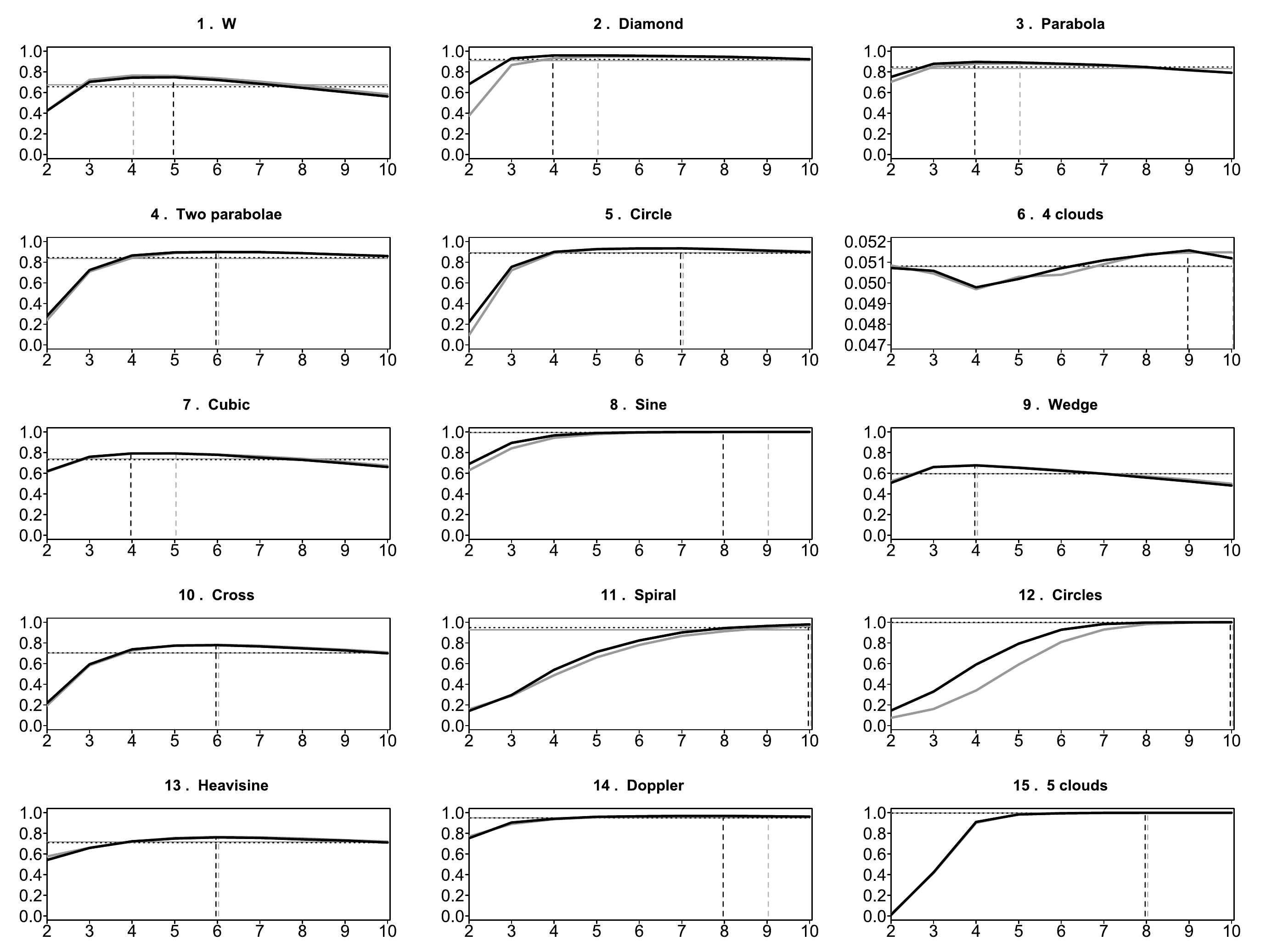}
  \caption{Estimated power as a function of partition size $m$, with $N=100$ sample points, for the DDP (black) and ADP (grey) summation variants using the likelihood ratio score for the setups of Figure~\ref{fig:setups}. The score per partition was the likelihood ratio test statitsic. For DDP (black) and ADP (grey), the horizontal dashed line is the power of the minimum $p$-value statistic, and the vertical lines is the optimal $m$.}
  \label{fig:pwr_100}
\end{figure}

Tables~\ref{tab:competitors} and~\ref{tab:max_variants}, and Figure~\ref{fig:pwr_100} show the power for the settings depicted in Figure~\ref{fig:setups}. We only considered the test based on the DDP minimum $p$-value statistic in Tables~\ref{tab:competitors} and~\ref{tab:max_variants}, since for the minimum $p$-value statistic the tests of ADP and DDP are almost identical.
These results provide strong evidence that for non-monotone noisy dependencies our tests have excellent power properties. Specifically, $S^{DDP}_{m\times m}$ with $m\in \{3,\ldots,10\}$ is more powerful than all other tests in Table~\ref{tab:competitors} in most settings. For example, it had greater power than all competitors in 9 settings with $m=4$ and in 11 settings with $m=5$, out of the 14 non-null settings. The test based on the minimum $p$-value has greater power than all competitors in 7 settings, and it is very close to the best competitor in most of the other settings.
 The MIC is best for the Sine example but performs poorly in all other examples. The minimum $p$-value statistic is a close second best in the Sine example, with a difference of only 0.005 from MIC, yet all other tests are more than 0.19 below MIC in power.  Overall, the HHG test is the best competitor, but its power is lower than that of the minimum $p$-value statistic when the optimal $m$ is greater than 4. 
 Table~\ref{tab:max_variants} shows that when aggregating by maximization, the choice of $m$ matters and the power is higher for $m>2$. However, the minimum $p$-value statistic, which is aggregated by summation and considers finer partitions, is more powerful for most settings, and is a close second in the remaining settings.

\begin{table}
  \caption{Power of competitors (columns 3--7), along with the DDP minimum $p$-value statistic for  $N=100$.  The standard error is at most $0.011$. The score per partition was the likelihood ratio test statitsic. The  DDP minimum $p$-value statistic performs very well in comparison to the other tests.  Although the competitors may have greater power in some examples, the advantage is usually small. By far the best competitor is HHG, yet it has a disadvantage when the relationship is more complex, thus benefiting from the finer partition of the minimum $p$-value test, especially in the Sine, Heavisine, Spiral and Circles examples. The highest power per row is underlined.}
  \label{tab:competitors}
  \scriptsize
  \centering
  \vspace{0.5cm}
  \begin{tabular}{lcccccc}
   Setup & $\min_{\{m\in 2,\ldots, 10\}}p_m$  & Spearman & Hoeffding & MIC & dCov & HHG  \\
  & & & & & &  \\
W & 0.655 & 0.000 & 0.414 & 0.526 &  0.361 & \underline{0.798} \\
  Diamond & 0.919 & 0.013 & 0.116 & 0.074 &  0.074 & \underline{0.965} \\
  Parabola & \underline{0.847} & 0.028 & 0.413 & 0.211  & 0.386  & 0.784 \\
  2Parabolas & \underline{0.844} & 0.095 & 0.135 & 0.048  & 0.124  & 0.723 \\
  Circle & \underline{0.886} & 0.000 & 0.033 & 0.046  & 0.002 &  0.850 \\
  Cubic & 0.731 & 0.344 & 0.655 & 0.515 &  0.627  & \underline{0.768} \\
  Sine & 0.995 & 0.368 & 0.494 & \underline{1.000} & 0.415 & 0.804 \\
  Wedge & 0.595 & 0.064 & 0.360 & 0.303 & 0.338 & \underline{0.673} \\
  Cross & 0.704 & 0.089 & 0.160 & 0.069 & 0.130 & \underline{0.706} \\
  Spiral &\underline{0.949} & 0.112 & 0.141 & 0.251 &  0.140  & 0.337 \\
  Circles & \underline{0.999} & 0.048 & 0.084 & 0.093 &  0.061 & 0.354 \\
  Heavisine & \underline{0.710} & 0.396 & 0.493 & 0.532  & 0.492  & 0.585 \\
  Doppler & 0.949 & 0.513 & 0.784 & \underline{0.975}  & 0.744  & 0.912 \\
  5Clouds & \underline{0.996} & 0.000 & 0.000 & 0.561  & 0.004  & 0.904 \\
  4Clouds & 0.051 & 0.050 & 0.057 & 0.050  & 0.051  & 0.050 \\
  \end{tabular}
\end{table}

\begin{table}
  \caption{The power of different variants aggregated by maximization (columns 3--6), along with the DDP minimum $p$-value statistic (column 2) for $N=100$. The standard error is at most $0.011$. The score per partition was the likelihood ratio test statitsic. Although maximization is better than summation in some examples, the advantage is usually small. The advantage of the minimum $p$-value statistic, which is based on aggregation by summation, is large in the Cubic, Cross, Spiral, and Circles relationships. In most examples, power increases with $m$. The power differences between the ADP and DDP variants are small.  We present only the maximum variants that take at most $O(N^3)$ to compute, therefore for ADP only results with $m=2$ are presented. The highest power per row is underlined.}
  \label{tab:max_variants}
  \scriptsize
  \centering
  \vspace{0.5cm}
  \begin{tabular}{l c ccc c}
   Setup & $\min_{\{m\in 2,\ldots, 10\}}p_m$& $M_{2\times 2}^{DDP}$ & $M_{3\times 3}^{DDP}$ & $M_{4\times 4}^{DDP}$ & $M_{2\times 2}^{ADP}$\\
  &&&&& \\
W & \underline{0.655} & 0.190 & 0.637 & 0.574 & 0.155 \\
  Diamond & 0.919 & 0.272 & \underline{0.931} & 0.912 & 0.247 \\
  Parabola & 0.847 & 0.533 & \underline{0.855} & 0.803 & 0.597 \\
  2Parabolas & 0.844 & 0.466 & \underline{0.907} & 0.897 & 0.578 \\
  Circle & \underline{0.886} & 0.222 & 0.880 & 0.884 & 0.170 \\
  Cubic & \underline{0.731} & 0.496 & 0.654 & 0.653 & 0.496 \\
  Sine & 0.995 & 0.768 & 0.958 & \underline{0.998} & 0.774 \\
  Wedge & \underline{0.595} & 0.410 & 0.536 & 0.478 & 0.455 \\
  Cross & \underline{0.704} & 0.268 & 0.680 & 0.673 & 0.341 \\
  Spiral & \underline{0.949} & 0.116 & 0.489 & 0.764 & 0.189 \\
  Circles & \underline{0.999} & 0.085 & 0.606 & 0.844 & 0.088 \\
  Heavisine & \underline{0.710} & 0.519 & 0.642 & 0.692 & 0.534 \\
  Doppler & 0.949 & 0.828 & 0.969 & \underline{0.977} & 0.833 \\
  5Clouds & 0.996 & 0.062 & \underline{0.999} & \underline{0.999} & 0.076 \\
  4Clouds & 0.051 & 0.052 & 0.051 & 0.051 & 0.052 \\
  \end{tabular}
\end{table}

\section{Application to real data}
\label{sec:real_data}

We examine the co-dependence between pairs of genes on chromosome 1 in
the yeast gene expression dataset from \citet{hughes2000functional}, which contained $N=300$ expression levels. After removing genes with missing values, we had $94$  genes and a family of $\ibinom{94}{2} = 4371$ pairs to examine simultaneously. Each pair was tested for independence by the tests of Spearman, Hoeffding, MIC, dCov and HHG on ranks, as well as by our new tests with $m$ ranging from 2 to $m_{\max} = 17$. The null tables were based on $20000$ permutations for  $N=300$.  The adjusted $p$-values from the Benjamini--Hochberg procedure~\citep{yoav1} were computed for each test statistic. 

\begin{figure}[htbp]
  \centering
  \includegraphics[width=0.5\textwidth, trim=0in 0in 0in 0in, clip]{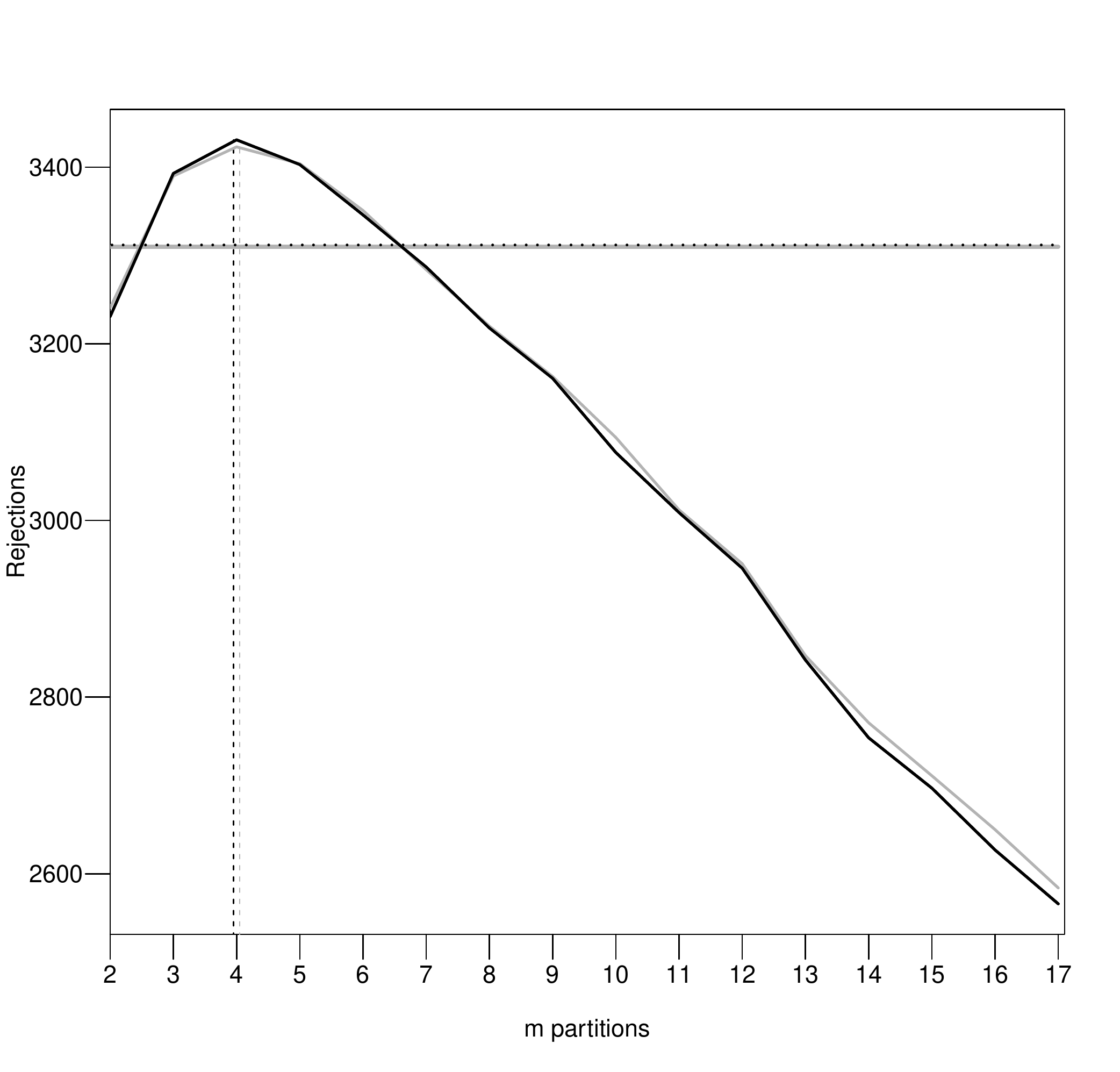}
  \caption{Number of discoveries of genes with associated expression patterns in the data of \citet{hughes2000functional}, by the Benjamini--Hochberg procedure at level $0.05$ using $S_{m\times m}^{DDP}$ (black) and $S_{m\times m}^{ADP}$ (gray) for $m=2,\ldots, 17$ . In addition, number of rejections using the minimum $p$-value statistic  with $m_{\max} = 17$, using the DDP (black horizontal line) and the ADP (gray horizontal line).}
  \label{fig:rosetta}
\end{figure}

Table~\ref{tbl:gene_expression} shows the pairwise agreements between the Benjamini--Hochberg procedure at level $0.05$ using the different test statistics considered in each row, with the minimum $p$-value statistic based on DDP. Clearly, a large number of pairwise associations are missed when testing is performed with Spearman's $\rho$ compared to the minimum $p$-value statistic, and only a small number of gene pairs detected with Spearman are missed by the  minimum $p$-value statistic (row 1 in Table~\ref{tbl:gene_expression}). These findings contradict an earlier examination of the data. \citet{steuer2002mutual} concluded that the most widely used approach for pairwise association testing, namely Spearman correlation, performs equivalently to a mutual information based testing approach. The authors speculated that actual dependencies, if any, are linear. The number of discoveries using MIC, Hoeffding, and dCov are much smaller than using the minimum $p$-value statistic. HHG on ranks also discovers less co-dependencies compared with the minimum $p$-value statistic. The agreement between the tests based on DDP and ADP was very high, as seen in the last row of Table~\ref{tbl:gene_expression} and in Figure~\ref{fig:rosetta}, which shows the number of rejections for $S_{m\times m}^{DDP}$ and $S_{m\times m}^{ADP}$ for $m=2,\ldots,17$. We conclude that in this dataset there are many nonlinear associations, but powerful tests are necessary in order to detect such associations in light of the large number of simultaneous tests that have to be carried out, and that the suggested tests can be valuable tools for this task.

Note that the data had ties due to low precision of the documented expression levels. Ties were broken randomly, see remark~\ref{rem:ties}. Repeated analysis with different seeds provided similar results. 

\begin{table}
  \caption{Benjamini--Hochberg rejections at level $0.05$ for the gene expression problem of \citet{hughes2000functional}. For different test statistics (rows), the number of rejections (column 2), and their intersection with the rejections using the minimum $p$-value statistic on DDP (column 3). The minimum $p$-value statistic on DDP had the highest number of rejections, 3312.}
  \label{tbl:gene_expression}
  \centering
  \footnotesize
  \begin{tabular}{lcc}
    \textbf{Test} & \textbf{Number of rejections} & \textbf{Number of intersections} \\ 
    Spearman & 2488 & 2445 \\
    MIC & 245 & 245 \\
    Hoeffding & 2890 & 2844 \\
    HHG on ranks & 3283 & 3199 \\
    dCov on ranks & 2889 & 2845 \\
    minimum $p$-value based on ADP & 3310 & 3294\\
  \end{tabular}
\end{table}

\section{Discussion}\label{sec:Discussion}
In this paper we proposed new partition-based test statistics for both the independence problem and the two-sample problem. We proved that the statistics are consistent for general alternatives and demonstrated in simulations that the power advantage of the tests based on finer partitions can be great.
We further showed that the power of our regularized statistics is very close to that of the statistics based on the optimal partition size. We recommend the test using  the minimum $p$-value statistic based on aggregation by summation, unless the alternative is suspected to be of very local nature. Specifically, in the $K$-sample problem if the difference between the distributions is on a very small range of the support, then the aggregation by maximization is preferred over aggregation by summation.

The algorithms described in Section \ref{subsec-sumalgs} for the test of independence based on regularized scores for a range of $m\times m$ partitions can easily be generalized to include $m_x\times m_y$ partitions, where $m_x\neq m_y$, with the same complexity for the ADP statistic (for the DDP statistic $m_x=m_y$). Considering unequal partition sizes for $X$ and $Y$ is expected to improve power when the (unknown) optimal partition has an $m_x$ value very different than the $m_y$ value. Moreover, when the (unknown) optimal partition has $m_x\approx m_y$, the power loss from considering the minimum $p$-value over all $m_x\times m_y$ values instead of over all $m\times m$ values is expected to be small.

The algorithms we suggested for the $K$-sample problem are $O(N^2)$ and therefore are feasible even for large $N$. For the test of independence, even though the complexity of our suggested algorithms is $O(N^4)$, for small $N$ these algorithms can be quite efficient in the following quite common multiple testing setting in modern studies.  If $M$ hypotheses are simultaneously examined with the same sample size, then the computational complexity of using our distribution-free tests is $O(M\times N^4)$ if null table  is available for this $N$, or $\max\{O(M\times N^4),O(B\times N^4)\}$ if the null table is generated by the user using $B$ Monte-Carlo replicates for the sample size $N$.  
If we needed to recompute the null distribution for every one of the $M$ hypotheses (as required for permutation tests that are not distribution-free, such as dCov and HHG), then the computational complexity would have been $O(M\times B\times N^4)$, which may be infeasible in modern studies where the number of hypotheses tests simultaneously examined can be several thousands or hundreds of thousands.  Since the null distribution needs to be generated only once in order to compute the significance of all the $M$ test statistics, due to the distribution-free property of our tests, they can be  feasible with today's computing power even for a few thousands samples. However, computing $O(N^4)$ test statistics is unfeasible for larger sample sizes. To reduce the computational complexity when $N$ is large, statistics which do not go over all partitions but rather just over a representative sample can be considered. This approach was used for example in  Jiang (2014) for the $K$-sample problem. A simple way of doing this is to divide the data into $\sqrt N \times \sqrt N$ bins and only consider partitions that do not break up these bins. We expect such statistics to also be consistent and the algorithms that accompany them to be computable in $O(N^2)$. 

If one expects relatively simple dependence structures, for large $N$, the $S^{DDP}_{3\times 3}$ is recommended, since it is both distribution-free and computable in $O(N^2)$ (see Remark \ref{rem-smallmalgs}). In our simulations it was as powerful as HHG and more powerful than dCov, and it has the advantage over HHG of being distribution-free. 

A thorough investigation of the suggested mutual information estimator in Section \ref{sec:new_tests} was outside the scope of this paper, but is of interest for future research. We suspect the asymptotic distribution of our mutual information estimator has a simple form. The bias of the estimator can be dealt with by modifying our estimator, and our algorithms accordingly, to only include partitions with cells of a minimum size, and by bias correction methods suggested in the literature, e.g., \cite{vu2007}. Although in this work we limited ourselves to a theoretical examination of the ADP summation statistic for mutual information estimation, we recognize that an estimator based only on the DDP may be useful, and we plan to explore it in the future.

\newpage

\appendix
\section{Proof of Theorem \ref{thm-consistency}}
\label{sec:test_consistency_proof}

\subsection{The DDP test}\label{subsec-proofDDPconsistency}
Denote $S_{m\times m} = S^{DDP}_{m\times m}$. For simplicity, we show the proof using Pearson's test statistic. The proof using the likelihood ratio test statistic is very similar and therefore omitted.
We want to show that for an arbitrary fixed $\alpha\in (0,1)$, if $H_0$ is false, then  $\lim_{N\rightarrow \infty} Pr(S_{m\times m}> S_{1-\alpha}^{tab})=1$, where $S_{1-\alpha}^{tab}$ denotes the $1-\alpha$ quantile of the null distribution of $S_{m\times m}$.

If $H_0$ is false, then without loss of generality $h(x_0,y_0)> f(x_0)g(y_0)$. Moreover, there exists a distance $R>0$ such that
$h(x,y)> f(x)g(y)$ for all points $(x,y)$ in the set $\Acal = \{(x,y):  x_0\leq x \leq x_0+R, y_0\leq y \leq y_0+R\}$. The set $\Acal$ has positive probability, and moreover $$\min_\Acal[h(x,y)- f(x)g(y)]>0.$$ Denote this minimum by $c>0$. Clearly, the following two subsets of $\Acal$ have positive probability as well:
\begin{eqnarray}
&& \Acal_1 = \{(x,y):  x_0\leq x \leq x_0+R/4, y_0\leq y \leq y_0+R/4\} \nonumber \\
&&  \Acal_2 = \{(x,y):  x_0+3R/4\leq x \leq x_0+R, y_0+3R/4\leq y \leq y_0+R\}. \nonumber
\end{eqnarray}
Denote the probabilities of $\Acal_1$ and $\Acal_2$ by $f_1$ and $f_2$, respectively. \\
 Let $\Gamma\{(x_1,y_1),\ldots,(x_N,y_N)\}$ be the set of partitions of size $m$
 based on at least one sample point in $\Acal_1$ and  on at least one sample point in $\Acal_2$. Let $N_i$ denote the number of sample points in $\Acal_i, i\in \{1,2\}$. Let $\Ical\in \Gamma\{(x_1,y_1),\ldots,(x_N,y_N)\}$ be such a (arbitrary fixed) partition. So for $\Ical$ there exists $(i,j)\subseteq \Ical$ such that $(x_i,y_i)\in \Acal_1$ and $(x_j,y_j)\in \Acal_2$. Consider the cell $C$ defined by the two points $(i,j)$.

 The fraction of observed counts in the cell $C$ is a linear combination of empirical cumulative distribution functions
 $$\frac{o_C}{N-(m-1)} = \hat F_{XY} (x_i,y_i) + \hat F_{XY} (x_j,y_j)-\hat F_{XY} (x_i,y_j)-\hat F_{XY} (x_j,y_i),  $$
and the expected fraction under the null is a function of the marginal cumulative  distributions
$$\frac{e_{C}}{N-(m-1)} = \{\hat F_{X} (x_j) -\hat F_{X} (x_i)\}\{\hat F_{Y} ( y_j) -\hat F_{Y} (y_i)\}. $$
where $\hat F$ denotes the empirical distribution function based on $N-(m-1)$ sample points.

By the Glivenko-Cantelli theorem, uniformly almost surely,
\begin{eqnarray}
&& \lim_{N\rightarrow \infty} \left(\frac{o_C}{N-(m-1)} -\int_{\{(x,y): x\in (x_i,x_j], y \in (y_i,y_j]\}} h(x,y)dx dy\right) = 0 , \nonumber \\
&& \lim_{N\rightarrow \infty} \left\{\frac{e_C}{N-(m-1)} -\left(\int_{\{x: x\in (x_i,x_j]\}} f(x)dx\right) \left(\int_{\{y: y\in (y_i,y_j]\}} g(y)dy\right)\right\} = 0. \nonumber \\
\end{eqnarray}

Therefore, by Slutsky's theorem and the continuous mapping theorem, we have that uniformly almost surely
\begin{eqnarray}
&& \lim_{N\rightarrow \infty}\frac{1}{N-(m-1)}\frac{(o_C-e_C)^2}{e_C}
= \lim_{N\rightarrow \infty}\frac{\left(\frac{o_C}{N-(m-1)} -\frac{e_C}{N-(m-1)}\right)^2 }{\frac{e_C}{N-(m-1)}} \nonumber \\
&& \geq  \lim_{N\rightarrow \infty}\left(\frac{o_C}{N-(m-1)} -\frac{e_C}{N-(m-1)}\right)^2  \nonumber \\
&&   = \lim_{N\rightarrow \infty}\left[\int_{\{(x,y): x\in (x_i,x_j], y \in (y_i,y_j]\}} \{h(x,y)-f(x)g(y)\}dx dy\right]^2 , \label{eq-GC1}
\end{eqnarray}
where the inequality follows from the fact that $\frac{e_C}{N-(m-1)}\leq 1$.

We shall show that this limit can be bounded from below by a positive constant that depends on $(x_0,y_0)$ and $R$ but not on $\Ical$. Since
\begin{eqnarray}
&& \{(x,y): x\in (x_0+R/4,x_0+3R/4], y \in (y_0+R/4,y_0+3R/4] \} \nonumber \\
&& \subseteq \{(x,y): x\in (x_i,x_j], y \in (y_i,y_j]\},\nonumber
\end{eqnarray}
a positive lower bound on expression (\ref{eq-GC1}) can be obtained:
\begin{eqnarray}
&&  \lim_{N\rightarrow \infty}\left[\int_{\{(x,y): x\in (x_i,x_j], y \in (y_i,y_j]\}} \{h(x,y)-f(x)g(y)\}dx dy\right]^2  \nonumber \\
&&\geq \left[\int_{\{(x,y): x\in (x_0+R/4,x_0+3R/4], y \in (y_0+R/4,y_0+3R/4] \}} \{h(x,y)-f(x)g(y)\}dx dy\right]^2 \nonumber \\
&&\geq c^2 \int_{\{(x,y): x\in (x_0+R/4,x_0+3R/4], y \in (y_0+R/4,y_0+3R/4] \}} dx dy = c^2R^2/4, \label{eq-auxGC1}
\end{eqnarray}
where the first inequality follows since $h(x,y)-f(x)g(y)>0$ in $\Acal$, and the second inequality follows since the minimum value is $c>0$.
Therefore, it follows that $\frac{1}{N-m+1}\frac{(o_C-e_C)^2}{e_C}$ converges uniformly almost surely to a positive constant greater than $c' = c^2 R^2/4,$
\begin{eqnarray}\label{eq-prunderaltcell}
 Pr\left(\lim_{N\rightarrow \infty} \frac{1}{N-m+1}\frac{(o_C-e_C)^2}{e_C}\geq c'\right) =1.
\end{eqnarray}

The partition $\Ical$ either contains the cell $C$, or a group of cells that divide $C$. By Jensen's inequality, it follows that if the partition $\Ical$ contains a group of cells that divide $C$, the score is made larger, since for any partition of the cell $C$, $C=\cup_lC_l$, $$\left(\frac{o_C-e_C}{e_C}\right)^2 = \left(\sum_l \frac{e_{C_l} \left(\frac{o_{C_l}}{e_{C_l}}-1\right)}{\sum_h e_{C_h}}\right)^2 \leq \frac{\sum_l e_{C_l} \left(\frac{o_{C_l}}{e_{C_l}}-1\right)^2}{\sum_le_{C_l}}=\frac{\sum_l\frac{(o_{C_l}-e_{C_l})^2}{e_{C_l}}}{e_C},$$
and therefore
\begin{equation}\label{eq-finerpartition}
\frac{(o_C-e_C)^2}{e_C}\leq \sum_l\frac{(o_{C_l}-e_{C_l})^2}{e_{C_l}}.
\end{equation}

Since $\sum_l\frac{(o_{C_l}-e_{C_l})^2}{e_{C_l}}$ or $\frac{(o_C-e_C)^2}{e_C}$ is part of the sum that defines $T^{\Ical}$, it follows from equations (\ref{eq-finerpartition}) and (\ref{eq-prunderaltcell}) that $\frac{T^{\Ical}}{N-m+1}$ converges uniformly almost surely to a positive constant greater than $c'$.

Let $|\Gamma|$ denote the cardinality of $\Gamma\{(x_1,y_1),\ldots,(x_N,y_N)\}$. Since  $\Ical\in \Gamma\{(x_1,y_1),\ldots,(x_N,y_N)\}$ was arbitrary fixed, and since the convergence for fixed $\Ical$ of $t_C/[N-(m-1)]$ to a limit bounded from below by a positive constant was uniform, 
it follows that $\frac{1}{|\Gamma|}\sum_{\Ical\in \Gamma}  \frac{T^\Ical}{N-m+1}$ converges almost surely to a positive constant at least as large as $c'$. To see this, note that from the uniform convergence in equation (\ref{eq-prunderaltcell}), it follows that for an arbitrary fixed $\epsilon>0$, there exists $N(\epsilon)$ (which does not depend on $C$) such that for all $N>N(\epsilon)$, $\frac{1}{N-(m-1)}\frac{(o_C-e_C)^2}{e_C}\geq c'-\epsilon$ for every $C$, and therefore that    $\frac{1}{|\Gamma|}\sum_{\Ical\in \Gamma}  \frac{T^\Ical}{N-m+1}\geq c'-\epsilon$ for all $N>N(\epsilon)$. 

Since $S_{m\times m}\geq \sum_{\Ical \in \Gamma}  T^\Ical$, it follows that almost surely
\begin{eqnarray}\label{eq-prunderalt}
 \lim_{N\rightarrow \infty} \frac{S_{m\times m}}{|\Gamma|(N-m+1)}> c'.
\end{eqnarray}

We shall show that $\lim_{N\rightarrow \infty} |\Gamma|/\binom{N}{m-1}$ is bounded below by a positive constant. First, we shall consider the case where $m$ is finite. Then, a subset of $\Gamma\{(x_1,y_1),\ldots, (x_N,y_N)\}$ is the set of all partitions with $m-2$ sample points  in $\Acal_1$, and one sample point in $\Acal_2$. Therefore,   $|\Gamma|\geq \binom{N_{1}}{m-2}N_{2}$.  Simple algebraic manipulations lead to the following expression for $\frac{\binom{N_{1}}{m-2}N_{2}}{\binom{N}{m-1}}$:
$$
(m-1)\frac{N_{2}}{N}\frac{N_{1}}{N-1}\cdots \frac{N_{1}-m+3}{N-m+2}.
$$
Since $N_{1}/N$ converges almost surely to $f_1$ , and similarly $N_{2}/N$ converges almost surely to $f_2$, then for $m\geq 3$ finite it follows that $\frac{\binom{N_{1}}{m-2}N_{2}}{\binom{N}{m-1}}$ converges almost surely to a positive constant.  Therefore, $|\Gamma|/\binom{N-1}{m-1}$ is bounded away from zero.
Second, we shall consider the case that $m\rightarrow \infty$. The complement of $\Gamma$, $\Gamma^C$,  is the set of contingency tables with no points in $\Acal_1$ or in $\Acal_2$. An upper bound for $|\Gamma^C|$ is
$$
\binom{N-N_{1}}{m-1}+\binom{N-N_{2}}{m-1}.
$$
Note that in order to show that $\lim_{N\rightarrow \infty} |\Gamma|/\binom{N}{m-1}$ is bounded below  by a positive constant,  since $|\Gamma| =\binom{N}{m-1}-|\Gamma^C|$, it is enough to show that   $|\Gamma^C|/\binom{N}{m-1}$ converges to zero as $N\rightarrow \infty$. Simple algebraic manipulations lead to the following expression for $\frac{\binom{N-N_{1}}{m-1}}{\binom{N}{m-1}}$:
$$
\left(1-\frac{N_{1}}{N}\right)\cdots\left(1-\frac{N_{1}}{N-(m-2)}\right)\leq \left(1-\frac{N_{1}}{N}\right)^m.
$$
Since $N_{1}/N$ converges almost surely to $f_1\in (0,1)$, it follows that $\frac{\binom{N-N_{1}}{m-1}}{\binom{N}{m-1}}$ converges almost surely to zero as $m\rightarrow \infty$. Similarly, since $N_{2}/N$ converges almost surely to $f_2\in (0,1)$, it follows that  $\frac{\binom{N-N_{2}}{m-1}}{\binom{N}{m-1}}$ converges almost surely to zero. Therefore, $|\Gamma^C|/\binom{N}{m-1}$ converges to zero as $N\rightarrow \infty$.

Since $\lim_{N\rightarrow \infty} |\Gamma|/\binom{N}{m-1}$ is bounded below by a positive constant, it follows from (\ref{eq-prunderalt}) that almost surely
\begin{eqnarray}\label{eq-prunderalt2}
 \lim_{N\rightarrow \infty} \frac{S_{m\times m}}{\binom{N}{m-1}(N-m+1)}\geq c'',
\end{eqnarray}
for some constant $c''>0$.

Consider now a random permutation $(\pi y_1),\ldots,(\pi y_N)$ of the $y$-values $y_1,\ldots,y_N$. Let $S_{m\times m}^\pi$ be the test statistic  that is computed from the data $(x_1,\pi y_1),\ldots,(x_N,\pi y_N)$.
Therefore, by Markov's inequality,
\begin{eqnarray}
&& Pr\left(S^{\pi}_{m\times m}\geq c''\binom{N}{m-1}(N-m+1) \,\mid\, \vec x, \vec y \right)\leq \frac{E(S^{\pi}_{m\times m} \vert \vec x, \vec y)}{ c''\binom{N}{m-1}(N-m+1)} \nonumber \\
& \approx & \frac{\binom{N}{m-1}(m-1)^2}{ c''\binom{N}{m-1}(N-m+1)}. \label{eq-prundernull}
\end{eqnarray}
where $\vec x  = (x_1,\ldots,x_N)$ and $\vec y = (y_1,\ldots,y_N)$.
The approximation in (\ref{eq-prundernull}) becomes more accurate the larger $N$ is, since each of the contingency tables is approximately $\chi^2$ with $(m-1)^2$ degrees of freedom. The right hand side of equation (\ref{eq-prundernull}) goes to 0 as $N\rightarrow \infty$, as long as $\lim_{N\rightarrow \infty} \frac{m}{\sqrt{N}}=0$.
Thus,
\begin{equation}\label{eq-UBundernull2}
\lim _{N\rightarrow \infty, m/\sqrt{N} \rightarrow 0}  Pr\left(S^{\pi}_{m\times m}\geq  c''\binom{N}{m-1}(N-m+1) \,\mid \, \vec x, \vec y \right) = 0.
\end{equation}

We now have all the necessary results to complete the proof. Specifically, 
$$
 \lim_{N\rightarrow \infty} Pr(S_{m\times m}\leq S_{1-\alpha}^{tab}) \leq \lim_{N\rightarrow \infty} Pr\left\{S_{m\times m} \leq c''\binom{N}{m-1}(N-m+1)\right\} =0,
 $$
 where the inequality follows from (\ref{eq-UBundernull2}), since $S_{1-\alpha}^{tab}$ is below $c''\binom{N}{m-1}(N-m+1)$ for $N$ large enough, and the equality follows from (\ref{eq-prunderalt2}), thus proving item 1 of Theorem \ref{thm-consistency}.

To prove item 2, we will use the following inequality for chi-square distributions, which appears in equation (4.3) of \citet{laurent2000adaptive}: for $U$ a $\chi^2$ statistic with $D$ degrees of freedom, for any positive $x$,
$Pr(U-D\geq 2\sqrt{Dx}+2x)\leq e^{-x}.$

Let $\Ical$ be a fixed arbitrary partition of size $m$. Since for $N$ large enough, under the null hypothesis, $T^{\Ical}$ is approximately a $\chi^2$ statistic with $D =  (m-1)^2$ degrees of freedom, it thus follows that for $x>D$,
\begin{equation}
Pr_{H_0}(T^{\Ical} - D \geq 4x)\leq e^{-x}.
\end{equation}
Let $x = \frac{c'}{8}(N-m+1)-D/4$. Then for $N$ large enough, $x>D$. It thus follows that
\begin{equation}
Pr_{H_0}(T^{\Ical} \geq \frac{c'}2(N-m+1))\leq e^{-(\frac{c'}{8}(N-m+1)-D/4)}\leq e^{-(\frac{c'}{8}(N-m+1)-(m-1)^2/4)} \label{eq-max1}
\end{equation}
By Bonferroni's inequality,
\begin{eqnarray}
&&Pr_{H_0}\left(M_{m\times m}^{DDP}\geq \frac{c'}2(N-m+1)\right)\leq \sum_{\Ical \in \Pi_m^{DDP}} Pr_{H_0}\left(T^{\Ical} \geq \frac{c'}2 (N-m+1)\right) \nonumber \\
&& \leq \binom{N}{m} e^{-(\frac{c'}{8}(N-m+1)-(m-1)^2/4)}, \label{eq-max2}
\end{eqnarray}
where the last inequality follows from (\ref{eq-max1}). Since $\binom{N}{m}$ is at most  $O(N^{\sqrt{N}})$, and since $$e^{-(\frac{c'}{8}(N-m+1)-(m-1)^2/4)}= O(e^{-(\frac{c'}{8}N)}),$$ it follows that the expression in (\ref{eq-max2}) goes to zero as $N\rightarrow \infty$.

Since we found  contingency tables for which under the alternative the test statistic $\frac{T^{\Ical}}{N-m+1}$ converges uniformly almost surely to a positive constant greater than $c'$ (\ref{eq-prunderaltcell}), it follows that $\frac{M^{DDP}_{m\times m}}{N-m+1}$ converges uniformly almost surely to a positive constant greater than $c'$ when the null is false.  From (\ref{eq-max2}) it follows that as $N\rightarrow \infty$, with $\lim_{N\rightarrow \infty}\frac m{\sqrt{N}}=0$,  the probability that the test statistics $\frac{M^{DDP}_{m\times m}}{N-m+1}$ will be above $\frac{c'}2$ goes to zero when the null is true. It follows that the null hypothesis will be rejected with asymptotic probability one when it is false.

\subsection{The ADP test}
We want to show that if $H_0$ is false, then for an arbitrary fixed $\alpha$, $\lim_{N\rightarrow \infty} Pr(S_{m\times m}^{ADP}> S_{1-\alpha}^{tab})=1$, where $S_{1-\alpha}^{tab}$ denotes the $1-\alpha$ quantile of the null distribution of $S_{m\times m}^{ADP}$. We use $\Acal, c, \Acal_1, \Acal_2, f_1, f_2$ as defined in the beginning of Appendix A of the main text.

For the ADP test, recall that the partitioning is based on selecting $m-1$ points from $1.5, \ldots, N-0.5$ for the partitions of the ranked $x$-values, and separately for the partitions of the ranked $y$-values. For a fixed rectangle, we say a grid point $(i+0.5, j+0.5)$ is in the rectangle if the two $x$-values with ranks $i$ and $i+1$, and the two $y$-values with ranks $j$ and $j+1$, are in the rectangle, for $(i,j)\in \{1,\ldots,N\}^2$.
Let $\Gamma\{(x_1,y_1),\ldots, (x_N,y_N)\}$ be the set of partitions of size $m$ with at least one grid point in $\Acal_1$ and at least one grid point in $\Acal_2$. Let $N_{ix}$ be the number of x-coordinates of the grid points in $\Acal_i, i\in \{1,2\}$, and $N_{iy}$ be the number of y-coordinates of the grid points in $\Acal_i, i\in \{1,2\}$.

Let $\Ical\in \Gamma\{(x_1,y_1),\ldots, (x_N,y_N)\}$ define an (arbitrary fixed) ADP partition in $\Gamma$. There exist two $x$-values in $\Acal_1$ that are separated by a grid point in $\Ical$, and two $x$-values in $\Acal_2$ that are separated by a grid point in $\Ical$, denote the average of these two $x$-values by $x^*_1$ and $x^*_2$. Let $y^*_1$ and $y^*_2$ be similarly defined for the $y$-values.

Let $C$ be the cell defined by the points $(x^*_i,y^*_i), i=1,2$. The fraction of observed counts in the cell $C$ is a  linear combination of empirical cumulative distribution functions
$$\frac{o_C}{N} = \hat F_{XY} (x^*_1,y^*_1) + \hat F_{XY} (x^*_2,y^*_2)-\hat F_{XY} (x^*_1,y^*_2)-\hat F_{XY} (x^*_2,y^*_1),  $$
and the expected fraction under the null, is a function of the cumulative marginal distributions
$$\frac{e_{C}}{N} = \{\hat F_{X} (x^*_2) -\hat F_{X} (x^*_1)\}\{\hat F_{Y} ( y^*_2) -\hat F_{Y} (y^*_1)\}, $$
where $\hat F$ denotes the empirical cumulative distribution function based on $N$ sample points.

By the Glivenko-Cantelli theorem, 
uniformly almost surely,
\begin{eqnarray}
&& \lim_{N\rightarrow \infty} \left(\frac{o_C}{N} -\int_{\{(x,y): x\in (x^*_1,x^*_2], y \in (y^*_1,y^*_2]\}} h(x,y)dx dy\right) = 0 , \nonumber \\
&& \lim_{N\rightarrow \infty} \left\{\frac{e_C}{N} -\left(\int_{\{x: x\in (x^*_1,x^*_2]\}} f(x)dx\right) \left(\int_{\{y: y\in (y^*_1,y^*_2]\}} g(y)dy\right)\right\} = 0. \nonumber \\
\end{eqnarray}

Therefore, by Slutsky's theorem and the continuous mapping theorem, we have that uniformly almost surely
\begin{eqnarray}
&& \lim_{N\rightarrow \infty}\frac{1}{N}\frac{(o_C-e_C)^2}{e_C}
= \lim_{N\rightarrow \infty}\frac{\left(\frac{o_C}{N} -\frac{e_C}{N}\right)^2 }{\frac{e_C}{N}} \nonumber \\
&& \geq  \lim_{N\rightarrow \infty}\left(\frac{o_C}{N} -\frac{e_C}{N}\right)^2  \nonumber \\
&&   = \lim_{N\rightarrow \infty}\left[\int_{\{(x,y): x\in (x^*_1,x^*_2], y \in (y^*_1,y^*_2]\}} \{h(x,y)-f(x)g(y)\}dx dy\right]^2 , \label{eq-GC2}
\end{eqnarray}
where the inequality follows from the fact that $\frac{e_C}{N}\leq 1$.

We shall show that the limit (\ref{eq-GC2})  can be bounded from below by a positive constant that depends on $(x_0,y_0)$ and $R$ but not on $\Ical$. Since
\begin{eqnarray}
&& \{(x,y): x\in (x_0+R/4,x_0+3R/4], y \in (y_0+R/4,y_0+3R/4] \} \nonumber \\
&& \subseteq \{(x,y): x\in (x^*_1,x^*_2], y \in (y^*_1,y^*_2]\},\nonumber
\end{eqnarray}
a positive lower bound can be obtained:
\begin{eqnarray}
&&  \lim_{N\rightarrow \infty}\left[\int_{\{(x,y): x\in (x^*_1,x^*_2], y \in (y^*_1,y^*_2]\}} \{h(x,y)-f(x)g(y)\}dx dy\right]^2  \nonumber \\
&&\geq \left[\int_{\{(x,y): x\in (x_0+R/4,x_0+3R/4], y \in (y_0+R/4,y_0+3R/4] \}} \{h(x,y)-f(x)g(y)\}dx dy\right]^2 \nonumber \\
&&\geq c^2 \int_{\{(x,y): x\in (x_0+R/4,x_0+3R/4], y \in (y_0+R/4,y_0+3R/4] \}} dx dy = c^2R^2/4, \nonumber
\end{eqnarray}
where the first inequality follows since $h(x,y)-f(x)g(y)>0$ in $\mathcal{A}$, and the second inequality follows since the minimum value is $c>0$.
Therefore, it follows that $\frac{1}{N}\frac{(o_C-e_C)^2}{e_C}$ converges uniformly almost surely to a positive constant greater than $c' = c^2 R^2/4,$
\begin{eqnarray}\label{eq-prunderaltcell2}
 Pr\left(\lim_{N\rightarrow \infty} \frac{1}{N}\frac{(o_C-e_C)^2}{e_C}\geq c'\right) =1.
\end{eqnarray}

The partition $\Ical$ either contains the cell $C$, or a group of cells that divide $C$. By Jensen's inequality, it follows that in the latter case the score is made larger,  see the arguments leading to expression (\ref{eq-finerpartition}). It thus follows that the score $T^{\Ical}/N$ converges uniformly almost surely to a positive constant greater than $c'$.

Let $|\Gamma|$ denote the number of $\Gamma\{(x_1,y_1),\ldots, (x_N,y_N)\}$. Since $\Ical \in \Gamma\{(x_1,y_1),\ldots, (x_N,y_N)\}$ was arbitrarily fixed, it follows that
$\frac{1}{|\Gamma|}\sum_{\Ical\in \Gamma}  \frac{T^\Ical}{N}$ converges almost surely to a positive constant greater than $c'/2$.
Since $S_{m\times m}\geq \sum_{\Ical \in \Gamma}  T^\Ical$, it follows that almost surely,
\begin{eqnarray}\label{eq-prunderalt22}
\lim_{N\rightarrow \infty} \frac{S_{m\times m}}{|\Gamma|N}\geq \frac{c'}2.
\end{eqnarray}

We shall show that $\lim_{N\rightarrow \infty} |\Gamma|/\{\binom{N-1}{m-1}\binom{N-1}{m-1}\}$ is bounded below  by a positive constant.
First, we shall consider the case that $m$ is finite. In this case, a subset of $\Gamma\{(x_1,y_1),\ldots, (x_N,y_N)\}$ is the set of all partitions with $m-2$ grid points  in $\Acal_1$, and one grid point in $\Acal_2$, for both $x$-values and $y$-values. Therefore,   $|\Gamma|\geq \binom{N_{1x}}{m-2}N_{2x} \binom{N_{1y}}{m-2}N_{2y}$.  Simple algebraic manipulations lead to the following expression for $\frac{\binom{N_{1x}}{m-2}N_{2x}}{\binom{N-1}{m-1}}$:
$$
(m-1)\frac{N_{2x}}{N-1}\frac{N_{1x}}{N-2}\cdots \frac{N_{1x}-m+3}{N-1-m+2}.
$$
Since $N_{1x}/N$ converges almost surely to $\int_{x_0}^{x_0+R/4} f(x)dx$, and similarly $N_{2x}/N$ converges almost surely to $\int_{x_0+3R/4}^{x_0+R} f(x)dx$, then for $m\geq 3$ finite it follows that $\frac{\binom{N_{1x}}{m-2}N_{2x}}{\binom{N-1}{m-1}}$ converges almost surely to a positive constant. Similarly, $\frac{\binom{N_{1y}}{m-2}N_{2y}}{\binom{N-1}{m-1}}$ converges almost surely to a positive constant. Therefore, $|\Gamma|/\{\binom{N-1}{m-1}\binom{N-1}{m-1}\}$ is bounded away from zero.

Second, we shall consider the case that $m\rightarrow \infty$. The complement of $\Gamma$, $\Gamma^C$,  is the set of contingency tables with no grid point in $\Acal_1$ or in $\Acal_2$. An upper bound for $|\Gamma^C|$ is:
$$
\binom{N-1}{m-1}\left\{\binom{N-1-N_{1x}}{m-1}+\binom{N-1-N_{2x}}{m-1}+\binom{N-1-N_{1y}}{m-1}+\binom{N-1-N_{2y}}{m-1}\right\}
$$
 Note that since $|\Gamma| =\{\binom{N-1}{m-1}\binom{N-1}{m-1}\}-|\Gamma^C|$, it is enough to show that   $|\Gamma^C|/\{\binom{N-1}{m-1}\binom{N-1}{m-1}\}$ converges to zero as $N\rightarrow \infty$. Simple algebraic manipulations lead to the following expression for $\frac{\binom{N-1-N_{1x}}{m-1}}{\binom{N-1}{m-1}}$:
$$
\left(1-\frac{N_{1x}}{N-1}\right)\cdots\left(1-\frac{N_{1x}}{N-1-(m-2)}\right)\leq \left(1-\frac{N_{1x}}{N-1}\right)^m.
$$
Since $N_{1x}/N$ converges almost surely to a positive fraction $\int_{x_0}^{x_0+R/4} f(x)dx$, it follows that $\frac{\binom{N-1-N_{1x}}{m-1}}{\binom{N-1}{m-1}}$ converges almost surely to zero. Similarly, since $N_{2x}/N$, $N_{1y}/N$ and $N_{2y}/N$ converge almost surely to positive fractions, it follows that  respectively, $\frac{\binom{N-1-N_{2x}}{m-1}}{\binom{N-1}{m-1}}$, $\frac{\binom{N-1-N_{1y}}{m-1}}{\binom{N-1}{m-1}}$, and $\frac{\binom{N-1-N_{2y}}{m-1}}{\binom{N-1}{m-1}}$  converge almost surely to zero. Thus $ |\Gamma^C|/\{\binom{N-1}{m-1}\binom{N-1}{m-1}\}$ converges almost surely to zero.

Since $\lim_{N\rightarrow \infty} |\Gamma|/\{\binom{N-1}{m-1}\binom{N-1}{m-1}\}$ is bounded below  by a positive constant, it follows from (\ref{eq-prunderalt22}) that almost surely,
\begin{eqnarray}\label{eq-prunderalt3}
\lim_{N\rightarrow \infty} \frac{S_{m\times m}}{\binom{N-1}{m-1}\binom{N-1}{m-1}N}\geq c'',
\end{eqnarray}
for some constant $c''>0$.

Consider now a random permutation $(\pi y_1),\ldots,(\pi y_N)$ of the $y$-values $y_1,\ldots,y_N$. Let $S_{m\times m}^\pi$ be the test statistic  that is computed from the data $(x_1,\pi y_1),\ldots,(x_N,\pi y_N)$.
By Markov's inequality,
\begin{eqnarray}
&& Pr\left(S^{\pi}_{m\times m}\geq c''\binom{N-1}{m-1}\binom{N-1}{m-1}N \mid \vec x, \vec y \right)\leq \frac{E(S^{\pi}_{m\times m} \mid \vec x, \vec y)}{c''\binom{N-1}{m-1}\binom{N-1}{m-1}N} \nonumber \\
& \approx & \frac{\binom{N-1}{m-1}\binom{N-1}{m-1}(m-1)^2}{c''\binom{N-1}{m-1}\binom{N-1}{m-1}N}, \label{eq-prundernull2}
\end{eqnarray}
where $\vec x  = (x_1,\ldots,x_N)$ and $\vec y = (y_1,\ldots,y_N)$.
The approximation in (\ref{eq-prundernull2}) becomes more accurate the larger $N$ is, since each of the contingency tables is approximately $\chi^2$ with $(m-1)^2$ degrees of freedom. The right hand side of equation (\ref{eq-prundernull2}) goes to zero as $N\rightarrow \infty$, as long as $\lim_{N\rightarrow \infty} \frac{m}{\sqrt{N}}=0$. Thus,

\begin{equation}\label{eq-UBundernull3}
\lim _{N\rightarrow \infty, m/\sqrt{N} \rightarrow 0}  Pr\left(S^{\pi}_{m\times m}\geq c''\binom{N-1}{m-1}\binom{N-1}{m-1}N \mid \vec x, \vec y \right) = 0.
\end{equation}

We now have all the necessary results to complete the proof. Specifically,
\begin{eqnarray}
&& \lim_{N\rightarrow \infty} Pr(S_{m\times m}\leq S_{1-\alpha}^{tab} ) \leq Pr\left(S^{\pi}_{m\times m}\leq c''\binom{N-1}{m-1}\binom{N-1}{m-1}N\right) = 0,
\end{eqnarray}
where the inequality follows from (\ref{eq-UBundernull3}), since $S_{1-\alpha}^{tab}$ is below $c''\binom{N-1}{m-1}\binom{N-1}{m-1}N$ for $N$ large enough, and the equality follows from (\ref{eq-prunderalt3}), thus proving item 1 of Theorem 1 for the ADP summation statistic.


\section{Proof of Theorem \ref{thm:mi}}
\label{sec:mi_proof}

We want to show that for all $\epsilon>0$, $\lim_{N\rightarrow \infty}Pr\left(\left|\frac{S_{m\times m}^{ADP}(L)}{N|\Pi|} - I_{XY}\right|>\epsilon\right) = 0$ if $\lim_{N\rightarrow \infty}\frac{m}{\sqrt{N}}=0$ and $\lim_{N\rightarrow \infty} m = \infty$, where $|\Pi| = \left(\binom{N-1}{m-1}\right)^2$ is the number of partitions. 

For continuous marginals, the copula function of the joint distribution of $(X,Y)$ is unique, denote it by $c(u,v)$. The mutual information is the negative copula entropy, $H_{UV} =  -\int c(u,v)\log c(u,v)dudv$,
\begin{eqnarray}
I_{XY} &=& \int c(F_X(x), F_Y(y))f(x)g(y)\log c(F_X(x),F_Y(y))dxdy \nonumber \\
&=& \int c(u,v)\log c(u,v)dudv = -H_{UV}. 
\end{eqnarray}

Consider an arbitrary fixed partition $\mathcal{I}= \{(i_1,j_1),\ldots, (i_{m-1},j_{m-1}) \}\subset \{1.5,\ldots, N-0.5 \}^2$. Recall that $\mathcal{C}(\mathcal{I})$ is the set of $m\times m$ cells that are defined by the partition. For a cell $C$, let $r_l(C)$ and $r_h(C)$ be, respectively, the lowest and highest $x$-grid integer values in $C$. Similarly, let $s_l(C)$ and $s_h(C)$ be, respectively, the lowest and highest $y$-grid integer values in $C$.

The entropy of the partition $\mathcal{I}$ is
{\scriptsize
$$
H_{UV}^{\mathcal{I}}= -\sum_{C\in \mathcal{C}(\mathcal{I})}Pr\left(\frac{r_l(C)}N\leq U \leq \frac{r_h(C)}N,  \frac{s_l(C)}N\leq V \leq \frac{s_h(C)}N\right)\log\left\{ Pr\left(\frac{r_l(C)}N\leq U \leq \frac{r_h(C)}N,  \frac{s_l(C)}N\leq V \leq \frac{s_h(C)}N\right)\right\}.
$$
}
The corresponding empirical (plug in) estimator is
$$\hat{H}^{\Ical}_{UV} = -\sum_{C\in \mathcal{C}(\mathcal{I})} \frac{o_C}N\log\left(\frac{o_C}N\right), \quad o_C = \sum_{i=1}^N I(r_l(C)\leq r_i \leq r_h(C), s_l(C)\leq s_i \leq s_h(C)).$$
Let $H_{U}^{\Ical}$ and $H_V^{\Ical}$ be the fixed marginal entropies of the partition $\mathcal{I}$:
\begin{eqnarray}
&& {H}^{\Ical}_{U} =  -\sum_{C_x\in \mathcal{C}_x(\mathcal{I})}\frac{r_h(C_x)-r_l(C_x)}N\log \left(\frac{r_h(C_x)-r_l(C_x)}N\right),  \nonumber \\
&&{H}^{\Ical}_{V} =  -\sum_{C_y\in \mathcal{C}_y(\mathcal{I})}\frac{r_h(C_y)-r_l(C_y)}N\log \left(\frac{r_h(C_y)-r_l(C_y)}N\right), \nonumber
\end{eqnarray}
where $\mathcal{C}_x(\mathcal{I})$ and $\mathcal{C}_y(\mathcal{I})$ are the intervals induced by $\Ical$  in $x$ and in $y$, respectively.
Note that given $\Ical$, the observed and expected margins of the partitions are fixed, and therefore 
\begin{eqnarray}
 {H}^{\Ical}_{U} &=& -\sum_{C\in \mathcal{C}(\mathcal{I})}o_C\log \left(\frac{r_h(C)-r_l(C)}N\right) \label{eq-margins1}\\
&=& -\sum_{C\in \mathcal{C}(\mathcal{I})}Pr(r_l(C)\leq U\leq r_h(C), s_l(C)\leq V\leq s_h(C))\log \left(\frac{r_h(C)-r_l(C)}N\right) \label{eq-margins2} \\
{H}^{\Ical}_{V} &=& -\sum_{C\in \mathcal{C}(\mathcal{I})}o_C\log \left(\frac{s_h(C)-s_l(C)}N\right)\label{eq-margins3} \\
&=& -\sum_{C\in \mathcal{C}(\mathcal{I})}Pr(r_l(C)\leq U\leq r_h(C), s_l(C)\leq V\leq s_h(C))\log \left(\frac{s_h(C)-s_l(C)}N\right). \label{eq-margins4}
\end{eqnarray}

The following simple derivation shows that the likelihood ratio score $T^{\Ical}$ is a linear combination of $\hat{H}^{\Ical}_{UV}, {H}^{\Ical}_{U}$ and ${H}^{\Ical}_{V} $:
\begin{eqnarray}
T^{\Ical} &=&\sum_{C\in \mathcal{C}(\mathcal{I})}o_C\log \frac{o_C}{e_C} \nonumber \\
&=& \sum_{C\in \mathcal{C}(\mathcal{I})}o_C\log \frac{o_C}N - \sum_{C\in \mathcal{C}(\mathcal{I})}o_C\log \frac{e_C}N  \nonumber \\
&=& -N\hat{H}^{\Ical}_{UV} - \sum_{C\in \mathcal{C}(\mathcal{I})}o_C\log \left(N \frac{r_h(C)-r_l(C)}N \frac{s_h(C)-s_l(C)}N \frac 1N \right) \nonumber \\
&=& -N\hat{H}^{\Ical}_{UV} + NH^{\Ical}_{U}+NH^{\Ical}_{V}, \nonumber
\end{eqnarray}
where the last equality follows from equations (\ref{eq-margins1}) and (\ref{eq-margins3}).

Let $E(\cdot)$ denote the expectation of a random variable. We bound from above our probability of interest by a sum of three probabilities as follows.

\begin{eqnarray}
&& Pr\left(\left|\frac{S_{m\times m}^{ADP}}{N|\Pi|} - I_{XY}\right|>\epsilon\right)  = Pr\left(\left|\frac{\sum_{\Ical} T^{\Ical}_{m\times m}(L)}{N|\Pi|} - I_{XY}\right|>\epsilon\right) \nonumber \\
&& =  Pr\left(\left|\frac{\sum_{\Ical} (H^{\Ical}_{U}+H^{\Ical}_{V}-\hat{H}^{\Ical}_{UV})}{|\Pi|} +H_{UV}\right|>\epsilon\right) \nonumber \\
&& =  Pr(|\sum_{\Ical}( H^{\Ical}_{U}+H^{\Ical}_{V}-\hat{H}^{\Ical}_{UV} +H_{UV})|>|\Pi|\epsilon) \nonumber \\
&& =  Pr(|\sum_{\Ical}(-\hat{H}^{\Ical}_{UV}+E(\hat{H}^{\Ical}_{UV})) +\sum_{\Ical}(-E(\hat{H}^{\Ical}_{UV})+H^{\Ical}_{UV}) \nonumber \\
&&+ \sum_{\Ical}(-H^{\Ical}_{UV}+H^{\Ical}_{U}+H^{\Ical}_{V}+H_{UV})|>|\Pi|\epsilon) \nonumber \\
&&\leq  Pr(|\sum_{\Ical}(-\hat{H}^{\Ical}_{UV}+E(\hat{H}^{\Ical}_{UV}))| +|\sum_{\Ical}(-E(\hat{H}^{\Ical}_{UV})+H^{\Ical}_{UV})| \nonumber \\
&&+ |\sum_{\Ical}(-H^{\Ical}_{UV}+H^{\Ical}_{U}+H^{\Ical}_{V}+H_{UV})|>|\Pi|\epsilon) \nonumber \\
&&\leq  Pr(|\sum_{\Ical}(-\hat{H}^{\Ical}_{UV}+E(\hat{H}^{\Ical}_{UV}))|>|\Pi|\epsilon/3)  \label{eq-MI-term1} \\
&& +Pr(|\sum_{\Ical}(-E(\hat{H}^{\Ical}_{UV})+H^{\Ical}_{UV})| >|\Pi|\epsilon/3)  \label{eq-MI-term2} \\
&&+ Pr(|\sum_{\Ical}(-H^{\Ical}_{UV}+H^{\Ical}_{U}+H^{\Ical}_{V}+H_{UV})|>|\Pi|\epsilon/3),  \label{eq-MI-term3}
\end{eqnarray}
where the last inequality follows from $\{|\sum_{\Ical}(-\hat{H}^{\Ical}_{UV}+E(\hat{H}^{\Ical}_{UV}))| +|\sum_{\Ical}(-E(\hat{H}^{\Ical}_{UV})+H^{\Ical}_{UV})|+ |\sum_{\Ical}(-H^{\Ical}_{UV}+H^{\Ical}_{U}+H^{\Ical}_{V}+H_{UV})|>|\Pi|\epsilon \}\subseteq \{(|\sum_{\Ical}(-\hat{H}^{\Ical}_{UV}+E(\hat{H}^{\Ical}_{UV}))|>|\Pi|\epsilon/3)\cup (|\sum_{\Ical}(-E(\hat{H}^{\Ical}_{UV})+H^{\Ical}_{UV})| >|\Pi|\epsilon/3)\cup|\sum_{\Ical}(-H^{\Ical}_{UV}+H^{\Ical}_{U}+H^{\Ical}_{V}+H_{UV})|>|\Pi|\epsilon/3\}$ and Bonferroni's inequality.

We will show that the three probabilities (\ref{eq-MI-term1})--(\ref{eq-MI-term3}) vanish as $N\rightarrow \infty , m\rightarrow \infty, \frac{m}{\sqrt{N}}\rightarrow 0$, thus proving the theorem.

The probability (\ref{eq-MI-term1}) can be upper-bounded as follows,
\begin{eqnarray}
&& Pr\left(|\sum_{\Ical} (\hat{H}^{\Ical}_{UV}-E(\hat{H}^{\Ical}_{UV} ))|\geq |\Pi|\epsilon /3 \right)\leq  \sum_{\Ical} Pr\left(| (\hat{H}^{\Ical}_{UV}-E(\hat{H}^{\Ical}_{UV} ))|\geq \epsilon /3 \right) \nonumber \\
&& \leq |\Pi|3e^{-\frac{N}2\epsilon^2/9\frac{1}{(\log N)^2}}
\end{eqnarray}
where the first inequality follows from the fact that $\{|\sum_{\Ical} (\hat{H}^{\Ical}_{UV}-E(\hat{H}^{\Ical}_{UV} ))|\geq |\Pi|\epsilon /3\}\subseteq\{\cup_{\Ical \in \Pi} (\hat{H}^{\Ical}_{UV}-E(\hat{H}^{\Ical}_{UV} )) \geq \epsilon/3 \}$ and  Bonferroni's inequality, and the second inequality follows from the upper bound (3.4) in \citet{paninski2003estimation}  for the plug in estimator for a given partition $\Ical$. This probability goes to zero as $N\rightarrow \infty$, since $|\Pi|$ is $O(N^{\sqrt{N}})$ and $\lim_{N\rightarrow \infty} O(N^{\sqrt{N}})e^{-\frac{N}2\epsilon^2/9\frac{1}{(\log N)^2}}=0$.

The event in the  second probability (\ref{eq-MI-term2}) is not random, so we need to show that $|\sum_{\Ical}(E(\hat{H}^{\Ical}_{UV})-H^{\Ical}_{UV})|< |\Pi|\epsilon /3$ for $N$ large enough.  Proposition 1 in  \citet{paninski2003estimation} states that $0\leq (H^{\Ical}_{UV}-E(\hat{H}^{\Ical}_{UV}))\leq \log (1+\frac{(m-1)^2-1}{N})$. Therefore,
\begin{equation}
 |\sum_{\Ical}(E(\hat{H}^{\Ical}_{UV})-H^{\Ical}_{UV})|\leq |\Pi| \log (1+\frac{(m-1)^2-1}{N})\nonumber
\end{equation}
Clearly, the RHS is below $|\Pi|\epsilon /3$ for $N$ large enough, if $\lim_{N\rightarrow \infty} \frac{m}{\sqrt{N}}=0$.

It remains to show that (\ref{eq-MI-term3}) vanishes as $N\rightarrow \infty$. This event is not random, so we will show that
$$\lim_{N\rightarrow \infty}\frac{|\sum_{\Ical}(-H^{\Ical}_{UV}+ H^{\Ical}_{U}+H^{\Ical}_{V}+H_{UV})|}{|\Pi|}< \epsilon /3.$$

By the mean value theorem, for cell $C$ there exists a point $(u_C,v_C)$ in $C$ such that $$Pr(r_l(C)\leq U\leq r_h(C), s_l(C)\leq V\leq s_h(C)) = c(u_C,v_C)\frac{r_h(C)-r_l(C)}N\frac{s_h(C)-s_l(C)}N.$$  Therefore,
\begin{eqnarray}
&& -H^{\Ical}_{UV} \nonumber\\
&&= \sum_{C\in \Ccal(\Ical)}c(u_C,v_C)\frac{r_h(C)-r_l(C)}N\frac{s_h(C)-s_l(C)}N \log\left(c(u_C,v_C)\frac{r_h(C)-r_l(C)}N\frac{s_h(C)-s_l(C)}N\right) \nonumber \\
&& = \sum_{C\in \Ccal(\Ical)}\frac{r_h(C)-r_l(C)}N\frac{s_h(C)-s_l(C)}N c(u_C,v_C) \log c(u_C,v_C) \label{eq-MI-term3-1} \\
&& + \sum_{C\in \Ccal(\Ical)} Pr(r_l(C)\leq U\leq r_h(C), s_l(C)\leq V\leq s_h(C)) \log \frac{r_h(C)-r_l(C)}N \nonumber \\
&& + \sum_{C\in \Ccal(\Ical)} Pr(r_l(C)\leq U\leq r_h(C), s_l(C)\leq V\leq s_h(C)) \log \frac{s_h(C)-s_l(C)}N \nonumber \\
&& = \sum_{C\in \Ccal(\Ical)}\frac{r_h(C)-r_l(C)}N\frac{s_h(C)-s_l(C)}N c(u_C,v_C) \log c(u_C,v_C) -H^{\Ical}_U - H^{\Ical}_V,  \label{eq-MI-term3-2}
\end{eqnarray}
where the last equality follows from equations (\ref{eq-margins2}) and (\ref{eq-margins4}).

By the definition of the Riemann integral, expression (\ref{eq-MI-term3-1}) can be made arbitrarily close to $-H_{UV}$. Specifically,  there exists a $0<d(\epsilon)<1$ such that if all cells satisfy $ \frac{r_h(C)-r_l(C)}N <d$ and  $\frac{s_h(C)-s_l(C)}N <d$, then
$$|\sum_{C\in \Ccal(\Ical)}\frac{r_h(C)-r_l(C)}N\frac{s_h(C)-s_l(C)}N c(u_C,v_C) \log c(u_C,v_C) + H_{UV}|< \epsilon/3.$$

Therefore, it follows that for any partition $\Ical \in \Pi$ for which all cells satisfy $\frac{r_h(C)-r_l(C)}N<d$ and  $\frac{s_h(C)-s_l(C)}N<d$, then we have
$$|(-H^{\Ical}_{UV}+H^{\Ical}_U + H^{\Ical}_V + H_{UV})|< \epsilon /3. $$

It remains to show that the contribution of the fraction of partitions that do not satisfy $\frac{r_h(C)-r_l(C)}N <d$ and  $\frac{s_h(C)-s_l(C)}N <d$ goes to zero as $N\rightarrow \infty$. Since the probability of selecting an $x$-value (or $y$-value) for a partition that will have a cell larger than $d$ is $1-d$, the fraction of ``bad" partitions is upper-bounded by
$$\frac{2m \binom{N(1-d)}{m-1}\binom{N}{m-1}}{\left(\binom{N-1}{m-1}\right)^2}\leq 2m(1-d)^{m-1}.$$
  Since $m\rightarrow \infty$ the fraction of bad partitions goes to zero.

 Note that $|(-H^{\Ical}_{UV}+H^{\Ical}_{U}+H^{\Ical}_{V} + H_{UV})|$ is at most $O(\log m^2)$ because by Jensen's inequality, $|H^{\Ical}_{UV}|\leq \log m^2, |H^{\Ical}_{U}|\leq \log m^2, |H^{\Ical}_{V}|\leq \log m^2,$ and $|H_{UV}| = I_{XY}$ is assumed to be bounded.
Therefore,
$$
\frac{|\sum_{\Ical}(-H^{\Ical}_{UV}+H^{\Ical}_{U}+H^{\Ical}_{V}+H_{UV})|}{|\Pi|}
\leq \epsilon/3 + O\left(\log m^2m(1-d)^{m-1}\right).
$$
Since the second term of the RHS goes to zero as $N\rightarrow \infty, m\rightarrow \infty, \frac{m}{\sqrt{N}}\rightarrow 0$, the proof is  complete.

\section{Proof of Theorem \ref{thm-consistencyminp}}\label{app:proofconsistencyminp}
We shall prove items 1 and 2  for the DDP statistic only, since the proof for the ADP statistic is very similar. We shall use the notation of Section \ref{subsec-proofDDPconsistency}. Let $\hat m$ be the value of $m$ with minimum $p$-value,
$$\hat m  = \arg \min_{m\in \{ 2,\ldots, m_{\max}\}} p_2,\ldots, p_m. $$

To prove item 1, we note that from equation (\ref{eq-prunderalt2}) it follows that under the alternative, 
$$\lim_{N\rightarrow \infty} \frac{S_{\hat m\times \hat m}}{\binom{N}{\hat m-1}(N-\hat m+1)}\geq c''. $$

Under the null,
\begin{eqnarray}
&& Pr\left(S^{\pi}_{\hat m\times \hat m}\geq c''\binom{N}{\hat m-1}(N-\hat m+1) \,\mid\, \vec x, \vec y \right) \nonumber \\
&& \leq \sum_{m=2}^{m_{\max}} Pr\left(S^{\pi}_{ m\times  m}\geq c''\binom{N}{m-1}(N-m+1) \,\mid\, \vec x, \vec y \right)  \nonumber \\
&&\leq  m_{\max} \max_{m \in \{2,\ldots, m_{\max}\}}\frac{\binom{N}{m-1}(m-1)^2}{ c''\binom{N}{m-1}(N-m+1)} \leq   m_{\max} \frac{(m_{\max}-1)^2}{ c''(N-m_{\max}+1)}, \nonumber
\end{eqnarray}
where the first inequality is the Bonferroni inequality, and the second inequality follows from  equation (\ref{eq-prundernull}).
Since the last term goes to 0 as $N\rightarrow \infty$ if $\lim_{N\rightarrow \infty} m/N^{1/3} = 0$, the proof of item 1 is complete.

The proof of item 2 is very similar to the proof in Section \ref{subsec-proofDDPconsistency}, the only modification is an additional application of Bonferroni's inequality under the null:
\begin{eqnarray}
&&Pr_{H_0}\left(M_{\hat m\times \hat m}^{DDP}\geq \frac{c'}2(N-\hat m+1)\right)\leq \sum_{m=2}^{m_{\max}}Pr_{H_0}\left(M_{ m\times m}^{DDP}\geq \frac{c'}2(N-m+1)\right) \nonumber \\
&&\leq m_{\max} \max_{m \in \{2,\ldots, m_{\max}\}} \binom{N}{m} e^{-(\frac{c'}{8}(N-m+1)-(m-1)^2/4)}\leq m_{\max}  \binom{N}{m_{\max}} e^{-(\frac{c'}{8}(N-m_{\max}+1)-(m_{\max}-1)^2/4)}, \nonumber
\end{eqnarray}
where the first inequality in the last row follows from (\ref{eq-max2}). Since $m_{\max}\binom{N}{m_{\max}}$ is at most  $O(N\times N^{\sqrt{N}})$, and since $$e^{-(\frac{c'}{8}(N-m_{\max}+1)-(m_{\max}-1)^2/4)}= O(e^{-(\frac{c'}{8}N)}),$$ it follows that the last expression goes to zero as $N\rightarrow \infty$.

\section{An example of mutual information estimation}\label{sec:mi example}
We examined our suggested estimator in the following setup. For $m=15$ and $N \in \{300,1000\}$ sample points drawn from a two--component Gaussian mixture, we simulated 50 datasets and computed the estimated mutual information using $S^{DDP}_{m\times m}$, $S^{ADP}_{m\times m}$, and the histogram estimator. The Gaussian mixture density was $$0.8 \times f_{\mathcal{N}}\left( \binom{0.5}{0.5}, \binom{0.05 \quad 0.025 }{0.025 \quad 0.05} \right) + 0.2 \times f_{\mathcal{N}} \left( \binom{0.125}{0.675} , \binom{0.01 \quad 0 }{0 \quad 0.01} \right),$$ where $f_{\mathcal{N}}(\mu, \Sigma)$ is the bivariate normal distribution with mean $\mu$ and covariance matrix $\Sigma$.
 For each partition, we applied the Miller--Madow correction~\citep{paninski2003estimation}, a simple and well-known modification that estimates the systematic error of the histogram estimators and improves the finite-sample properties of these estimators. We compared it to the histogram estimator, as well as to the estimator based on the DDP statistic that considers only a subset of all possible partitions.
  Table~\ref{tbl:mi} shows that the variability and the bias decrease as $N$ increases for all estimators, and that the ADP estimator is the least variable, as is intuitively expected since it is the average over many partitions.  In practice, it is difficult to identify the optimal $m$: it should not be too small so that the local dependence structure is not missed, causing large bias, nor should it be too large so that the grid created is too sparse, causing large variance.
\begin{table}
  \caption{The average (standard error) of the mutual information estimates using random samples of size 300 (column 2) and 1000 (column 3) from the two component Gaussian mixture, by the following methods: the naive histogram estimator that partitions each axis to $15$ intervals of equal count, $S^{ADP}_{15\times 15} / ({N|\Pi^{ADP}_{15}|})$, and $S^{DDP}_{15\times 15} / ({N|\Pi^{DDP}_{15}|})$. The true mutual information value was $0.1784$.}
  \label{tbl:mi}
  \centering
  \footnotesize
  \vspace{0.5cm}
  \begin{tabular}{lcc}
                            & $\mathbf{N=300}$    & $\mathbf{N=1000}$   \\ 
    Histogram               & $0.3165$ ($0.0052$) & $0.1854$ ($0.0029$) \\
    Data derived partitions & $0.3010$ ($0.0030$) & $0.1879$ ($0.0022$) \\
    All derived partitions  & $0.2954$ ($0.0028$) & $0.1860$ ($0.0021$) \\
  \end{tabular}
\end{table}

\section{Algorithm for the DDP statistic}\label{app-DDPalgorithm}
\begin{figure}[htbp]
  \centering
  \includegraphics[page=1, width=0.3\textwidth, clip]{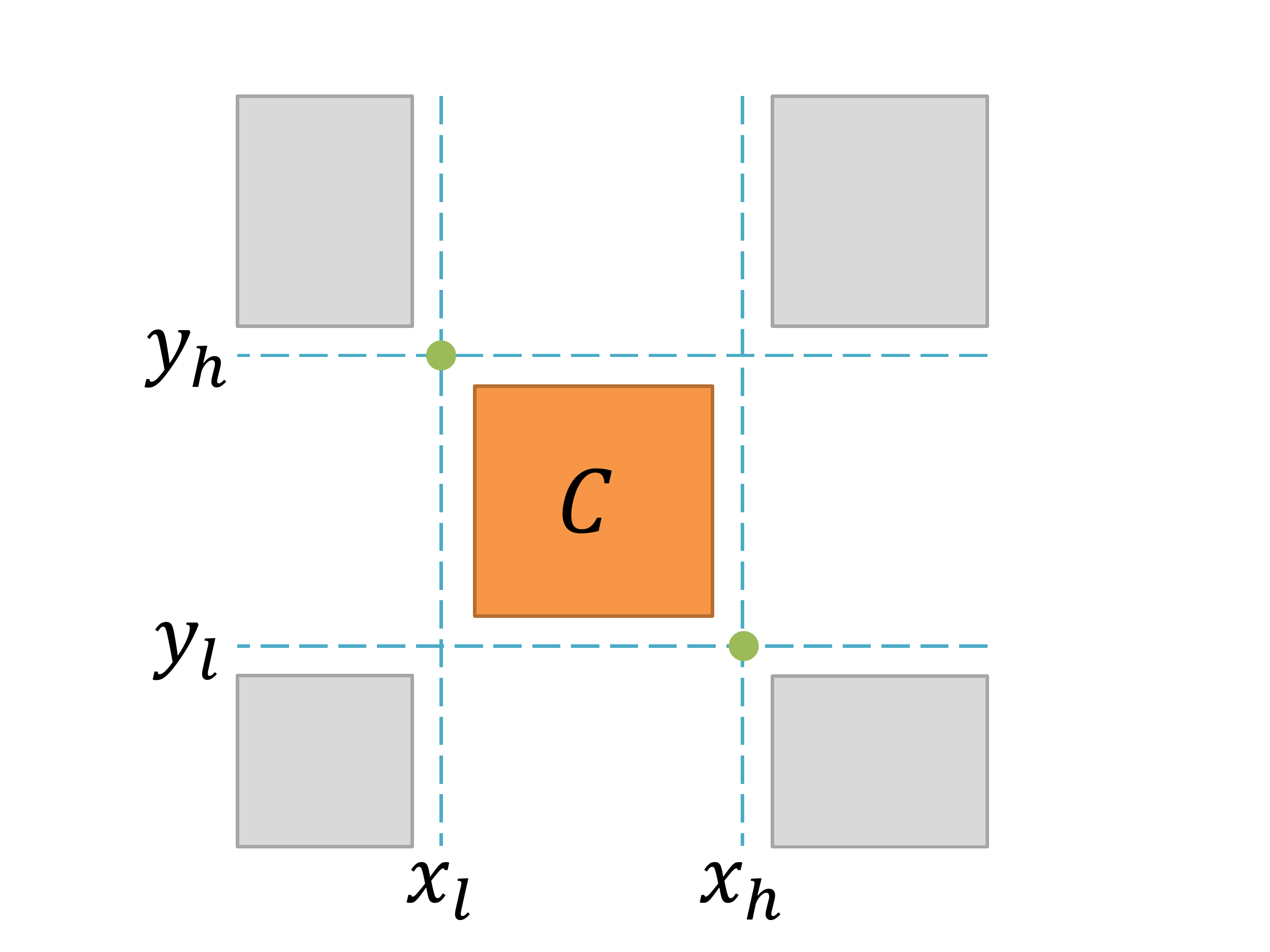}
  \includegraphics[page=2, width=0.3\textwidth, clip]{Figures.pdf}
  \includegraphics[page=3, width=0.3\textwidth, clip]{Figures.pdf}
  \caption[Cell type definition in fast DDP/ADP algorithm]{Inner cell example $C$ (orange), and the sample points that define a partition where $C$ is a cell (green). If any of the sample points ranked on a boundary column or row of $C$ is not on the corner of $C$, then $C$ can never be a DDP cell (offending point in purple, middle panel). An inner cell can be defined either by two data points (left), by three data points (middle), or by four data points (right).}
  \label{fig:ddp_cell}
\end{figure}

For the DDP statistic, the algorithm is very similar to that for the ADP statistic, except that only DDP are considered  on the grid of ranked data $\{1, \ldots, N\}^2$. 
The algorithm is slightly more complex because the number of partitions that include a cell $C$ depends on the partition size $m$, on the data, and  on the type of cell, with four possible types. Specifics follow for internal cells.
The first type is a cell $C$ for which there is a sample point that falls on the boundary of $C$ but not on one of its corners. Then no DDP can ever have $C$ as a cell, and therefore the number of DDP that include $C$ is zero. For example, in Figure~\ref{fig:ddp_cell} (middle panel), if the open circle is an observation, and therefore the filled circle with the same $y$ value is not, since there are no ties, then any DDP with this $y$ value will necessarily partition $C$ at the $x$ value of the open circle observation, and thus $C$ cannot be a cell in any DDP. For the remaining three types of cells, if there are sample points that fall on the boundary of $C$ they are necessarily on the corners of $C$. These types of cells differ by the number of observations that determine the cell.
The second type is a cell $C$  defined by two observed points. Then, the number of DDP that include $C$ is the number of ways to choose $m-3$ points from the points in the four outer areas defined by $(0,r_l)\times(0,s_l)$, $(0,r_l)\times(s_h,N]$, $(r_h,N]\times(0,s_l)$, and $(r_h,N]\times(s_h,N]$, see Figure~\ref{fig:ddp_cell} (left panel) for illustration. The number of points in the four areas is calculable in $O(1)$ using $A$, as defined in equation (\ref{eq-agg-sum-ind-A}). Specifically, the count of samples that fall strictly inside any cell with rank ranges $r \in [r_l, r_h]$ and $s \in [s_l, s_h]$ is: $$o_C = A(r_h - 1, s_h - 1) - A(r_l, s_h - 1) - A(r_h - 1, s_l) + A(r_l, s_l).$$ 

The third and fourth type are cells defined by three or four observed points, respectively, see Figure~\ref{fig:ddp_cell} (middle and right panels). Now the number of DDP that include $C$ is exactly the number of ways to choose $m-4$ and $m-5$ points, respectively, from the points in the four outer areas. Again, the number of points in the four outer areas is calculable in $O(1)$ using $A$.  Since all cells are defined by two, three, or four points, there are no additional types of cells.

Let us denote the number of sample points in the four outer areas of a cell $C$ by $OUT$.
Further denote by $C(OUT)$ the group of all cells which have exactly $OUT$ points in the outer areas.
For the sake of brevity, lets consider only cells of type 2. Similarly to equation (\ref{eq-agg-sum-ind-2}) for the ADP statistic, the contribution to the score $S^{DDP}_{m\times m}$ of internal cells of type 2 can be written as  
\begin{equation}\label{app-eq-algddp}
\sum_{OUT=2}^{N-2}n(OUT,m) \sum_{C\in C(OUT)}t_C  = \sum_{OUT=2}^{N-2}n(OUT,m) T(OUT).
\end{equation}

The algorithm proceeds as follows. First in a preprocessing phase we perform two computations:
1)go over all cells and calculate $t_C$ and $OUT$ for each cell and update $T(OUT)=T(OUT)+t_C$. This stage takes $O(N^4)$; 
2) for all $u, v\in \{0, \ldots,N\}$ we calculate and store all $\binom{u}{v}$  in $O(N^2)$ steps  using Pascal's triangle method.

 Now for each $m$, since $n(OUT,m)= \ibinom{OUT}{m-3}$, given $T(OUT)$, clearly equation (\ref{app-eq-algddp}) can be calculated in $O(N)$ for a fixed $m$ and in $O(N^2)$ for all $m$s.
Therefore the total complexity is again $O(N^4)$ due to the preprocessing phase.

\section{A fast algorithm for computing the HHG statistic}\label{supp:sec-fastHHG}
Here we describe a fast algorithm for computing the univariate original (distribution-dependent) HHG test statistic.

If $D_X(x_1, x_2)$ and $D_Y(y_1, y_2)$ are distance metrics in the variables tested for independence (for which, again, we have a paired sample with $N$ i.i.d samples), the HHG test requires computing $N(N-1)$ different $2 \times 2$ contingency tables according to the following partitions of the distance-distance plane. For every ``origin'' sample $i$, and for every ``radius'' sample $j$, the remaining $N-2$ samples are classified according to whether their $X$ and $Y$ distances from $i$ are both smaller than the $X$ and $Y$ distances from $j$ to $i$, or if only $X$, only $Y$, or neither are smaller than the respective $j$ to $i$ distances. In the general case, generating contingency tables for all pairs can be done in $N^2 \log N$, as described in~\citep{Heller12}. In the univariate case, an $O(N^2)$ algorithm proceeds as follows.

Instead of working in the distance-distance plane, the algorithm is specified in terms of the $(x, y)$, i.e., the sample plane. It is sufficient to consider the discrete grid expanded from unique $X$ samples and unique $Y$ samples actually observed (these can be identified in $O(N \log N)$). The double cumulative sum over this $N \times N$ grid is computed as in Section \ref{subsec-sumalgs} of the main text, with the only difference being that after $A$ is initialized to all zeros, it is updated sequentially with $A(r_i, s_i) = A(r_i, s_i) + 1$ for every sample of ranks in $x$ and ranks in $y$, $(r_i,s_i), \ i = 1, 2, \ldots, N$, to account for possible ties.

\begin{figure}[htbp]
  \centering
  \includegraphics[page=1, width=1.0\textwidth, trim=0in 6.5in 0in 0in, clip]{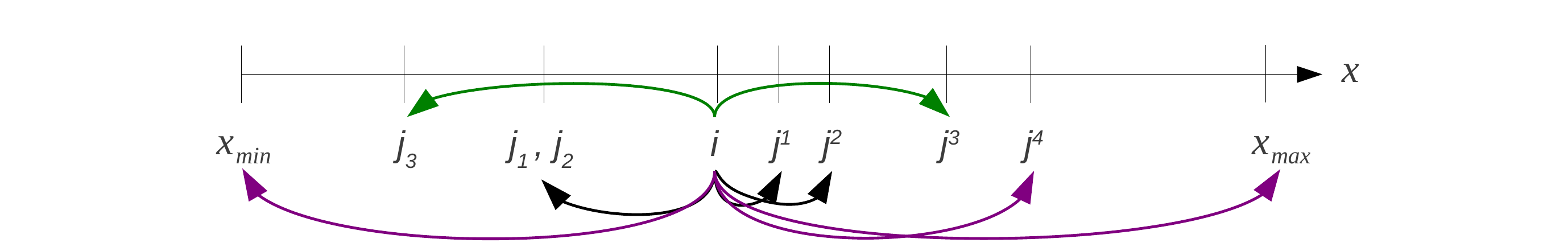}
  \caption[Coordinates in the fast HHG algorithm]{Finding the grid coordinates to sample for the $O(N^2)$ univariate (distribution-dependent) HHG algorithm.}
  \label{fig:hhg_n2_alg}
\end{figure}

Since, in the univariate case,  $D_X(x_i, x_k) <D_X(x_i, x_j)$ is equivalent to $|x_k - x_i| <|x_j - x_i|$, and similarly for $y$, partition cells are simply axis-aligned rectangles, as in the distribution-free test. Here, however, only sample $j$ is a vertex, and sample $i$ is the center of mass. The diagonal-opposing vertex from $j$ may not even be a point in the sample, and thus is not directly captured by $A$. Still, the appropriate point to sample $A$ in, for computing the contingency table cell in $O(1)$, can be found in $O(1)$ additional amortized time, as follows:

\begin{enumerate}
  \item
  Sort once the unique values of $x$, and do the same for $y$.
  \item
  When traversing all pairs,  first traverse $i$.
  \begin{enumerate}
    \item
    For every $i$, traverse the sorted values of $x$ with two concurrent iterators starting from $x_i$, one moving right (i.e., from low to high $x$) and one advancing left.
    \item
    Subsequently to each step taken with the right iterator, arriving at an $x_j$, advance the left iterator until a value is encountered which is farther from $x_i$ than $x_j$ is, and this is the opposing vertex $x$ coordinate value for the rectangle for the pair $i, j$.
  \end{enumerate}
  The process, which is depicted in Figure~\ref{fig:hhg_n2_alg}, takes $O(N)$ time for $N$ values, and is repeated for the $y$ axis.
\end{enumerate}

\section{Additional simulations for the two-samples problem}\label{app:additional simulations}
Table \ref{tab:FisherVsMinP2sample} shows a comparison of the performance of the minimum $p$-value and Fisher-combined test statistics. The variant with highest power in most setups (specifically, setups 2,4,5,6,8,9,10,11,12) is the Fisher-combined $p$-value using aggregation by summation and $m_{\max} = 50$. However, the Fisher-combined $p$-value test is sensitive to the choice of $m_{\max}$, which is unknown in practice. Since we view this as a significant weakness of the Fisher-combined statistic, and since the minimum $p$-value does not have this weakness and has very good power when compared with Fisher as well as when compared with other tests (with a large range of $m_{\max}$ values examined), we recommend the minimum $p$-value test statistic.

\begin{table}[ht]
\caption{Power of the minimum $p$-value test statistic (columns 2--5) as well as the Fisher-combined $p$-value test statistic (columns 6--9),  using the different aggregation methods and two values of $m_{\max}$, for $N=500$ and the setups of Figure \ref{fig:setups2sample}. The difference between $m_{\max} = 50$ and $m_{\max} = 149$ is much larger for the Fisher-combined than for the minimum $p$-value test statistic when aggregated by summation (columns 8--9 versus columns 4--5). Moreover, the power is typically lower for the Fisher-combined that for the minimum $p$-value test statistic when the aggregation is by maximization (columns 6--7 versus columns 2--3).
 }\label{tab:FisherVsMinP2sample}
\scriptsize
\centering
\begin{tabular}{rrrrrrrrr}
  \hline
   &\multicolumn{4}{c}{Minimum $p$-value} & \multicolumn{4}{c}{Fisher combined $p$-value}  \\
      &\multicolumn{2}{c}{Max aggregation} & \multicolumn{2}{c}{Sum aggregation}       &\multicolumn{2}{c}{Max aggregation} & \multicolumn{2}{c}{Sum aggregation} \\
Setup &   $m_{\max}=50$ &  $m_{\max}=149$ &  $m_{\max}=50$ & $m_{\max}=149$  &  $m_{\max}=50$ &  $m_{\max}=149$ &  $m_{\max}=50$ & $m_{\max}=149$  \\
  \hline
1 & 0.836 & 0.825 & 0.500 & 0.491 & 0.631 & 0.424 & 0.549 & 0.560 \\ 
2 & 0.810 & 0.799 & 0.883 & 0.873 & 0.819 & 0.701 & 0.917 & 0.898 \\ 
3 & 0.783 & 0.785 & 0.553 & 0.733 & 0.867 & 0.847 & 0.502 & 0.782 \\ 
4 & 0.836 & 0.827 & 0.945 & 0.937 & 0.749 & 0.567 & 0.952 & 0.882 \\ 
5 & 0.607 & 0.592 & 0.708 & 0.686 & 0.531 & 0.381 & 0.760 & 0.655 \\ 
6 & 0.829 & 0.818 & 0.836 & 0.820 & 0.749 & 0.567 & 0.882 & 0.821 \\ 
7 & 0.358 & 0.339 & 0.516 & 0.492 & 0.201 & 0.135 & 0.430 & 0.287 \\ 
8 & 0.770 & 0.752 & 0.795 & 0.775 & 0.542 & 0.361 & 0.814 & 0.672 \\ 
9 & 0.531 & 0.512 & 0.641 & 0.613 & 0.344 & 0.227 & 0.635 & 0.470 \\ 
10  & 0.728 & 0.711 & 0.824 & 0.806 & 0.569 & 0.393 & 0.849 & 0.721 \\ 
11 & 0.556 & 0.540 & 0.707 & 0.686 & 0.509 & 0.380 & 0.756 & 0.657 \\ 
12 & 0.402 & 0.390 & 0.599 & 0.577 & 0.399 & 0.293 & 0.655 & 0.557 \\ 
13 & 0.852 & 0.844 & 0.777 & 0.764 & 0.809 & 0.663 & 0.823 & 0.822 \\ 
   \hline
\end{tabular}
\end{table}

Table \ref{tab:priors2sample} shows the power using different priors for regularization. The priors are as follows: $\pi(m) = \sqrt{N}^m exp(-\sqrt N)/m!$ for Poisson; $\pi(m) = \ibinom{N-1}{m-1}p^m(1-p)^{(N-m)}$ for Binomial, with $p=0.119$ so that the penalty becomes that of the Akaike information criterion (i.e., $\log \pi(\Ical|m)\pi(m) = -2m$); $\pi(m) =1/K$ for fixed $K$ for Uniform. We also considered the prior in \cite{Jiang14}, resulting in   the DS test with penalty $-\lambda_0\log N (m-1)$. According to the recommendation in \cite{Jiang14}, we chose  the value of $\lambda_0$ so that the level under the null is as close as possible to 0.05 from below, so $\lambda_0 = 1.11088$ for $N=100$, and $\lambda_0 = 0.904$ for $N=500$.  The Poisson prior (with rate $\sqrt N$) was by far the best among all priors considered, and its power was comparable to that of the minimum $p$-value displayed in Table \ref{tab:FisherVsMinP2sample}.

\begin{table}[ht]
\caption{Power using different priors , for $N=500$ and the setups of Figure \ref{fig:setups2sample}.
 }\label{tab:priors2sample}
\scriptsize
\centering
\begin{tabular}{rrrrrr}
  \hline
    & Poisson prior & Poisson prior & Uniform Prior  & Binomial Prior &  DS prior \\
 Setup & Max aggregation & Sum aggregation & Max aggregation & Max aggregation & Max aggregation \\
  \hline
1 & 0.829 & 0.597 & 0.848 & 0.450  & 0.845 \\
2 & 0.855 & 0.924 & 0.641 & 0.745  & 0.565 \\
3 & 0.715 & 0.639 & 0.282 & 0.874  & 0.239 \\
4 & 0.861 & 0.920 & 0.813 & 0.593  & 0.794 \\
5 & 0.659 & 0.746 & 0.592 & 0.411  & 0.573 \\
6 & 0.854 & 0.874 & 0.804 & 0.597  & 0.787 \\
7 & 0.291 & 0.310 & 0.430 & 0.139  & 0.432 \\
8 & 0.719 & 0.760 & 0.835 & 0.371  & 0.836 \\
9 & 0.489 & 0.545 & 0.599 & 0.232  & 0.600 \\
10  & 0.709 & 0.793 & 0.749 & 0.406  & 0.744 \\
11 & 0.607 & 0.720 & 0.466 & 0.390  & 0.433 \\
12 & 0.460 & 0.629 & 0.298 & 0.318  & 0.268 \\
13 & 0.860 & 0.857 & 0.802 & 0.697  & 0.776 \\
   \hline
\end{tabular}
\end{table}

\begin{table}[ht]
\caption{Power of competitors (columns 4--9), along with the minimum $p$-value using the $M_m$ $p$-values (column 1) and the $S_m$ $p$-values (column 2).}
\scriptsize
\centering
\begin{tabular}{cccccccccc}
  \hline
 & \multicolumn{2}{c}{Minimum $p$-value} & & &  &  &  &  \\
N &Setup&  Max aggregation & Sum Aggregation & Wilcoxon & KS & CVM & AD & HHG & DS \\
  \hline
500 & Normal shift & 0.59 & 0.82 &  \underline{0.91} & 0.81 & 0.89 & 0.90 & 0.85 & 0.73 \\ 
500&   Normal scale & 0.76 & 0.85 & 0.05 & 0.36 & 0.46 & 0.77 &  \underline{0.88} & 0.83 \\ 
500 &   Normal shift \& scale & 0.83 & 0.90 & 0.46 & 0.69 & 0.74 & 0.88 &  \underline{0.92} & 0.90 \\ 
   \hline
100&   Normal shift & 0.39 & 0.58 & \underline{0.68} & 0.53 & 0.65 & 0.67 & 0.60 & 0.49 \\ 
100&  Normal scale & 0.51 & 0.59 & 0.06 & 0.20 & 0.23 & 0.41 & \underline{0.63} & 0.54 \\ 
100&  Normal shift \& scale & 0.54 & 0.64 & 0.49 & 0.51 & 0.58 & 0.64 & \underline{0.68} & 0.62 \\ 

\end{tabular}
\end{table}

\begin{figure}[htbp]
  \centering
  \includegraphics[page=1, width=1\textwidth, trim=0in 0in 0in 0in, clip]{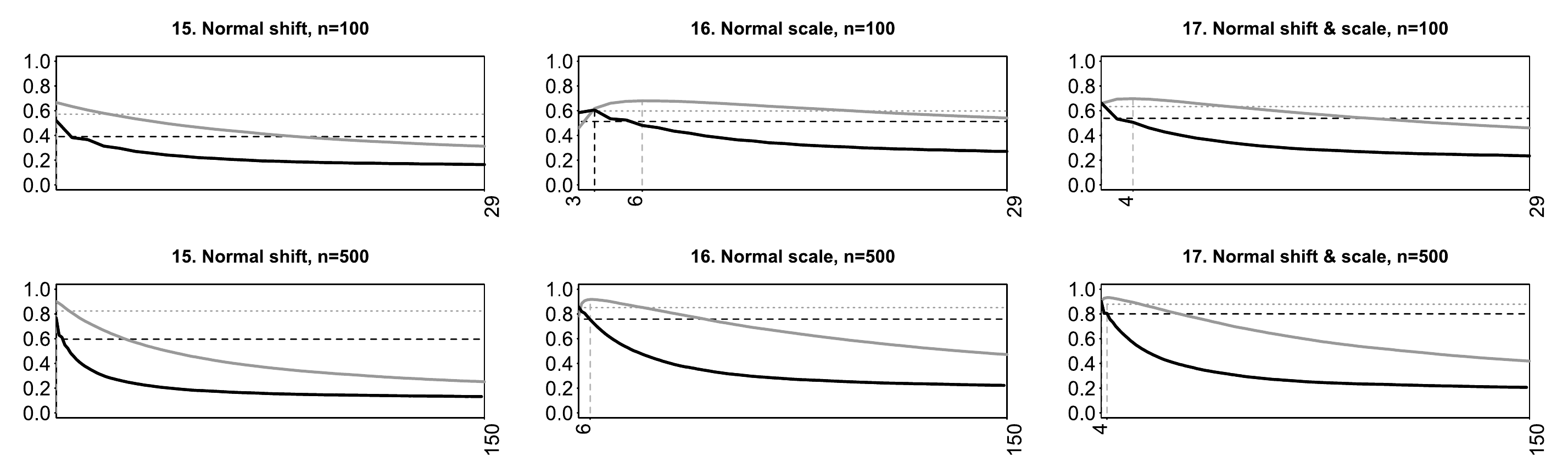}
  \caption{Estimated power  for the $M_m$ (black) and $S_m$ (grey) statistics for $m\in \{2,\ldots, 149\}$  for the Gaussian shift difference (first column, $N(0,1)$ versus $N(0.5,1)$ for $N=100$, and versus $N(0.3,1)$ for $N=500$), Gaussian scale difference (second column, $N(0,1)$ versus $N(0, 0.6^2)$ for $N=100$, and versus $N(0, 0.75^2)$ for $N=500$), and for the Gaussian shift and scale difference (third column, $N(0,1)$ versus $N(0.36, 0.7^2)$ for $N=100$, and versus $N(0.2, 0.8^2)$ for $N=500$). The power of the minimum $p$-value is the horizontal dashed black line when it combines the $p$-values based on $M_m$, and the horizontal dotted grey line when it combines the $p$-values based on $S_m$. The vertical lines show the optimal $m$ for $M_m$ (grey) and $S_m$ (black).}
  \label{fig:pwr2sampleGaussian}
\end{figure}

\begin{figure}[htbp]
  \centering
  \includegraphics[page=1, width=0.8\textwidth, trim=0in 0in 0in 0in, clip]{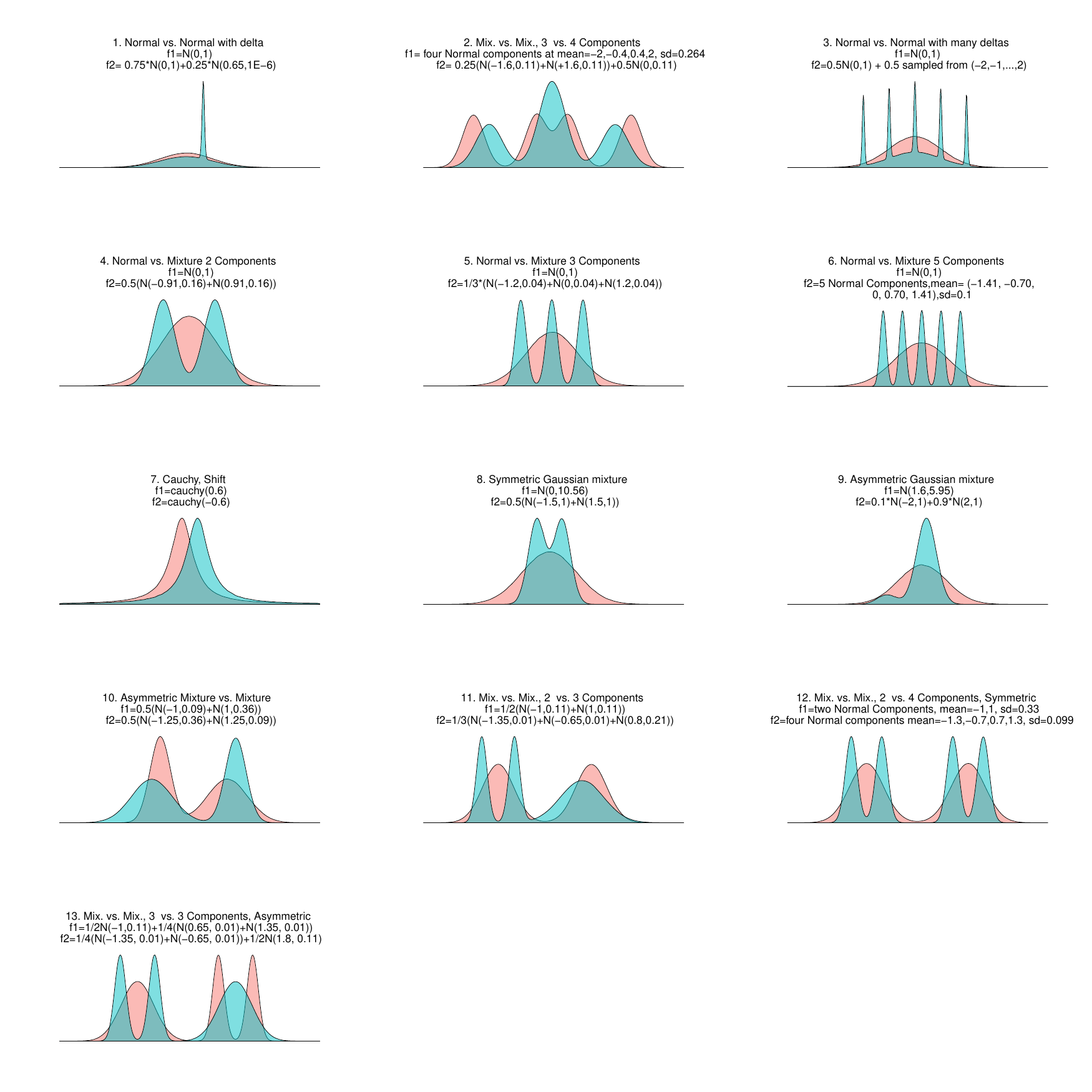}
  \caption{The two distributions in 13 different setups considered for $N=100$, which differ in the number of intersections of the densities, the range of support where the differences lie, and the whether they are symmetric or not.}
  \label{fig:setups2sampleN100}
\end{figure}
\begin{figure}[htbp]
  \centering
  \includegraphics[page=1, width=1\textwidth, trim=0in 0in 0in 0in, clip]{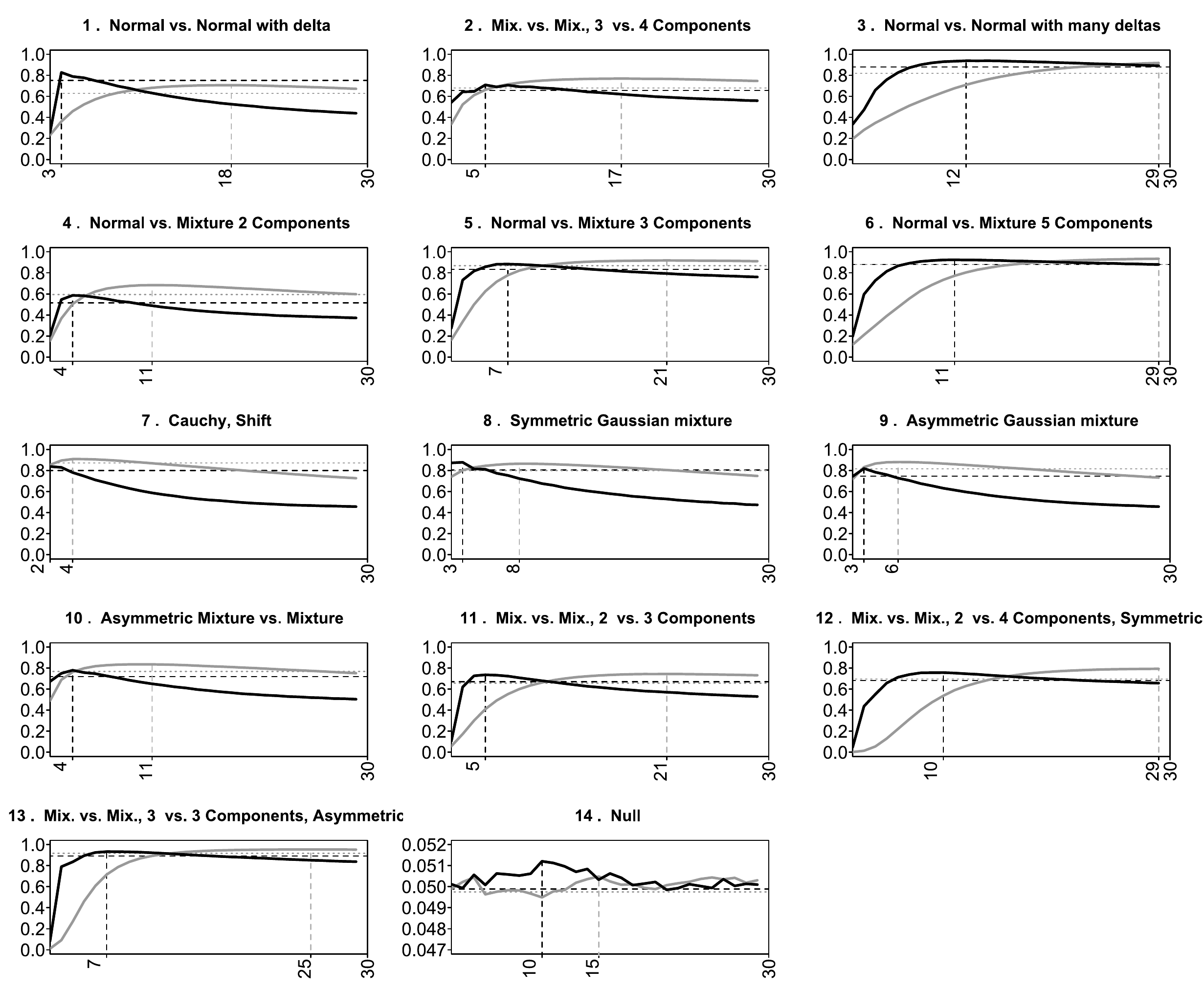}
  \caption{Estimated power with $N=100$ sample points for the $M_m$ (black) and $S_m$ (grey) statistics for $m\in \{2,\ldots, 29\}$  for the setups of Figure~\ref{fig:setups2sampleN100}. The power of the minimum $p$-value is the horizontal dashed black line when it combines the $p$-values based on $M_m$, and the horizontal dotted grey line when it combines the $p$-values based on $S_m$. The vertical lines show the optimal $m$ for $M_m$ (grey) and $S_m$ (black).}
  \label{fig:pwr2sampleN100}
\end{figure}

For sample size $N=100$, we examined the distributions depicted in Figure~\ref{fig:setups2sampleN100}, and we used 20000 simulated data sets, in each of the configurations. 
Table \ref{tab:competitors2sampleN100} and Figure \ref{fig:pwr2sampleN100} show the power for the setups in Figure \ref{fig:setups2sampleN100}. 
These results concur with the results in Section \ref{subsec-sim2sample} for $N=500$,  and show that if the number of intersections of the two distributions is at least four, tests statistics with $m\geq 4$ have an advantage.

\begin{table}[ht]
\caption{Power of competitors (columns 4--9), along with the minimum $p$-value statistic using the $M_m$ $p$-values (column 1) and the $S_m$ $p$-values (column 2), for $N=100$. The standard error was at most 0.0035. The advantage of the test based on the minimum $p$-value is large when the number of intersections of the two densities is at least four (setups of rows 2,3, 4,5,6,10,11,12, and 13). The best competitors are HHG and DS, but HHG is essentially an $m\leq 3$ test, and DS penalizes large $m$s severely, therefore in setups where $m\geq 4$ partitions are better they can perform poorly. Among the two variants in columns 1 and 2, the better choice clearly depends on the range of support in which the differences in distributions occur:  aggregation by maximum has better power when the difference between the distributions is very local (setups of rows 1 and 3), and aggregation by summation has better power otherwise. The highest power per row is underlined. }\label{tab:competitors2sampleN100}
\scriptsize
\centering
\begin{tabular}{llcccccccc}
  \hline
 & &\multicolumn{2}{c}{Min $p$-value aggreg. } &  &  &  &  &  &  \\
& Setup &by Max & by Sum& Wilcoxon & KS & CVM & AD & HHG & DS \\
  \hline
1 & Normal vs. Normal with delta & \underline{0.752} & 0.628  & 0.187 & 0.331 & 0.294 & 0.265 & 0.433 & 0.660 \\ 
2 & Mix. Vs. Mix., 3  Vs. 4 Components  & 0.656 & \underline{0.680}  & 0.000 & 0.003 & 0.000 & 0.063 & 0.465 & 0.505 \\ 
3 & Normal vs. Normal with many deltas  & \underline{0.878} & 0.819  & 0.053 & 0.125 & 0.100 & 0.165 & 0.281 & 0.342 \\ 
4 & Normal vs. Mixture 2 Components  & 0.515 & \underline{0.595}  & 0.052 & 0.231 & 0.153 & 0.154 & 0.397 & 0.382 \\ 
5 & Normal vs. Mixture 3 Components  & 0.834 & \underline{0.869}  & 0.053 & 0.177 & 0.106 & 0.134 & 0.317 & 0.534 \\ 
6 & Normal vs. Mixture 5 Components & 0.879 & \underline{0.880}  & 0.051 & 0.123 & 0.084 & 0.100 & 0.200 & 0.373 \\ 
7 & Cauchy, Shift & 0.799 & 0.871  & 0.847 & 0.917 & 0.920 & 0.893 & \underline{0.933} & 0.846 \\ 
8 & Symmetric Gaussian mixture & 0.803 & 0.798  & 0.035 & 0.182 & 0.191 & 0.491 & 0.727 & \underline{0.812} \\ 
9 & Asymmetric Gaussian mixture & 0.747 & 0.816  & 0.046 & 0.369 & 0.407 & 0.593 & \underline{0.845} & 0.740 \\ 
 10 & Asymmetric Mixture vs. Mixture  & 0.718 & \underline{0.769}  & 0.000 & 0.157 & 0.128 & 0.306 & 0.655 & 0.652 \\ 
11 & Mix. Vs. Mix., 2  Vs. 3 Components& \underline{0.670} & 0.656  & 0.000 & 0.032 & 0.011 & 0.031 & 0.129 & 0.441 \\ 
12 & Mix. Vs. Mix., 2  Vs. 4 Components, Symmetric & 0.682 & \underline{0.696}  & 0.000 & 0.000 & 0.000 & 0.000 & 0.005 & 0.238 \\ 
13 & Mix. Vs. Mix., 3  Vs. 3 Components, Asymmetric & 0.891 & \underline{0.917}  & 0.000 & 0.000 & 0.000 & 0.000 & 0.053 & 0.575 \\ 
14 & Null & 0.050 & 0.050  & 0.049 & 0.039 & 0.049 & 0.050 & 0.050 & 0.041 \\    
\hline   
\end{tabular}
\end{table}

\section{Simulations of monotone relationships}
\label{sec:mono_simulations}
\begin{figure}[htbp]
  \centering
  \includegraphics[page=1, width=1.0\textwidth, trim=0in 0in 0in 0in, clip]{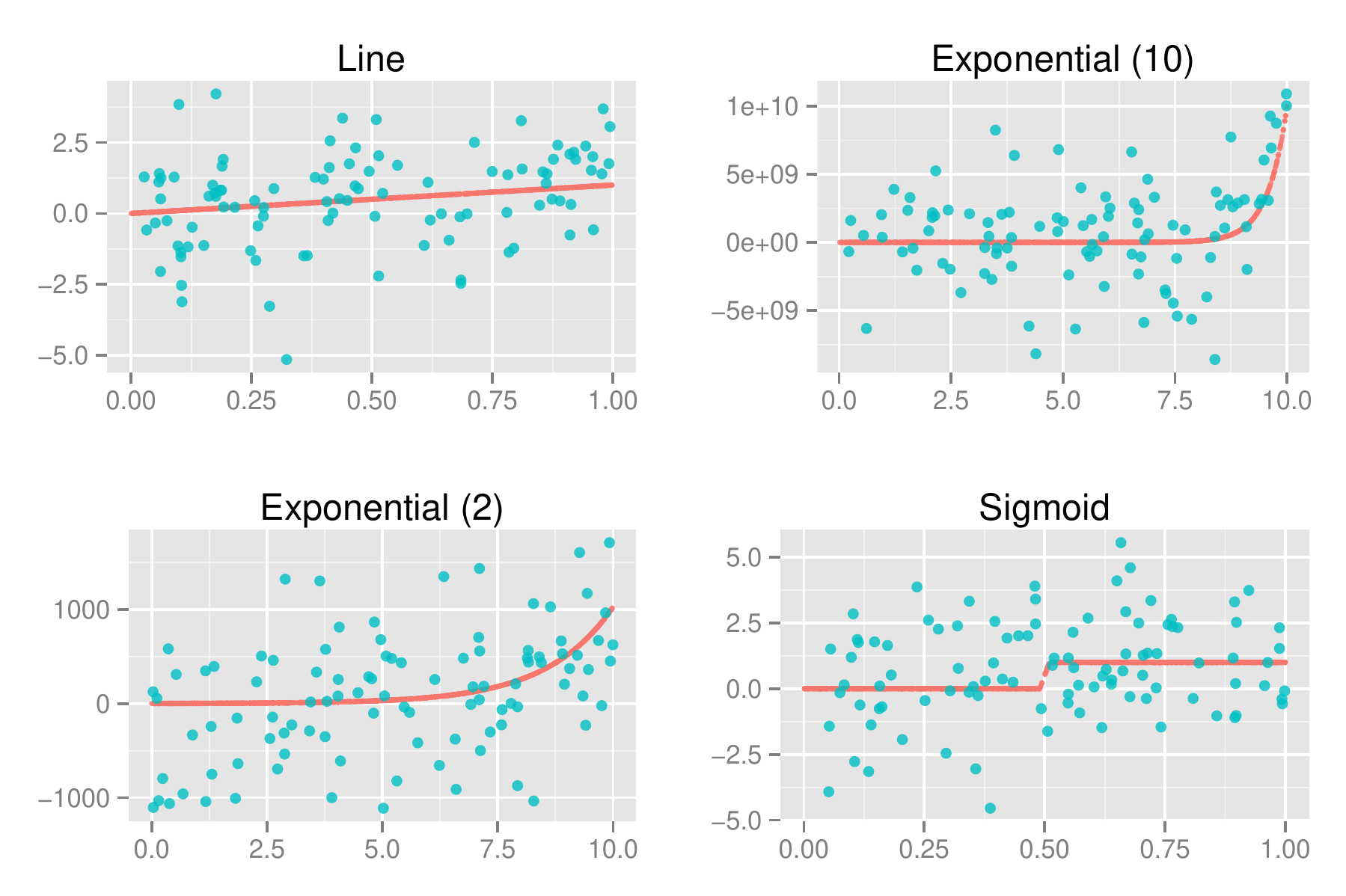}
  \caption[Monotonic simulation setups]{Bivariate monotone relationships (in red), along with a sample of $N=100$ noisy observations (in blue).}
  \label{fig:mic_setups_mono}
\end{figure}

\begin{figure}[htbp]
  \centering
  \includegraphics{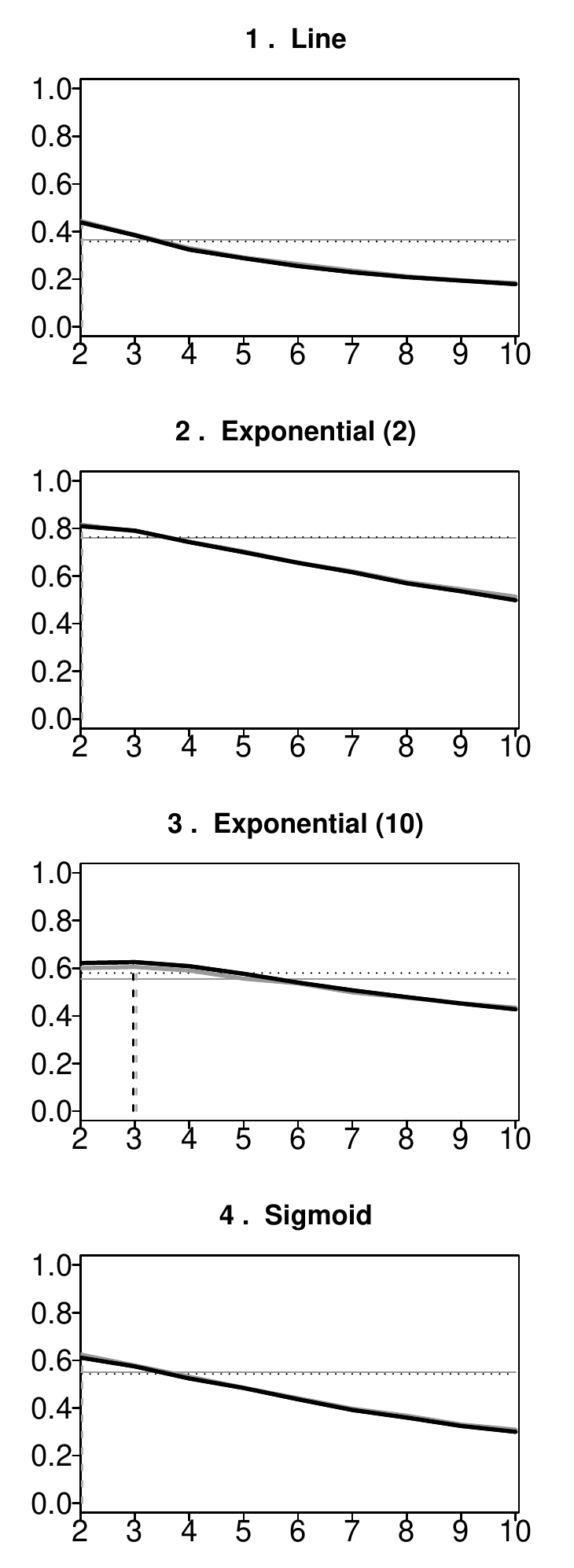}
  \caption[Monotonic simulation results]{Estimated power for the DDP (black) and ADP (gray) summation variants, using the likelihood ratio score, for the setups from Figure~\ref{fig:mic_setups_mono}, for a sample of size $N=100$. The horizontal lines are the power of $\min_{m \in \{2,\ldots, m_{\max} \}}p_m$ using DDP (dotted black) and using ADP (gray)
   .}
  \label{fig:mic_results_mono}
\end{figure}

Four monotone relationships are presented in Figure~\ref{fig:mic_setups_mono}. Figure~\ref{fig:mic_results_mono} shows that the differences for the summation variants between ADP and DDP are negligible, and that the power decreases with $m$ for all of these variants. Similar conclusions hold for the maximum variants in Table~\ref{tbl:mono}. Table~\ref{tbl:mono} further shows that
for a linear relationship, Pearson, Spearman, Hoeffding and dCov are clearly superior, but for the remaining three monotone relationships $\min_{m \in \{2,\ldots, m_{\max} \}}p_m$ has quite good power properties.

\begin{table}[ht]
  \caption[Monotonic simulation results]{The power of competitors (rows 3--7), along with the minimum  $p$-value statistic based on DDP (row 1) and on ADP (row 2)  and the different maximum variants (rows 8--11), for $N=100$. The standard error is at most $0.011$.}
  \label{tbl:mono}
  \centering
  \scriptsize
  \begin{tabular}{rrrrrr}
Test   & Line & Exp2x & Exp10x & Sigmoid \\
  \hline
$\min_{m \in \{2,\ldots, 10 \}}p_m$ using DDP & 0.358 & 0.763 & 0.580 & 0.543 \\
$\min_{m \in \{2,\ldots, 10 \}}p_m$ using ADP & 0.365 & 0.760 & 0.555 & 0.550 \\
  Spearman & 0.459 & 0.758 & 0.396 & 0.630 \\
  Hoeffding & 0.446 & 0.750 & 0.409 & 0.637 \\
  MIC & 0.282 & 0.198 & 0.312 & 0.130 \\
  dCov & 0.433 & 0.746 & 0.395 & 0.637 \\
  HHG & 0.337 & 0.706 & 0.509 & 0.545 \\
 $M_{2\times 2}^{DDP}$ & 0.287 & 0.688 & 0.678 & 0.438 \\
  $M_{3\times 3}^{DDP}$ & 0.203 & 0.579 & 0.569 & 0.355 \\
 $M_{4\times 4}^{DDP}$ & 0.177 & 0.511 & 0.479 & 0.301 \\
 $M_{2\times 2}^{ADP}$ & 0.294 & 0.715 & 0.746 & 0.440 \\
  \hline
  \end{tabular}
\end{table}

\section{Comparisons of dCov/HHG on ranks and on data}\label{app-dcovHHGondata}
In Section \ref{sec:simulation} we only considered distribution-free competitors. Therefore, we applied dCov and HHG on ranks rather than on data, even though they were designed as permutation tests on data. Usually, the computational advantage of performing the tests on $(rank(X), rank(Y))$ instead of on $(X,Y)$, due to the distribution-free property of the tests on $(rank(X), rank(Y))$, comes at a cost of lower power, as noted by \cite{Szekely07}. Table \ref{tab:competitorsOndata} shows a power comparison of these two permutation tests on ranks and on data. The results on ranks are not identical numerically (though very close) to those of Table \ref{tab:competitors} due to the use of a different seed to generate the data for these same settings. The power in most settings is indeed greater when the tests are used on data, and the maximal difference is almost 30\% (in the Spiral setting for HHG) in favour of using the data. However, in some settings the power is actually larger for the test on ranks, e.g., in the  Heavisine and 5Clouds settings for HHG, where the difference is 10\% and 
16\%, respectively, in favour of using the ranked observations. Comparing the power of the HHG and dCov tests on data (in Table \ref{tab:competitorsOndata}) to the power of our suggested minimum $p$-value statistic (in Table \ref{tab:competitors}), we see that although dCov on data may have more power than dCov on ranks, it has far less power than HHG on data, and that HHG on data has less power than our test when the relationship is more complex, especially in the Sine, Heavisine, Spiral and Circles examples. 

\begin{table}
 \caption{Power of the competitors dCov and HHG on ranks as well as on data for $N=100$.   The tests on data have greater power than on ranks in most, but not all, examples. HHG on data has the best power out of these four competitors, but like the HHG on ranks it has a disadvantage in comparison with our novel  minimum $p$-value test when the relationship is more complex,  especially in the Sine, Heavisine, Spiral, Circles, and 5Clouds examples. }
  \label{tab:competitorsOndata}
  \scriptsize
  \centering
  \vspace{0.5cm}
  \begin{tabular}{lcccc}
   & \multicolumn{2}{c}{dCov} & \multicolumn{2}{c}{HHG}\\
   Setup & on data  & on ranks & on data & on ranks   \\
  & & & &   \\
W & 0.467 &  0.351 & 0.876 & 0.813 \\
  Diamond & 0.136 &  0.076 & 0.997 & 0.968 \\
  Parabola &  0.418 & 0.369 & 0.727 & 0.795 \\
  2Parabolas & 0.183 & 0.120 & 0.873 & 0.726 \\
  Circle &  0.001 & 0.003 & 0.791 & 0.858 \\
  Cubic &  0.631 &  0.612 & 0.660 & 0.742 \\
  Sine & 0.412 & 0.427 & 0.803 & 0.788 \\
  Wedge & 0.471 & 0.327 & 0.755 & 0.661 \\
  Cross & 0.184  & 0.138 & 0.795 & 0.704 \\
  Spiral & 0.182 &  0.130 & 0.598 & 0.335 \\
  Circles & 0.054 &  0.059 & 0.479 & 0.357 \\
  Heavisine &  0.476 & 0.470 & 0.471 & 0.570 \\
  Doppler &  0.747 & 0.736 & 0.901 & 0.914 \\
  5Clouds & 0.000 & 0.001 & 0.738 & 0.903 \\
  4Clouds &  0.050 & 0.050 & 0.050 & 0.050 \\
  \end{tabular}
\end{table}


\begin{thebibliography}{35}
\providecommand{\natexlab}[1]{#1}
\providecommand{\url}[1]{\texttt{#1}}
\expandafter\ifx\csname urlstyle\endcsname\relax
  \providecommand{\doi}[1]{doi: #1}\else
  \providecommand{\doi}{doi: \begingroup \urlstyle{rm}\Url}\fi

\bibitem[Anderson and Darling(1952)]{Anderson52}
T.W. Anderson and D.A. Darling.
\newblock Asymptotic theory of certain "goodness of fit" criteria based on
  stochastic processes.
\newblock \emph{The Annals of Mathematical Statistics}, 23\penalty0
  (2):\penalty0 193--212, 1952.

\bibitem[Baringhaus and Franz(2004)]{Baringhaus04}
L.~Baringhaus and C.~Franz.
\newblock On a new multivariate two-sample test.
\newblock \emph{Journal of Multivariate Analysis}, 88:\penalty0 190--206, 2004.

\bibitem[Benjamini and Hochberg(1995)]{yoav1}
Y.~Benjamini and Y.~Hochberg.
\newblock Controlling the false discovery rate - a practical and powerful
  approach to multiple testing.
\newblock \emph{J. Roy. Stat. Soc. B Met.}, 57 (1):\penalty0 289--300, 1995.

\bibitem[Blum et~al.(1961)Blum, Kiefer, and Rosenblatt]{blum1961distribution}
J.~Blum, J.~Kiefer, and M.~Rosenblatt.
\newblock Distribution free tests of independence based on the sample
  distribution function.
\newblock \emph{The Annals of Mathematical Statistics}, pages 485--498, 1961.

\bibitem[Darling(1957)]{Darling57}
D.A. Darling.
\newblock The kolmogorov-smirnov, cramer-von mises tests.
\newblock \emph{The Annals of Mathematical Statistics}, 28\penalty0
  (4):\penalty0 823--838, 1957.

\bibitem[Donoho and Johnstone(1995)]{Donoho95}
D.~Donoho and I.~Johnstone.
\newblock Adapting to unknown smoothness via wavelet shrinkage.
\newblock \emph{Journal of the American Statistical Association}, 90\penalty0
  (432):\penalty0 1200--1224, 1995.

\bibitem[Feuerverger(1993)]{Feuerverger93}
A.~Feuerverger.
\newblock A consistent test for bivariate dependence.
\newblock \emph{International Statistical Review}, 61\penalty0 (3):\penalty0
  419--433, 1993.

\bibitem[Gorfine et~al.(2011)Gorfine, Heller, and Heller]{Gorfine11}
M.~Gorfine, R.~Heller, and Y.~Heller.
\newblock Comment on ‘detecting novel associations in large data sets’ by
  reshef et al., science.
\newblock \emph{Available: \url{http://iew3 technion ac
  il/~gorfinm/files/science6.pdf}}, 2011.

\bibitem[Gretton and Gyorfi(2010)]{Gretton10}
A.~Gretton and L.~Gyorfi.
\newblock Consistent nonparametric tests of independence.
\newblock \emph{Journal of Machine Learning Research}, 11:\penalty0 1391--1423,
  2010.

\bibitem[Gretton et~al.(2007)Gretton, Bogwardt, Rasch, Scholkopf, and
  Smola]{Gretton06}
A.~Gretton, K.M. Bogwardt, M.J. Rasch, B.~Scholkopf, and A.~Smola.
\newblock A kernel method for the two-sample problem.
\newblock \emph{Advances in Neural Information Processing Systems (NIPS)}, 19,
  2007.

\bibitem[Gretton et~al.(2008)Gretton, Fukumizu, TEO, Song, Scholkopf, and
  Smola]{Gretton08}
A.~Gretton, K.~Fukumizu, C.H. TEO, L.~Song, B.~Scholkopf, and A.~Smola.
\newblock A kernel statistical test of independence.
\newblock \emph{Advances in Neural Information Processing Systems},
  20:\penalty0 585--592, 2008.

\bibitem[Gretton et~al.(2012)Gretton, Borgwardt, Rasch, Sch{�}lkopf, and
  Smola]{Gretton12b}
A.~Gretton, K.M. Borgwardt, M.J. Rasch, Sch{�}lkopf, and A.~Smola.
\newblock A kernel two-sample test.
\newblock \emph{The Journal of Machine Learning Research}, 13:\penalty0
  723--773, 2012.

\bibitem[Harchaoui et~al.(2008)Harchaoui, Bach, and Moulines]{Harouchi08}
Z~Harchaoui, F.~Bach, and E.~Moulines.
\newblock Testing for homogeneity with kernel fisher discriminant analysis.
\newblock \emph{Advances in Neural Information Processing Systems (NIPS), long
  version: arXiv:0804.1026v1,}, 20:\penalty0 609--616, 2008.

\bibitem[Heller et~al.(2013)Heller, Heller, and Gorfine]{Heller12}
R.~Heller, Y.~Heller, and M.~Gorfine.
\newblock A consistent multivariate test of association based on ranks of
  distances.
\newblock \emph{Biometrika}, 100(2):\penalty0 503--510, 2013.

\bibitem[Hoeffding(1948)]{hoeffding1948non}
W.~Hoeffding.
\newblock A non-parametric test of independence.
\newblock \emph{The Annals of Mathematical Statistics}, 19\penalty0
  (4):\penalty0 546--557, 1948.

\bibitem[Hughes et~al.(2000)Hughes, Marton, Jones, Roberts, Stoughton, Armour,
  Bennett, Coffey, Dai, He, et~al.]{hughes2000functional}
T.R. Hughes, M.J. Marton, A.R. Jones, C.J. Roberts, R.~Stoughton, C.D. Armour,
  H.A. Bennett, E.~Coffey, H.~Dai, Y.D. He, et~al.
\newblock Functional discovery via a compendium of expression profiles.
\newblock \emph{Cell}, 102\penalty0 (1):\penalty0 109--126, 2000.

\bibitem[Jiang et~al.(2014)Jiang, Ye, and Liu]{Jiang14}
B.~Jiang, C.~Ye, and J.~Liu.
\newblock Non-parametric k-sample tests via dynamic slicing.
\newblock \emph{Journal of the American Statistical Association},
  DOI:10.1080/01621459.2014.920257, 2014.

\bibitem[Kinney and Atwal(2014)]{kinney2013equitability}
J.~Kinney and G.~Atwal.
\newblock Equitability, mutual information, and the maximal information
  coefficient.
\newblock \emph{Proceedings of the national academy of sciences of the USA},
  doi:10.1073, 2014.

\bibitem[Laurent and Massart(2000)]{laurent2000adaptive}
B.~Laurent and P.~Massart.
\newblock Adaptive estimation of a quadratic functional by model selection.
\newblock \emph{The Annals of Statistics}, 28\penalty0 (5):\penalty0
  1302--1338, 2000.

\bibitem[Lehmann and Romano(2005)]{Lehmann05}
E.L. Lehmann and J.P. Romano.
\newblock \emph{Testing Statistical Hypotheses, 3rd Edition}.
\newblock Springer, New York, 2005.

\bibitem[Newton(2009)]{Newton09}
M.~Newton.
\newblock Introducing the discussion paper by {S}zekely and {R}izzo.
\newblock \emph{The Annals of Applied Statistics}, 3 (4):\penalty0 1233--1235,
  2009.

\bibitem[Paninski(2003)]{paninski2003estimation}
L.~Paninski.
\newblock Estimation of entropy and mutual information.
\newblock \emph{Neural Computation}, 15\penalty0 (6):\penalty0 1191--1253,
  2003.

\bibitem[Pettitt(1976)]{Pettitt76}
A.N. Pettitt.
\newblock A two-sample anderson--darling rank statistic.
\newblock \emph{Biometrika}, 63\penalty0 (1):\penalty0 161--168, 1976.

\bibitem[Reshef et~al.(2011)Reshef, Reshef, Finucane, Grossman, McVean,
  Turnbaugh, Lander, Mitzenmacher, and Sabeti]{Reshef11}
D.N. Reshef, Y.A. Reshef, H.K. Finucane, S.R. Grossman, G.~McVean, P.J.
  Turnbaugh, E.S. Lander, M.~Mitzenmacher, and P.C. Sabeti.
\newblock Detecting novel associations in large data sets.
\newblock \emph{Science}, 334\penalty0 (6062):\penalty0 1518--1524, 2011.

\bibitem[Scholz and Stephens(1987)]{scholz87}
F.W. Scholz and M.A. Stephens.
\newblock K-sample anderson-darling tests.
\newblock \emph{Journal of the American Statistical Association}, 82\penalty0
  (399):\penalty0 918--924, 1987.

\bibitem[Sejdinovic et~al.(2013)Sejdinovic, Sriperumbudur, Gretton, and
  Fukumizu]{Sejdinovic12}
D.~Sejdinovic, B.~Sriperumbudur, A.~Gretton, and K.~Fukumizu.
\newblock Equivalence of distance-based and rkhs-based statistics in hypothesis
  testing.
\newblock \emph{Annals of Statistics}, 41 (5):\penalty0 2263--2291, 2013.

\bibitem[Simon and Tibshirani(2011)]{Simon11}
N.~Simon and R.~Tibshirani.
\newblock Comment on ‘detecting novel associations in large data sets’ by
  reshef et al., science.
\newblock \emph{arXiv:1401.7645}, 2011.

\bibitem[Steuer et~al.(2002)Steuer, Kurths, Daub, Weise, and
  Selbig]{steuer2002mutual}
R.~Steuer, J.~Kurths, C.~Daub, J.~Weise, and J.~Selbig.
\newblock The mutual information: detecting and evaluating dependencies between
  variables.
\newblock \emph{Bioinformatics}, 18\penalty0 (suppl 2):\penalty0 S231--S240,
  2002.

\bibitem[Sz{\'e}kely and Rizzo(2004)]{Szekely05}
G.~Sz{\'e}kely and M.~Rizzo.
\newblock Testing for equal distributions in high dimensions.
\newblock \emph{InterStat}, 2004.

\bibitem[Sz{\'e}kely and Rizzo(2009)]{Szekely09}
G.~Sz{\'e}kely and M.~Rizzo.
\newblock Brownian distance covariance.
\newblock \emph{The Annals of Applied Statistics}, 3 (4):\penalty0 1236--1265,
  2009.

\bibitem[Sz{\'e}kely et~al.(2007)Sz{\'e}kely, Rizzo, and Bakirov]{Szekely07}
G.~Sz{\'e}kely, M.~Rizzo, and N.~Bakirov.
\newblock Measuring and testing dependence by correlation of distances.
\newblock \emph{The Annals of Statistics}, 35:\penalty0 2769--2794, 2007.

\bibitem[Thas and Ottoy(2007)]{Thas07}
O.~Thas and J.~Ottoy.
\newblock An extension of the anderson-darling k-sample test to arbitrary
  sample space partition sizes.
\newblock \emph{Journal of Statistical Computation and Simulation}, 74\penalty0
  (9):\penalty0 651--665, 2007.

\bibitem[Thas and Ottoy(2004)]{Thas04}
O.~Thas and J.P. Ottoy.
\newblock A nonparamteric test for independence based on sample space
  partitions.
\newblock \emph{Communcations in Statistics - Simulation and Computation}, 33
  (3):\penalty0 711--728, 2004.

\bibitem[Vu et~al.(2007)Vu, Yu, and Kass]{vu2007}
Q.~Vu, B.~Yu, and R.~Kass.
\newblock Coverage-adjusted entropy estimation.
\newblock \emph{Statistics in medicine}, 26:\penalty0 4039--4060, 2007.

\bibitem[Yu et~al.(2011)Yu, Liang, Ciampa, and Chatterjee]{yu2011efficient}
K.~Yu, F.~Liang, J.~Ciampa, and N.~Chatterjee.
\newblock Efficient p-value evaluation for resampling-based tests.
\newblock \emph{Biostatistics}, 12\penalty0 (3):\penalty0 582--593, 2011.

\end{thebibliography}
\end{document}